\pdfoutput=1
\documentclass[usenatbib]{mn2e}
\usepackage{hyperref}
\usepackage{apjfonts}
\usepackage{amssymb}
\usepackage{amsmath}
\usepackage{ctable}
\usepackage{url}
\usepackage{breakurl}
\usepackage{fixltx2e} 

\newcommand{\be}{\begin{equation}}
\newcommand{\ee}{\end{equation}}

\newcommand{\msun}{M_{\sun}}

\newcommand{\paperone}{Paper {\small I}}

\newcommand{\gizmourl}{\burl{http://www.tapir.caltech.edu/~phopkins/Site/GIZMO.html}}
\newcommand{\vspacerpostplot}{\vspace{-0.4cm}}
\newcommand{\NNb}{N_{\rm NGB}}
\newcommand{\Ddim}{\nu}
\newcommand{\divB}{\nabla\cdot{\bf B}}
\newcommand{\divb}{\divB}
\newcommand{\comment}[1]{}

\newcommand\plotonesize[2]
 {\centering \leavevmode \includegraphics[width={#2\columnwidth}]{#1}}
\newcommand{\plotsidesize}[2]
 {\centering \leavevmode \includegraphics[width={#2\textwidth}]{#1}}
\newcommand{\acknowledgments}{\begin{small}\section*{Acknowledgments}\end{small}}
\newcommand\altaffilmark[1]{$^{#1}$}
\newcommand\altaffiltext[1]{$^{#1}$}
\title[Meshless MHD]{Accurate, Meshless Methods for Magnetohydrodynamics\vspace{-0.5cm}}

\author[Hopkins \&\ Raives]{
\parbox[t]{\textwidth}{ 
Philip F. Hopkins\altaffilmark{1}\thanks{E-mail:phopkins@caltech.edu} \&\ Matthias J.\ Raives\altaffilmark{1}
} 
\vspace*{6pt} \\
\altaffiltext{1}{TAPIR \&\ The Walter Burke Institute for Theoretical Physics, Mailcode 350-17, California Institute of Technology, Pasadena, CA 91125, USA\vspace{-1.1cm}} \\
}

\date{Submitted to MNRAS, April, 2015\vspace{-0.6cm}}
\begin{document}
\maketitle
\label{firstpage}

\begin{abstract}
Recently, we explored new, meshless finite-volume Lagrangian methods for hydrodynamics: the ``meshless finite mass'' (MFM) and ``meshless finite volume'' (MFV) methods; these capture advantages of both smoothed-particle hydrodynamics (SPH) and adaptive mesh-refinement (AMR) schemes. We extend these to include ideal magnetohydrodynamics (MHD). The MHD equations are second-order consistent and conservative. We augment these with a divergence-cleaning scheme, which maintains $\nabla\cdot{\bf B}\approx0$. We implement these in the code {\small GIZMO}, together with state-of-the-art SPH MHD. We consider a large test suite, and show that on all problems the new methods are competitive with AMR using constrained transport (CT) to ensure $\nabla\cdot{\bf B}=0$. They correctly capture the growth/structure of the magnetorotational instability (MRI), MHD turbulence, and launching of magnetic jets, in some cases converging more rapidly than state-of-the-art AMR. Compared to SPH, the MFM/MFV methods exhibit convergence at fixed neighbor number, sharp shock-capturing, and dramatically reduced noise, divergence errors, \&\ diffusion. Still, ``modern'' SPH can handle most test problems, at the cost of larger kernels and ``by hand'' adjustment of artificial diffusion. Compared to non-moving meshes, the new methods exhibit enhanced ``grid noise'' but reduced advection errors and diffusion, easily include self-gravity, and feature velocity-independent errors and superior angular momentum conservation. They converge more slowly on some problems (smooth, slow-moving flows), but more rapidly on others (involving advection/rotation). In all cases, we show divergence-control beyond the Powell 8-wave approach is necessary, or all methods can converge to unphysical answers even at high resolution.
\end{abstract}

\begin{keywords}
methods: numerical --- hydrodynamics --- instabilities --- turbulence --- cosmology: theory
\vspace{-1.0cm}
\end{keywords}

\vspace{-1.1cm}
\section{Introduction: The Challenge of Existing Numerical Methods}
\label{sec:intro}

Magnetic fields are an essential component in astrophysical hydrodynamics, and for many astrophysical problems can be reasonably approximated by ideal (infinite conductivity) magnetohydrodynamics (MHD). The MHD equations are inherently non-linear, however, and most problems require numerical simulations. But this poses unique challenges, especially for numerical methods which are Lagrangian (i.e.\ the mesh elements follow the fluid), rather than Eulerian (solved on a fixed grid). 

In most discretizations \citep[although see][]{kawai:divb.free.cell.centered}, evolving the MHD equations in time will lead to a violation of the ``divergence constraint'' (the requirement that $\divb=0$). Unfortunately, this cannot simply be ignored to treated as a ``standard'' numerical error term which should converge away with increasing resolution, because certain errors introduced by a non-zero $\divb$ are numerically unstable: they will eventually destroy the correct solution (even at infinite resolution) and/or produce unphysical results (e.g.\ negative pressures). Arguably the most elegant solution is the so-called ``constrained transport'' (CT) method of \citet{evans:1988.constrained.transport}, which maintains $\divB$ to machine precision; however, while there is no obvious barrier {\em in principle} to implementing this in meshless and unstructured mesh methods \citep[see recent developments by][]{mocz:2014.constrained.transport.mhd}, it has thus far only been practical to implement for real problems in regular, Cartesian grid (or adaptive-mesh refinement; AMR) codes. But for many problems in astrophysics, Lagrangian, mesh-free codes have other advantages: they minimize numerical diffusion and over-mixing, move with the fluid so automatically provide enhanced resolution with the mass (in a continuous manner, which avoids low-order errors necessarily introduced by AMR refinement boundaries), couple simply and accurately to cosmological expansion and $N$-body gravity codes, easily handle high Mach numbers, conserve angular momentum and naturally handle orbiting disks without prior knowledge of the disk geometry, avoid ``grid alignment'' and carbuncle instabilities (where the grid imprints preferred directions on the gas), and feature errors which are independent of the fluid bulk velocity (so can converge more rapidly when the fluid moves). 

A variety of approaches have been developed to deal with these errors. The simplest commonly-used method, the so-called ``Powell 8-wave cleaning,'' simply subtracts the unstable error terms resulting from a non-zero $\divB$ from the equation of motion \citep{powell:1999.8wave.cleaning}. This removes the more catastrophic numerical instabilities, but does {\em not} solve the convergence problem -- many studies have shown that certain types of problems, treated with only this method, will simply converge to the wrong solution \citep{toth:2000.divB.constraint,mignone:2010.ctu.mhd.divb.constraint,mocz:2014.constrained.transport.mhd}. And the subtraction necessarily violates momentum conservation, so one would ideally like the subtracted terms (the $\divB$ values) to remain as small as possible. Therefore more sophisticated ``cleaning'' schemes have been developed, the most popular of which have been variants of the \citet{dedner:2002.divb.cleaning.scheme} method: this adds source terms which transport the divergence away (in waves) and then damp it. This has proven remarkably robust and stable.  

However, applications of these techniques in Lagrangian codes in astrophysics have remained limited. The most popular Lagrangian method, smoothed-particle hydrodynamics (SPH), suffers from several well-known errors that make MHD uniquely challenging. The SPH equations are not consistent at any order (meaning they contain zeroth-order errors; \citealt{morris:1996.sph.stability,dilts:1999.sph.stability,read:2010.sph.mixing.optimization}); this introduces errors which converges away very slowly and causes particular problems for divergence-cleaning. Also, naive implementations of the equations are vulnerable to the tensile and particle pairing instabilities. And artificial diffusion terms, with ad-hoc parameters, are required in SPH to deal with discontinuous fluid quantities. As such, many previous implementations of MHD in SPH were unable to reproduce non-trivial field configurations, were extremely diffusive, or were simply unable to numerically converge; in turn key qualitative phenomena such as the magneto-rotational instability (MRI) and launching of magnetic jets could not be treated \citep[see][]{swegle:1995.sph.stability,monaghan:2000.sph.tensile.instability,borve:2001.sph.mhd.regularization,maron:2003.gradient.particle.mhd,price:2006.sph.mhd.neutron.star.mergers,rosswog:2007.sph.mhd,price:2008.sph.mhd.star.clusters,dolag:2009.mhd.gadget}. 

Recently, however, a number of breakthroughs have been made in Lagrangian hydrodynamics, with the popularization of moving-mesh and mesh-free finite-volume Godunov methods. \citet{springel:arepo,duffell:2011.TESS,gaburov:2012.public.moving.mesh.code} have developed moving-mesh MHD codes, which capture many of the advantages of both AMR and SPH, using the \citet{dedner:2002.divb.cleaning.scheme} cleaning method. Meanwhile, \citet{lanson.vila:2008.meshfree.consistency,lanson.vila:2008.meshfree.convergence,gaburov:2011.meshless.dg.particle.method,hopkins:gizmo} have developed a class of new, mesh-free finite volume methods which are both high-order consistent (convergent) and fully conservative. These are very similar to moving-mesh codes (in fact, Voronoi moving-meshes are technically a special case of the method). In \citealt{hopkins:gizmo}, these are developed for hydrodynamics in the multi-method, hydrodynamics+gravity+cosmology code {\small GIZMO}, which is an extension of the $N$-body gravity and domain decomposition algorithms from {\small GADGET-3} \citep{springel:gadget} to include a variety of new hydrodynamic methods.\footnote{A public version of this code, including the full MHD implementation used in this paper, is available at \gizmourl. Users are encouraged to modify and extend the capabilities of this code; the development version of the code is available upon request from the author.} In \citet{hopkins:gizmo}, a broad range of test problems are considered, and it is shown that these also capture most of the advantages of AMR and SPH, while avoiding many of their disadvantages. Particularly important, these eliminate the low-order errors, numerical instabilities, and artificial diffusion terms which have plagued SPH. \citet{gaburov:2011.meshless.dg.particle.method} considered a range of MHD test problems and found very encouraging preliminary results; they showed that these mesh-free methods could handle complicated non-linear problems like the MRI with accuracy comparable to state-of-the-art grid codes. Meanwhile, tremendous improvements have also been made in SPH \citep{ritchie.thomas:2001.egy.wtd.sph,price:2008.sph.contact.discontinuities,wadsley:2008.sph.mixing.cosmology,cullen:2010.inviscid.sph,read:2012.sph.w.dissipation.switches,saitoh:2012.dens.indep.sph,hopkins:lagrangian.pressure.sph,tricco:2012.sphmhd.methods,tricco:2013.sphmhd.methods,tricco:thesis}. 

Therefore, in this paper, we extend the mesh-free MFM and MFV Lagrangian hydrodynamics in {\small GIZMO} to include MHD, and consider a systematic survey of a wide range of test problems, and compare state-of-the-art grid-based (AMR) codes, MFM, MFV, and SPH MHD methods, implemented in the same code. This includes problems such as the MRI and MHD jets which have been historically challenging. We show in all cases that the new meshfree methods exhibit good convergence and stable behavior, and are able to capture all of the important behaviors, even at low resolution. On some problem classes, we show they converge faster than state-of-the-art AMR codes using CT.

\vspace{-0.5cm}
\section{Numerical Methodology}
\label{sec:methods}

\subsection{Review of the New Meshless Methods}
\label{sec:methods:outline}

\paperone\ derives and describes the pure-hydrodynamic version of the numerical methods here in detail, including self-gravity and cosmological integration. This is almost entirely identical in MHD; therefore we will not repeat it. However we will very briefly review the new numerical methods. 

The equations we solve are the standard finite-volume Godunov-type equations: the fundamental difference between our meshless methods and a moving-mesh is simply that the definition of the volume partition (how the volume is divided among different mesh-generating points or ``cells/particles'') is distinct. The further difference between this and a fixed-grid code is, of course, that the mesh-generating points/cells move, and that their arrangement can be irregular (as opposed to a Cartesian grid). 

In a frame moving with velocity ${\bf v}_{\rm frame}$, the homogeneous Euler equations in ideal MHD (and pure hydrodynamics, which forms the special case ${\bf B}=0$) can be written as a set of hyperbolic partial differential conservation equations of the form 
\begin{align}
\label{eqn:conservation}
\frac{\partial {\bf U}}{\partial t} + \nabla\cdot({\bf F}-{\bf v}_{\rm frame}\otimes{\bf U}) &= {\bf S}
\end{align}
where $\nabla\cdot {\bf F}$ refers to the inner product between the gradient operator and tensor ${\bf F}$, $\otimes$ is the outer product, ${\bf U}$ is the ``state vector'' of conserved (in the absence of sources) variables, the tensor ${\bf F}$ is the flux of conserved variables, and ${\bf S}$ is the vector of source terms
\begin{align}
{\bf U}  &= 
\left(
\begin{array}{c}
\rho \\
\rho\,{\bf v} \\
\rho\,e \\
{\bf B} \\ 
\rho\,\psi
\end{array}
\right)
&{\bf F}  = 
\left(
\begin{array}{c}
\rho\,{\bf v} \\
\rho\,{\bf v}\otimes{\bf v} + P_{T}\,\mathcal{I} - {\bf B}\otimes {\bf B} \\
(\rho\,e + P_{T})\,{\bf v} - ({\bf v}\cdot {\bf B})\,{\bf B} \\
{\bf v}\otimes{\bf B} - {\bf B}\otimes {\bf v} \\ 
\rho\,\psi\,{\bf v}
\end{array}
\right)
\end{align}
where $\rho$ is mass density, $e = u + |{\bf B}|^{2}/2\rho + |{\bf v}|^{2}/2$ is the total specific energy ($u$ the internal energy), $P_{T} = P + |{\bf B}|^{2}/2$ is the sum of thermal and magnetic pressures, and $\psi$ is a scalar field defined below.

The meshless equations of motion are derived in \paperone\ in standard Galerkin fashion beginning from the integral form of the conservation laws, after multiplying Eq.~\ref{eqn:conservation} by a test function $\phi$ 
\begin{align}
\label{eqn:integral}
0 &= \int_{\Omega}\left(\frac{d{{\bf U}}}{dt}\,\phi - {\bf F}\cdot\nabla\phi -{\bf S}\,\phi \right)\,d\Omega + \int_{\partial \Omega} ({\bf F}\,\phi)\cdot \hat{{\bf n}}_{\partial \Omega}\,d\,\partial \Omega
\end{align}
Here the domain $\Omega$ is such that $d\Omega = d^{\Ddim}{\bf x}\,d t$, where $\Ddim$ is the number of spatial dimensions, $d{f}/dt \equiv \partial f/\partial t + {\bf v}_{\rm frame}({\bf x},\,t)\cdot\nabla f$ is the co-moving derivative of any function $f$, and  $\hat{{\bf n}}_{\partial \Omega}$ is the normal vector to the surface $\partial \Omega$, and the test function is an arbitrary differentiable Lagrangian function. 

To transform this into a {\em discrete} set of equations, must chose how to partition the volume (for the ``averaging/integration'' step). If we choose a uniform Cartesian grid between uniformly spaced points, then we will recover the standard Godunov-type finite-volume grid-based equations of motion (like that in {\small ATHENA} and many popular AMR codes). If we choose a Voronoi tesselation between moving mesh-generating points, we recover a moving-mesh method similar to {\small AREPO}. For the new methods in \paperone, we partition the volume according to a {\em continuous} weighting function $f$, such that the fraction of the differential volume $d^{\Ddim}{\bf x}$ at the point ${\bf x}$ associated with the mesh-generating point at ${\bf x}={\bf x}_{i}$ is given by 
\begin{align}
\label{eqn:wt.fn.kernel}
f_{i}({\bf x}) &\equiv \frac{W({\bf x}-{\bf x}_{i},\,h({\bf x}))}{\sum_{j} W({\bf x}-{\bf x}_{j},\,h({\bf x}))} 
\end{align}
Where $W$ is any kernel/weight function and $h$ is a ``kernel length.'' Note that this guarantees a ``partition of unity'' (the volume is perfectly divided into cell-volumes $V_{i}$), and leads to a Voronoi-like partition, but with slightly smoothed boundaries between cells (which leads to some advantages, and some disadvantages, compared to moving meshes, where the boundaries are strict step functions). In the limit where $W$ goes to a delta function, the method becomes exactly a Voronoi-type moving-mesh method.\footnote{Here and throughout this paper, we will define the kernel size $h_{i}$ at the location of cell/particle $i$ as the effective cell side-length, based on the cell volume $V_{i}$. In 1D/2D/3D, this is $h_{i}=V_{i}$, $h_{i}=(V_{i}/\pi)^{1/2}$, $h_{i}=(3\,V_{i}/4\pi)^{1/3}$, respectively. The volume $V_{i}$ is calculated directly from the neighbor positions defining the volume partition (see \paperone). This {\em exactly} reproduces the grid spacing if the particles are arranged in a Cartesian grid. Note that, in principle, $W({\bf x},\,h)$ can be non-zero at $|{\bf x}|>h$. In MFM/MFV methods this has {\em nothing} to do with the effective cell/particle volume/size (conserved quantities are {\em not} ``smoothed'' over the kernel, but only averaged inside the a single cell of volume $V_{i}$ just like in a grid code), but instead reflects the size of the stencil (number of neighbor cells) between which fluxes are computed. As in grid codes, increasing the stencil size can increase diffusion, but does not directly alter the resolution scale.}

This choice is combined with a second-order accurate moving-least squares matrix-based gradient operator, which has been utilized in many other methods (including grid-based codes; see \citealt{maron:2012.phurbas.algorithm,tiwari:2003.finite.pointset.method,liu:2005.finite.particle.method,luo:2008.compressible.flow.galerkin,lanson.vila:2008.meshfree.consistency,lanson.vila:2008.meshfree.convergence,mocz:2014.galerkin.arepo}). Eq.~\ref{eqn:integral} can then be expanded and analytically integrated to yield a second-order accurate set of discrete evolution equations: 
\begin{align}
\label{eqn:final}
\frac{d}{dt}(V\,{\bf U})_{i} + \sum_{j}\,\tilde{{\bf F}}_{ij}\cdot {\bf A}_{ij} = (V\,{\bf S})_{i}
\end{align}
This is identical to the standard Godunov-type finite-volume equations. The term $V_{i}\,{\bf U}_{i}$ is simply the cell-volume integrated value of the conserved quantity to be carried with cell/particle $i$ (e.g.\ the total mass $m_{i}=V_{i}\,\rho_{i}$, momentum, or energy associated with the cell $i$); its time rate of change is given by the sum of the fluxes $\tilde{{\bf F}}_{ij}$ into/out of an ``effective face area'' ${\bf A}_{ij}$, plus the volume-integrated source terms. The full mathematical derivation and expression for the ${\bf A}_{ij}$ is given in \paperone\ (\S~2.1). 

We then use a standard MUSCL-Hancock type scheme for finite-volume Godunov methods to solve Eq.~\ref{eqn:final}. This is commonly used in grid and moving-mesh codes \citep{vanleer:1984.slopelimiters,toro:1997.reimann.solver.book,teyssier:2002.RAMSES,fromang:2006.ramses.amr.schemes,mignone:2007.pluto.code.methods,cunningham:2009.ct.mhd.code,springel:arepo}; it involves a slope-limited linear reconstruction of face-centered quantities from each mesh generating point (cell ``location''), with a first-order drift/predict step for evolution over half a timestep, and then the application of a Riemann solver to estimate the time-averaged inter-cell fluxes for the timestep. See \paperone (\S~2 and Appendices~A \&\ B) for details. The points then move with the center-of-mass gas velocity.

In \paperone, we derive two variants of this method and implement them in {\small GIZMO}. First, the meshless finite-volume (MFV) method. This solves the Riemann problem between cells assuming the effective faces move with the mean cell velocity; this is analogous to a moving-mesh code, and includes mass fluxes between cells. Second, the meshless finite-mass (MFM) method. This solves the Riemann problem assuming the face deforms in a fully Lagrangian fashion; in this case there are no mass fluxes. The two are formally identical up to a difference in the non-linear (second-order) error terms in the fluxes, provided the cells move with the gas velocity. In practice, each has some advantages and disadvantages, discussed below.

\vspace{-0.5cm}
\subsection{Code Modifications for MHD}
\label{sec:methods:mods}

Everything described above is identical in hydodynamics and MHD; and all details of the code (except those specifically described below) are unchanged from \paperone. 

As usual for finite-volume Godunov schemes, we explicitly evolve the conservative variables $(V\,{\bf B})_{i}$ (integrated magnetic field over the volume partition corresponding to a mesh-generating point) and $(m\,\psi)_{i} = \int \rho\,\psi\,dV_{i}$; primitive variables and gradients are then constructed from these (e.g.\ ${\bf B}_{i}\equiv (V\,{\bf B})_{i}/V_{i}$ as in \paperone).

\vspace{-0.5cm}
\subsubsection{The Riemann Solver}

As in \paperone, we solve the 1D, un-split Riemann problem in the rest-frame of the effective face between the cells. However we require a Riemann solver that allows ${\bf B}\ne 0$. Since in the hydro case we use the HLLC solver, here we adopt the widely-used HLLD solver \citep{miyoshi:2005.hlld.reimann.solver}. This is accurate at the order required, and extremely well-tested (see e.g.\ \citealt{miyoshi:2005.hlld.reimann.solver,stone:2008.athena} and modern versions of e.g.\ {\small RAMSES} and {\small ENZO}). 

The frame motion is calculated for both the MFM and MFV methods as in \paperone. We emphasize that for our MFM method, it is {\em required} that we use a solver which explicitly includes the contact wave (i.e.\ the contact discontinuity, on either side of which mass is conserved). This is because in MFM we always solve the Riemann problem with frame moving exactly with the contact wave; attempting to simply find a frame where the mass flux vanishes in a simpler (such as the HLL or Rusanov) approximation leads to incorrect, and in many cases unphysical solutions.

\vspace{-0.5cm}
\subsubsection{Signal Velocities and Time-Stepping}

As in \paperone, we limit timesteps with a standard local Courant-Fridrisch-Levy (CFL) timestep criterion; in MHD we must replace the sound speed with the fast magnetosonic speed: 
\begin{align}
\label{eqn:CFL} \Delta t_{{\rm CFL},\,i} &= 2\,C_{\rm CFL}\,\frac{h_{i}}{|v_{{\rm sig},\,i}|} \\ 
v_{{\rm sig},\,i}^{\rm MAX} &= {\rm MAX}_{j}{\Bigl[} v_{{\rm f},\,ij} + v_{{\rm f},\,ji} - 
{\rm MIN}{\Bigl(}0,\,\frac{({\bf v}_{i}-{\bf v}_{j})\cdot ({\bf x}_{i}-{\bf x}_{j})}{|{\bf x}_{i}-{\bf x}_{j}|}{\Bigr)}
{\Bigr]} \\ 
(v_{{\rm f},\,ij})^{2} \equiv& \frac{1}{2}\,\left[{c_{s,\,i}^{2} + v_{A,\,i}^{2}} + 
\sqrt{\left( c_{s,\,i}^{2} + v_{A,\,i}^{2} \right)^{2} - {4\,c_{s,\,i}^{2}\,v_{A,\,i}^{2}\,(\hat{{\bf B}}_{i}\cdot \hat{{\bf x}}_{ij})^{2}}}\right] 
\end{align}
Here ${\bf x}_{ij} \equiv {\bf x}_{i}-{\bf x}_{j}$ is the separation between two points, $\hat{\bf x}$ is the unit vector ${\bf x}/|{\bf x}|$,  $c_{s,\,i}$ is the sound speed, $v_{A}$ is the Alfven speed, $h_{i}$ is the effective cell size defined above, ${\rm MAX}_{j}$ refers to the maximum over all interacting neighbors $j$ of $i$,  and $v_{\rm sig}$ is the signal velocity \citep{whitehurst:1995.moving.mesh.protocode,monaghan:1997.sph.solvers.viscosities}.

\vspace{-0.5cm}
\subsubsection{Divergence-Cleaning}

Ideally, this would complete the method; but the above method cannot ensure the divergence constraint. To do this, we must add the following source terms:
\begin{align}
\label{eqn:source} {\bf S}  =& {\bf S}_{\rm Powell} + {\bf S}_{\rm Dedner} \\ 
=& -\nabla\cdot {\bf B}\,
\left(
\begin{array}{c}
0 \\
{\bf B} \\
{\bf v}\cdot{\bf B} \\
{\bf v} \\ 
0 
\end{array}
\right)
- 
\left(
\begin{array}{c}
0 \\
0 \\
{\bf B}\cdot (\nabla\psi) \\
\nabla\psi \\ 
(\nabla\cdot {\bf B})\,\rho\,c_{h}^{2} + \rho\,\psi/\tau
\end{array}
\right)
\end{align}
The first term (${\bf S}_{\rm Powell}$) represent the \citet{powell:1999.8wave.cleaning} or ``8-wave'' cleaning, and subtracts the numerically unstable terms from non-zero $\nabla\cdot {\bf B}$. This is necessary to ensure numerical stability and Galilean invariance -- most problems will crash or converge to incorrect solutions without this. The second term (${\bf S}_{\rm Dedner}$) follows the method of \citet{dedner:2002.divb.cleaning.scheme}, who introduce a conservative scalar field $\psi$ which transports divergence away from the source and damps it. This is necessary to keep $\nabla\cdot{\bf B}$ low, minimizing the resulting errors.

Following \citet{dedner:2002.divb.cleaning.scheme} and \citet{gaburov:2011.meshless.dg.particle.method}, it is straightforward to show that this leads to the following form for the discrete terms in our equation of motion (Eq.~\ref{eqn:final}):\footnote{Eq.~\ref{eqn:source} is the continuum equation; we stress that we are {\em not} free to choose how we discretize it. To actually ensure numerical stability, the form of the $\nabla\cdot {\bf B}$ terms must exactly match those terms from our Riemann solver solution which are unstable (e.g.\ the tensile terms). Likewise, the divergence cleaning must act specifically to reduce $\nabla \cdot {\bf B}$ {\em defined in the same manner}, or it does not serve any useful purpose. It might be tempting, for example, to use the value of $\nabla\cdot {\bf B}$ calculated from our second-order accurate matrix gradient estimator for the Powell terms, or to construct a pair-wise symmetrized version of Eq.~\ref{eqn:source.discrete} which manifestly maintains momentum conservation. However, these will not actually eliminate the unstable terms, and we have confirmed that they lead to catastrophic errors in our tests.}
\begin{align}
\nonumber (V{\bf S})_{i}  
=& -(V\nabla\cdot {\bf B})^{\ast}_{i}\,
\left(
\begin{array}{c}
0 \\
{\bf B}_{i} \\
{\bf v}_{i}\cdot{\bf B}_{i} \\
{\bf v}_{i} \\ 
0 
\end{array}
\right)\\
\label{eqn:source.discrete} &-
\left(
\begin{array}{c}
0 \\
0 \\
{\bf B}_{i}\cdot (V\nabla\psi)_{i}^{\ast} \\
(V\nabla\psi)_{i}^{\ast} \\ 
(V\nabla\cdot {\bf B})_{i}^{\ast}\,\rho_{i}\,c_{h,\,i}^{2} + (m\psi)_{i}/\tau_{i}
\end{array}
\right) \\ 
\nonumber \\
(V\nabla\cdot {\bf B})^{\ast}_{i} \equiv & -\sum_{j}\,\bar{B}^{\prime}_{x,\,ij}\,|{\bf A}_{ij}| \\ 
(V\nabla\psi)^{\ast}_{i} \equiv & -\sum_{j}\,\bar{\psi}_{ij}\,{\bf A}_{ij} 
\end{align}
where the $\bar{B}^{\prime}$ and $\bar{\psi}$ terms are defined below.

We have some freedom to choose $c_{h,\,i}$ and $\tau_{i}$. Following \citet{tricco:2012.sphmhd.methods}, we take $c_{h,\,i} = \sigma_{h}^{1/2}\, v_{{\rm sig},\,i}^{\rm MAX} / 2$, where $v_{{\rm sig},\,i}^{\rm MAX} $ is the maximum signal velocity as described above ($\sigma_{h}$ is simply a convenience parameter which re-normalizes the characteristic speed). The value $v_{{\rm sig},\,i}^{\rm MAX}/2$ is close to the fast magnetosonic wave speed, but also accounts for super-sonic cell approach velocities, which is critical for good behavior in highly supersonic compressions. We have experimented with variations in the dimensionless parameter $\sigma_{h}$, and find the best results for $\sigma_{h}=1$ (values $\sigma_{h}\ll 1$ produce ineffective cleaning, values $\gg 1$ lead to numerical instability). We take $\tau_{i} = h_{i} / (\sigma_{p}\,c_{\tau,\,i} )$ (where $h_{i}$ is the effective cell length defined above).\footnote{In timestepping, we update $(m\psi)_{i}$ for the $\tau_{i}$ term with the implicit solution $(m\psi)_{i}\propto \exp{(-\Delta t_{i} / \tau_{i})}$; this allows us to take larger timesteps without numerical instability.} For $c_{\tau,\,i}$ (the damping speed), we have considered several choices, all of which give very similar results. These are detailed in Appendix~\ref{sec:damping.speed}; our default choice is Eq.~\ref{eqn:ctau}, which is closely related to the local fastest possible signal velocity. We have also experimented extensively with $\sigma_{p}$, and find a best compromise between stability and diffusivity for values $\sigma_{p}\sim 0.05-0.3$; we adopt $\sigma_{p}=0.1$ as our default in all problems here.

Note that, unlike the original \citet{dedner:2002.divb.cleaning.scheme} formulation (and implementations in codes like {\small AREPO} and {\small PLUTO}), this means $c_{h,\,i}$ and $\tau_{i}$ are spatially variable. This allows us to maintain hierarchical timestepping (while forcing a constant $c_{h}$ imposes a global maximum timestep, a severe CPU cost penalty), and has many other advantages (see Appendix~\ref{sec:non.constant.dedner}). But it is not {\em a priori} obvious that this will maintain stability. However, in Appendix~\ref{sec:non.constant.dedner} we discuss this in detail and derive a rigorous stability criterion, which should be satisfied by our choices above, designed so that $c_{h}$ and $\tau$ are {\em locally} smooth (on the kernel scale). We confirm this stability in our numerical tests. 

We caution that the discrete source terms, particularly the Powell terms which subtract $i$-centered quantities, are not manifestly antisymmetric between cell pairs $ij$. This means that momentum and energy conservation in MHD are only accurate up to integration accuracy, times a term proportional to $\nabla \cdot {\bf B}$ (unlike in hydrodynamics, where conservation can be ensured at machine accuracy).\footnote{{\em Some} of the Dedner terms can be made anti-symmetric without destroying the numerical stability of the scheme; for example the $\psi$-flux correction described below, and the $(V\nabla\cdot{\bf B})\,c_{h}^{2}$ term (by using a single wavespeed $c_{h,\,i} = c_{h}$ for the whole problem). However in all our tests the conservation errors from those terms are always sub-dominant to the error from the (inescapable) Powell $(V\nabla\cdot{\bf B})_{i}^{\ast}\,{\bf v}_{i}$ term (and all these errors vanish when $\nabla\cdot{\bf B}\rightarrow 0$). So we find the best overall conservation properties result from using the most accurate possible cleaning scheme, rather than a partially-conservative (but less accurate) cleaning.} Controlling $\nabla \cdot {\bf B}$ is critical to minimize these errors. 

These source terms also modify the Riemann problem. When we perform the reconstruction to obtain the left and right states ${\bf U}_{L}$ ($j$-side) and ${\bf U}_{R}$ ($i$-side), we can define a convenient coordinate system where $\hat{x}^{\prime} = \hat{\bf A}_{ij}$ (i.e.\ the $x$-axis is normal to the effective face between cells $i$ and $j$). In this coordinate system, the normal-component of the $B$-fields $B_{x}^{\prime}$, will in general not be equal. But equal values (i.e.\ non-zero $\nabla \cdot {\bf B}$ in the 1D problem) are required for a physical solution. Without divergence-cleaning (Powell-only), this is handled by simply replacing $B_{x,\,L}^{\prime}$ and $B_{x,\,R}^{\prime}$ with the mean value $\bar{B}_{x,\,ij}^{\prime} \equiv (B_{x,\,L}^{\prime} + B_{x,\,R}^{\prime})/2$. \citet{dedner:2002.divb.cleaning.scheme} showed that with the source terms of Eq.~\ref{eqn:source}, the infinitely-sharp discontinuity leads to a physical solution $B_{x,\,L}^{\prime}\rightarrow B_{x,\,R}^{\prime} \rightarrow \bar{B}_{x,\,ij}^{\prime}$, $\psi_{L}\rightarrow \psi_{R} \rightarrow \bar{\psi}_{ij}$, in infinitesimally small time: 
\begin{align}
\label{eqn:b.normal}\bar{B}_{x,\,ij}^{\prime} &= \frac{1}{2} \left( B_{x,\,L}^{\prime} + B_{x,\,R}^{\prime} \right) + \frac{1}{2\,\tilde{c}_{h,\,ij}}\,\left( \psi_{L} - \psi_{R} \right) \\ 
\bar{\psi}_{ij} &= \frac{1}{2} \left( \psi_{L} + \psi_{R} \right) + \frac{\tilde{c}_{h,\,ij}}{2}\,\left( B_{x,\,L}^{\prime} - B_{x,\,R}^{\prime} \right) \\ 
\tilde{c}_{h,\,ij} &= {\rm MAX}\left[ v_{{\rm f},\,L}\, , \,v_{{\rm f},\,R}  \right] \\ 
\label{eqn:vfast.interface} v_{{\rm f},\,L}^{2} &= \frac{1}{2}\,\left[{c_{s,\,L}^{2} + v_{A,\,L}^{2}} + 
\sqrt{\left( c_{s,\,L}^{2} + v_{A,\,L}^{2} \right)^{2} - {4\,c_{s,\,L}^{2}\,B_{x,\,L}^{\prime\,2}/\rho_{L}}}\right] 
\end{align}
here $\tilde{c}_{h,ij}$ is the fastest wave speed {\em in the local 1D problem}, which can be computed only from the $i$ and $j$ values (it does not necessarily correspond to $c_{h,\,i}$ in Eq.~\ref{eqn:source} above).\footnote{We have also explored an alternative, two-wavespeed formulation of the $\bar{B}$ and $\bar{\psi}$ terms, discussed in Appendix~\ref{sec:appendix:twowave.bbar}. For all tests here, the difference is small.} This is separable from the full Riemann solution. So, in the Riemann problem, we first update $B_{x}^{\prime}$ and $\psi$ according to the above, {\em then} compute the full Riemann solution using the updated values (and usual $B_{y,\,L,R}^{\prime}$ and $B_{z,\,L,R}^{\prime}$). The flux of $B_{x}^{\prime}$ is then $\tilde{F}_{ij,\,B_{x}^{\prime}} = v_{x,\,{\rm face}}^{\prime}\,\bar{B}_{x,\,ij}^{\prime}$, where $v_{x,\,{\rm face}}$ is the normal face velocity in the boosted frame (in which we solve the Riemann problem). Because $\psi$ is advected with the fluid, we follow \citet{gaburov:2011.meshless.dg.particle.method} and simply take the $\psi$ flux to be $\tilde{F}_{ij,\,\psi} = \tilde{F}_{ij,\,\rho}\,\psi_{L}$ for $\tilde{F}_{ij,\,\rho}>0$, and $\tilde{F}_{ij,\,\psi} = \tilde{F}_{ij,\,\rho}\,\psi_{R}$ for $\tilde{F}_{ij,\,\rho}<0$, where $\tilde{F}_{ij,\,\rho}$ is the mass flux (so this vanishes for our MFM method).\footnote{We have experimented with using $\bar{\psi}_{ij}$ for the $\psi$-flux. However, this yields no improvement on any test problem, and the $( B_{x,\,L}^{\prime} - B_{x,\,R}^{\prime} )$ term from $\bar{\psi}_{ij}$ can introduce numerical instability under some circumstances. Therefore we use the simpler $\psi$-flux.} 

The HLLD solver requires an initial guess for the left and right wavespeeds, to compute a solution. If we define $v_{x}^{\prime}$ as the normal-component of the reconstructed velocities, then we use $S_{L} = {\rm MIN}[ v_{x,\,L}^{\prime} \, , \, v_{x,\,R}^{\prime} ] - {\rm MAX}[ v^{ij}_{{\rm f},\,L} \, , \, v^{ij}_{{\rm f},\,R} ]$, where $v^{ij}_{{\rm f}}$ is the {\em updated} fast magnetosonic wavespeed using the updated normal $B_{x,\,ij}^{\prime}$: 
\begin{align}
(v_{{\rm f},\,L}^{ij})^{2} &= \frac{1}{2}\,\left[{c_{s,\,L}^{2} + v_{A,\,L}^{2}} + 
\sqrt{\left( c_{s,\,L}^{2} + v_{A,\,L}^{2} \right)^{2} - {4\,c_{s,\,L}^{2}\,\bar{B}_{x,\,ij}^{\prime\,2}/\rho_{L}}}\right] 
\end{align}
and $S_{R} = {\rm MAX}[ v_{x,\,L}^{\prime} \, , \, v_{x,\,R}^{\prime} ] + {\rm MAX}[ v^{ij}_{{\rm f},\,L} \, , \, v^{ij}_{{\rm f},\,R} ]$.\footnote{The HLLD solver can fail in some very rare circumstances if ``bad'' guesses are used. We therefore check whether our first estimate produces a solution where the pressure is everywhere positive. If this fails, then we instead compute the Roe-averaged velocity $v_{\rm Roe}$ and fast magnetosonic speed $c_{\rm Roe}$, and try $S_{L} = {\rm MIN}[ v_{x,\,L}^{\prime} - v^{ij}_{{\rm f},\,L}\, , \, v_{\rm Roe} - c_{\rm Roe} ]$ and $S_{R} = {\rm MAX}[ v_{x,\,R}^{\prime} + v^{ij}_{{\rm f},\,R}\, , \, v_{\rm Roe} + c_{\rm Roe} ]$. We then check again; if this fails (which does not occur in any test problem here) we test a Lax-Friedrich estimate ($S_{L}=-S_{R}$, with $S_{R}$ from our first guess). If this somehow fails still, we go back and re-compute the interface using a piecewise-constant (first-order) approximation, then check the series of wavespeeds again. If this fails, the code exits with an error (this only occurs when unphysical values are input into the solver).}

Finally, as in \paperone, we solve the Riemann problem in the boosted frame ${\bf v}_{\rm frame}$ corresponding to the mean motion of the quadrature point between mesh-generating points. We must therefore boost back to the simulation frame. The de-boosted fluxes for cells follow \paperone, for the hydro terms, but with the additional terms for ${\bf B}$ (for the fluxes to cell-$i$): 
\begin{align}
\label{eqn:deboost.B} \left( \tilde{{\bf F}}_{ij}({\bf B}) \cdot {\bf A}_{ij}\right)_{i} \rightarrow & \tilde{{\bf F}}_{ij}({\bf B}) \cdot {\bf A}_{ij} - \bar{B}_{x,\,ij}^{\prime}\,|{\bf A}_{ij}|\,{\bf v}_{\rm frame} \\ 
\left( \tilde{{\bf F}}_{ij}(e) \cdot {\bf A}_{ij}\right)_{i} \rightarrow & \tilde{{\bf F}}_{ij}(e) \cdot {\bf A}_{ij} - \bar{B}_{x,\,ij}^{\prime}\,|{\bf A}_{ij}|\,{\bf v}_{\rm frame} \cdot {\bf B}_{i} 
\end{align}
The second equation just accounts for the energy flux associated with the corrected ${\bf B}$-flux.

\begin{figure}
    \plotonesize{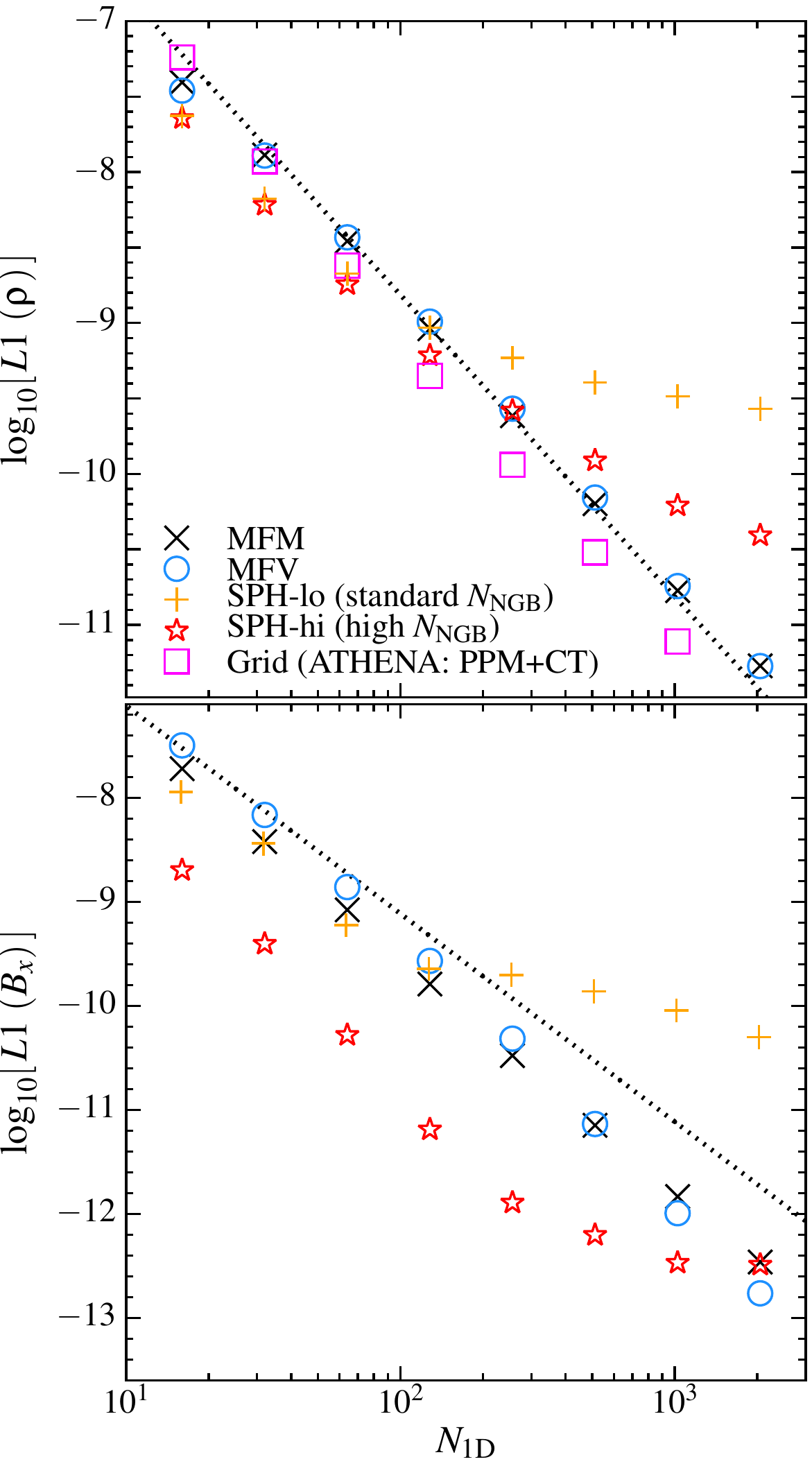}{0.95}
    \vspace{-0.25cm}
    \caption{Linear magnetosonic wave test problem (see \S~\ref{sec:wave}). Here a traveling, one-dimensional fast magnetosonic wave is propagated one wavelength; we then define the $L1$ norm as the mean absolute error relative to the known analytic solution in density ({\em top}) or magnetic field $B_{x}$ ({\em bottom}; error here is equivalent to the numerical $\divB\ne0$ errors). We compare our new, meshless Lagrangian finite-volume Godunov methods (``meshless finite-mass'' or MFM, and ``meshless finite-volume'' or MFV) from \paperone\ (see \S~\ref{sec:methods:outline}), to the best current implementation of SPH MHD (see \S~\ref{sec:methods:sph}), and to state of the art grid codes, here {\small ATHENA} run as a third-order PPM code using constrained transport (CT) to ensure $\divB=0$ to machine precision. Dotted line shows second-order convergence ($L1\propto N^{-2}$); MFM/MFV and grid/PPM methods converge at this rate, as expected. Convergence is also good ($L1\propto N^{-2.3}$) in MFM/MFV for the divergence errors ($\langle B_{x} \rangle = 1$, so these are fractionally very small). SPH shows some (slower) convergence until its known zeroth-order errors dominate; then errors flatten with resolution. This is reduced in SPH by increasing the kernel size. ``SPH-lo'' (standard $\NNb$) uses the equivalent of $\NNb=32$ in 3D (our default choice in all MFM/MFV runs). ``SPH-hi'' uses a 3D-equivalent $\NNb=120$.
        \vspacerpostplot 
    \label{fig:wave}}
\end{figure}

\vspace{-0.5cm}
\subsection{The SPH MHD Implementation in GIZMO}
\label{sec:methods:sph}

As described in \paperone, {\small GIZMO} is a multi-method code: users can run with the MFM or MFV hydrodynamic methods, or SPH, if desired. We therefore update our SPH implementation to include MHD. The exact SPH equations are given in Appendix~\ref{sec:spmhd.numerics}. 

Briefly, the non-magnetic implementation of SPH follows the ``modern'' P-SPH method developed in \citet{hopkins:lagrangian.pressure.sph} and extended in \paperone. This includes state-of-art re-formulations of the SPH hydrodynamics equations to eliminate the known ``surface tension'' errors \citep{saitoh:2012.dens.indep.sph,hopkins:lagrangian.pressure.sph}, Lagrangian-derived terms to account for variable smoothing lengths \citep{springel:entropy}, addition of artificial diffusion terms for thermal energy \citep{price:2008.sph.contact.discontinuities,wadsley:2008.sph.mixing.cosmology}, higher-order switches for artificial diffusion terms to minimize unnecessary dissipation \citep{cullen:2010.inviscid.sph}, the use of higher-order kernel functions to allow larger SPH neighbor numbers and reduce the zeroth-order SPH errors \citep{dehnen.aly:2012.sph.kernels}, switches to prevent disparate time-stepping between neighbor particles \citep{saitoh.makino:2009.timestep.limiter}, introduction of more accurate matrix-based gradient estimators \citep{garciasenz:2012.integral.sph}, and conservative, more accurate coupling of SPH to gravity \citep{price:2007.lagrangian.adaptive.softening,barnes:2012.softening.is.smoothing}. 

The SPH MHD implementation combines these improvements to SPH with the MHD algorithms from the series of papers by \citet{tricco:2012.sphmhd.methods,tricco:2013.sphmhd.methods,tricco:thesis}. These introduce artificial diffusion for magnetic fields (artificial resistivity), with a similar ``switch'' to reduce unnecessary dissipation, and re-discretize the MHD equations from the particle Lagrangian included the \citet{dedner:2002.divb.cleaning.scheme} and \citet{powell:1999.8wave.cleaning} terms, so that the divergence cleaning actually acts on the tensile-unstable terms and these terms are properly subtracted (unlike most previous SPH-MHD implementations). We make some further improvements: following \citet{price:2012.sph.mhd.jets,bate:2014.core.collapse.rad.mhd.sph} and directly evolving the conserved quantities $(V{\bf B})_{i}$ and $(m\psi)_{i}$, and slightly modifying the artificial resistivity terms to allow the method to capture cosmological field growth and MHD fluid-mixing instabilities.

\vspace{-0.5cm}
\section{Test Problems}
\label{sec:tests}

We now consider a series of test problems. To make comparison as fair as possible, we will consider MFM/MFV/SPH implementations in the same code ({\small GIZMO}). Unless otherwise stated, we compare to fixed-grid results from {\small ATHENA} \citep{stone:2008.athena}; this is representative of the state-of-the-art in finite-volume, non-moving mesh codes. Because we are interested in methods which can be applied to complicated, multi-physics systems, we use the {\em same} numerical implementation and identical values of purely numerical parameters (e.g.\ $\sigma_{p}$ and $C_{\rm CFL}$), within each method, for all problems. It is of course possible to improve performance in any simple test problem by customizing/tweaking the method, but this usually entails a (sometimes serious) loss of accuracy on other problems. The implementations used here are therefore our attempts at a ``best compromise'' across all problems.

\subsection{Linear Magnetosonic Waves: Testing Convergence}
\label{sec:wave}

We begin by considering a simple linear one-dimensional magnetosonic wave.\footnote{See \burl{http://www.astro.princeton.edu/~jstone/Athena/tests/linear-waves/linear-waves.html}} The problem is trivial, but since virtually all numerical schemes become first-order at discontinuities, smooth linear problems with known analytic solutions are necessary to measure formal convergence. Following \citet{stone:2008.athena}, we initialize a unit-length (in the $x$-direction), periodic domain, with polytropic $\gamma=5/3$ gas, density $\rho=1$, pressure $P=1/\gamma$, and magnetic field ${\bf B}/\sqrt{4\pi} = (1,\,\sqrt{2}, 1/2)$. We add to this a traveling fast magnetosonic wave\footnote{For the grid and MFM/MFV methods, the perturbation is initialized by keeping the particles equidistant and modifying the conserved quantities. In SPH however, significantly better performance is obtained by initializing the density perturbation by using exactly equal particle masses with perturbed locations (this allows apparent convergence in SPH to extend further). We therefore show this case for SPH (as it corresponds to the more likely case in real problems). We note that this initial condition also improves our MFM/MFV performance, but by a smaller amount. The SPH performance is also substantially improved if we ``turn off'' artificial viscosity (or set its minimum to zero), but this is numerically unstable and produces catastrophic errors in most of our tests \citep[see e.g.][]{hopkins:lagrangian.pressure.sph,hopkins:gizmo,hu:2014.psph.galaxy.tests,rosswog:2014.sph.accuracy}.} with amplitude $\delta \rho / \rho = 10^{-6}$, and allow the wave to propagate one wavelength; we then define the L1 error norm for the density (or any other variable) $L1(\rho) = N^{-1}\, \sum_{i} |\rho(x_{i},\,t) - \rho(x_{i},\,t=0)|$. We consider $N=16,\,32,\,64,\,128,\,256,\,512,\,1024,\,2048$.

All methods we consider are able to evolve the wave. Fig.~\ref{fig:wave} plots the L1 norm for density (the velocity variables look similar), and $B_{x}$. In $\rho$, we see that both our MFM and MFV methods converge -- as expected -- with second-order accuracy. We compare to a state-of-the-art grid code, here {\small ATHENA}, run in the most accurate possible mode: PPM (formally a third-order reconstruction method), with the CTU integrator, and CT used to ensure $\divB=0$. Despite the higher order of {\small ATHENA} and the fact that it uses CT to ensure $\divB$ errors remain at the machine-error level, the errors are nearly identical to our MFM/MFV results. The $L1$ norm for $B_{x}$ directly measures the divergence errors; since we do not use CT, these are non-zero. However they are (1) very small, and (2) converge away appropriately -- in fact, we see super-convergence ($L1 \propto N^{-2.3}$) for our MFM/MFV methods. As in \paperone, the convergence rate in all variables is {\em independent} of the kernel neighbor number in MFM/MFV: the choice only controls the {\em normalization} of the errors (larger neighbor numbers reduce noise, but increase diffusion). Based on our experiments in \paperone, we find roughly optimal results using a 3D-equivalent neighbor number $\NNb=32$ ($\NNb\approx4$ in 1D). 

For SPH, we see slower convergence in $\rho$, but it is reasonable at low resolution; but at high-resolution, the SPH zeroth-order errors (here, the ``kernel bias'' error in density estimation, and the systematic gradient error which results from imperfect particle order in the kernel) begin to dominate, and the errors flatten with resolution. This is more severe if we use a lower neighbor number; going to higher-order kernels and higher neighbor numbers suppresses the errors, although they still eventually appear. For true convergence, $\NNb$ must increase with $N$, as is well-known \citep[see][and references therein]{zhu:2014.sph.convergence}. Here we compare 3D-equivalent $\NNb=32$ (our default MFM/MFV choice used in {\em all} runs in this paper), henceforth referred to as ``SPH-lo,'' to 3D $\NNb=120$ (henceforth ``SPH-hi'').\footnote{Increasing the neighbor number $\NNb$ with particle number $N$ must be done carefully in SPH, as discussed in e.g.\ \citet{price:2012.sph.review,dehnen.aly:2012.sph.kernels,zhu:2014.sph.convergence}, since one must avoid the pairing instability and also account for the fact that simply increasing neighbor number in SPH changes the resolution length. For SPH-lo we use the popular \citet{schoenberg:1946.smoothing.kernels} cubic spline kernel, so $\NNb=32$ in 3D corresponds to an ``effective'' resolution scale of $h\approx1.1$ in units of the mean inter-particle spacing (following \citealt{dehnen.aly:2012.sph.kernels}, $h=2\,\sigma$, where $\sigma$ is the standard deviation of the kernel). For SPH-hi we use the \citet{schoenberg:1946.smoothing.kernels} quintic spline, so $\NNb=120$ in 3D corresponds to $h\approx 1.4$. Therefore we caution that the ``effective resolution'' of SPH-hi is slightly larger ($\approx20\%$) than SPH-lo at fixed $N$; however the difference is much smaller than might naively be expected based on $\NNb$ alone.}

\begin{figure*}
    \plotsidesize{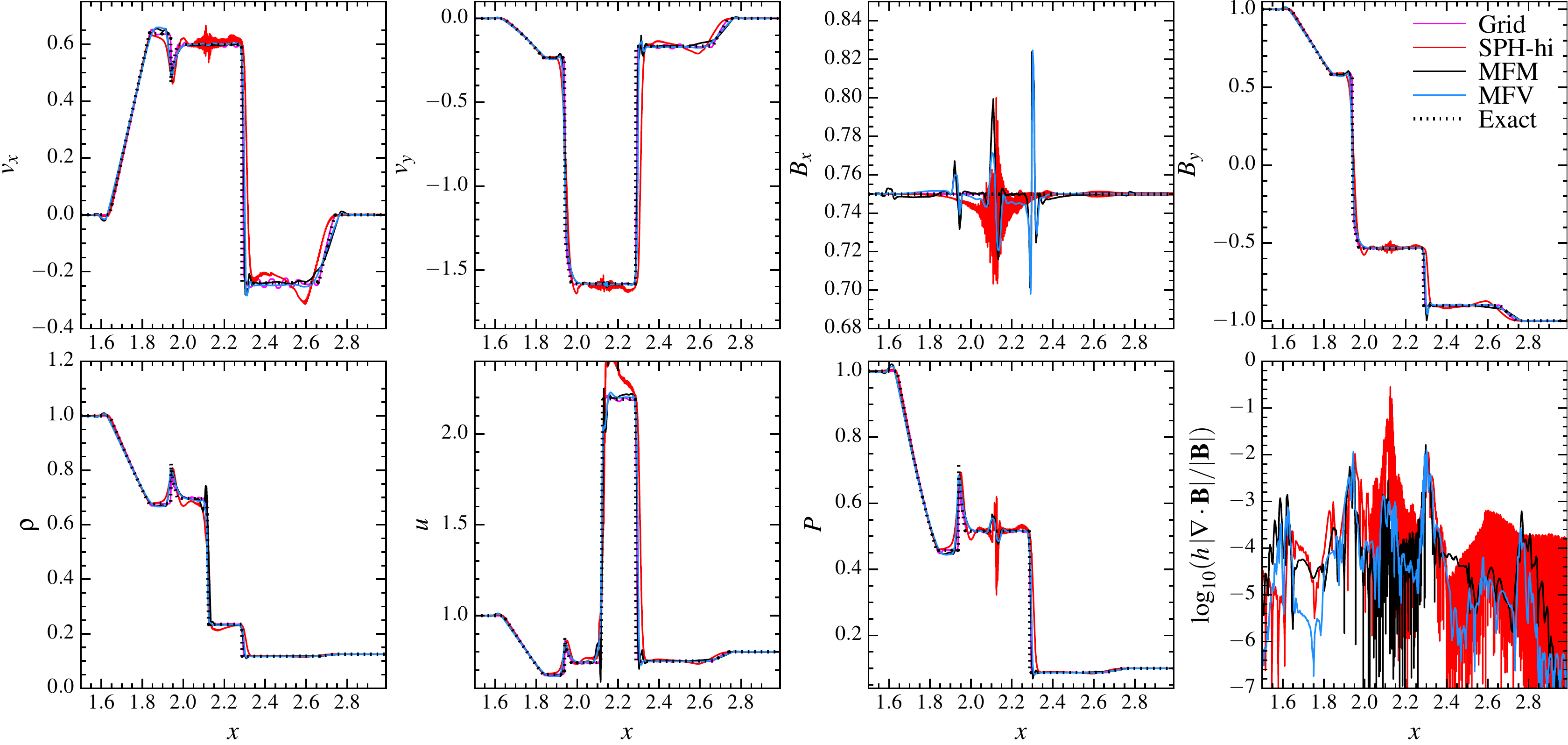}{0.98}
    \vspace{-0.25cm}
    \caption{Brio-Wu shocktube (\S~\ref{sec:briowu}), at time $t=0.2$; we compare the exact solution to that computed at finite resolution (plotted region contains $\sim 200$ elements across the $x$-direction) with different methods. 
    High-order grid methods have converged well at this resolution, except for post-shock ringing in $v_{x}$. MFM/MFV methods also show good convergence; but at this resolution, MFM still shows some small ``overshoot'' in the jumps at $x\approx2.1,\,2.3$ (more sensitive to our slope-limiter than the method itself), and both show some small (percent-level) errors in $B_{x}$ owing to the $\divB$ errors; however the fractional magnitude of $\divB$ is controlled well by our cleaning scheme (typical errors $\sim 10^{-4}$ at this resolution; still below $10^{-2}$ at jumps). Discontinuities are well-captured across $\sim 2$ cells/particles in the linear direction. SPH (with high $\NNb$) captures all the key features, but at this resolution shows larger noise, $\sim 20\%$-level overshoots in $v_{x},\,v_{y},B_{y},\,P$ at rarefactions ($x\approx 1.9,\,2.6$), some suppression of internal energy around $x\approx 2.3$ (owing to more smeared-out dissipation from divergence-cleaning), and significantly larger $\divB$ errors (reaching $\sim10\%$); these do converge away but more slowly. 
        \vspacerpostplot 
    \label{fig:briowu}}
\end{figure*}

\begin{figure*}
    \plotsidesize{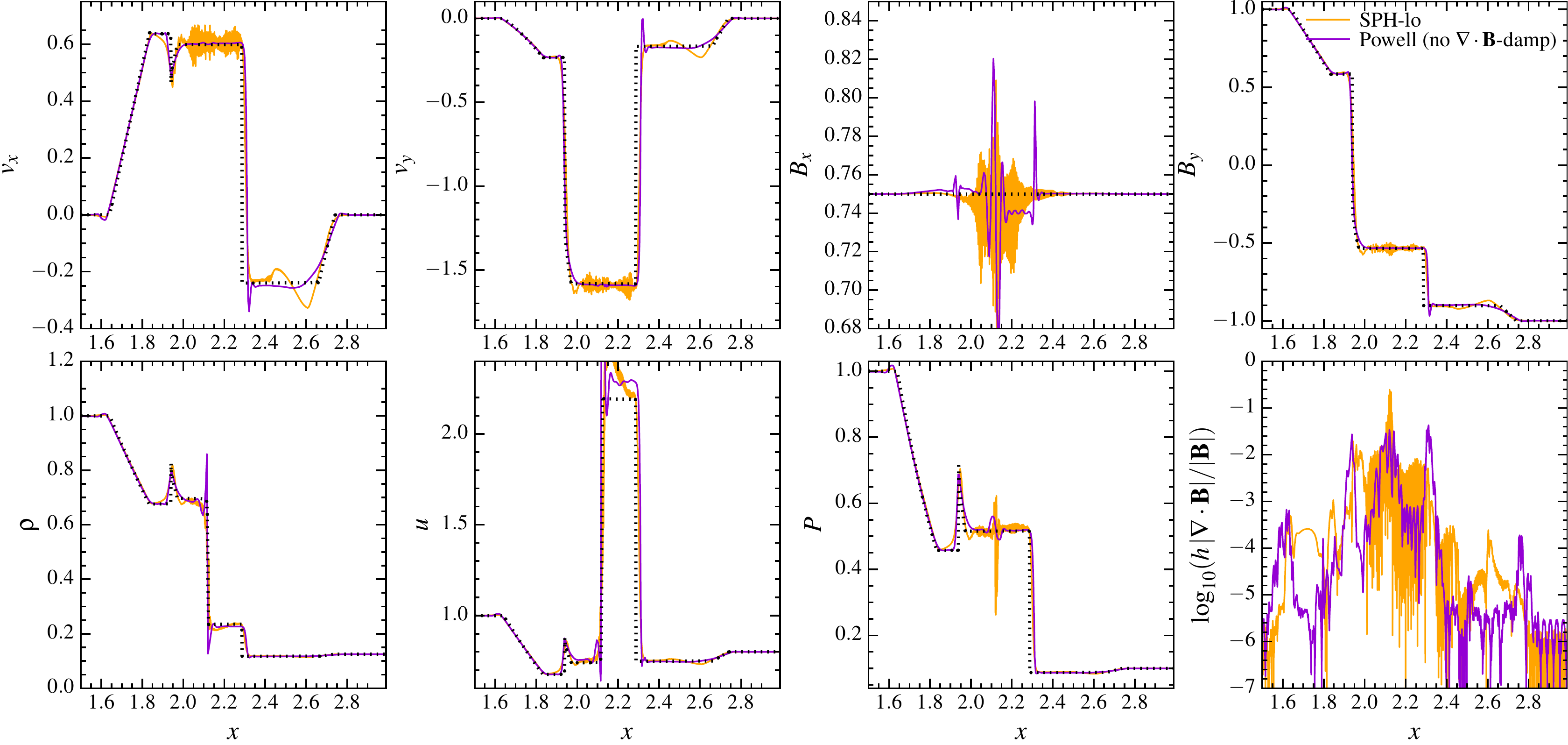}{0.95}
    \vspace{-0.25cm}
    \caption{Brio-Wu shocktube, as Fig.~\ref{fig:briowu}, for additional methods. If we consider MFM/MFV using only the \citet{powell:1999.8wave.cleaning} source terms to stabilize MHD, but no \citet{dedner:2002.divb.cleaning.scheme} divergence-control (as has often been done in the literature), we obtain incorrect shock jumps; most noticeably in $u$ \&\ $B_{x}$. This error {\em does not converge away}. This is despite the fact that the formal $\divB$ errors are still small; the key is the terms in the \citet{dedner:2002.divb.cleaning.scheme} scheme that enter the Riemann problem and act specifically at discontinuities. We also compare SPH run with the same (lower) neighbor number as our MFM/MFV methods; here the noise is larger (as expected).
    \vspacerpostplot 
    \label{fig:briowu.bad}}
\end{figure*}

\begin{figure}
    \plotonesize{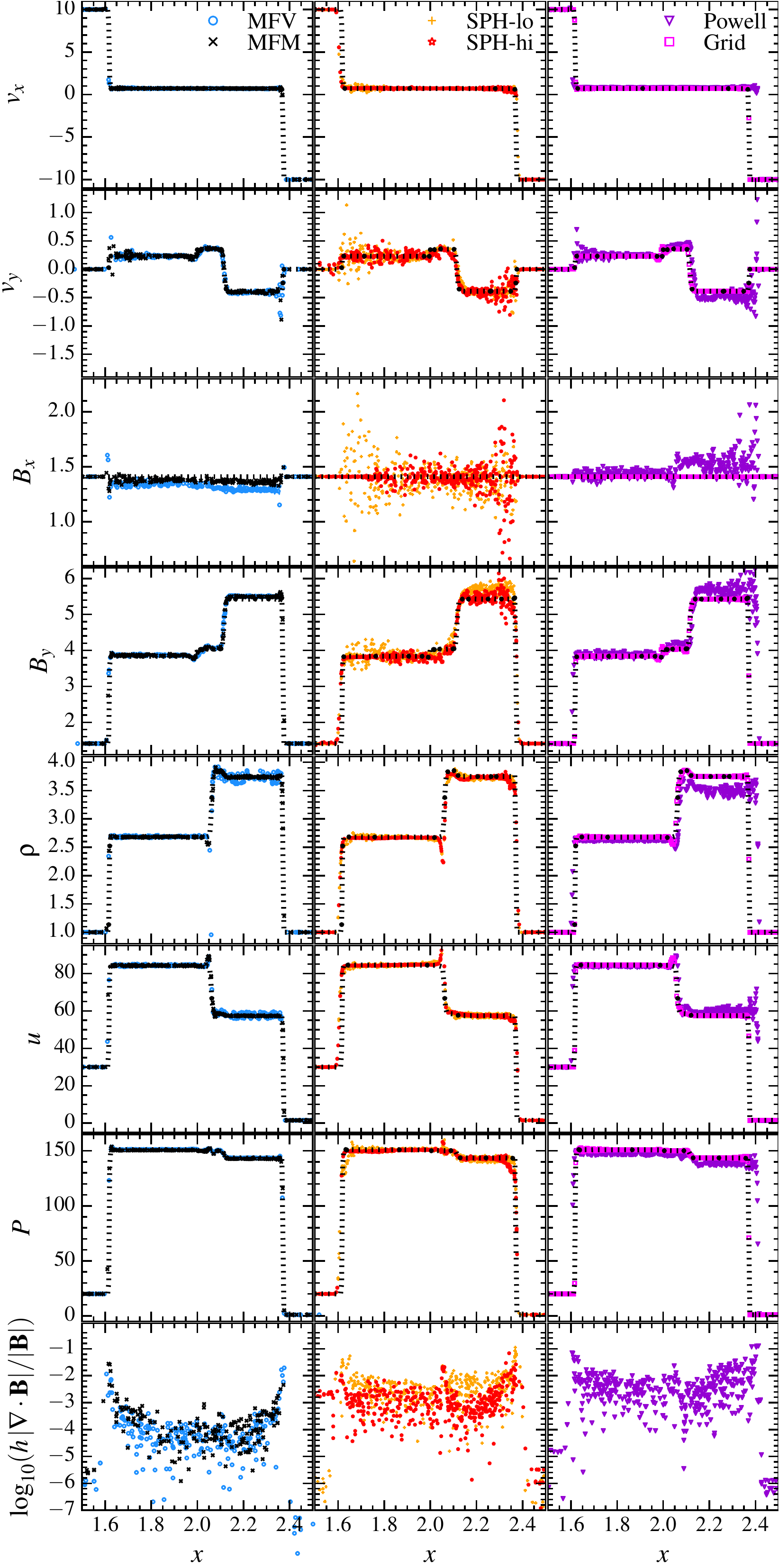}{0.97}
    \vspace{-0.25cm}
    \caption{The Toth super-sonic shocktube (\S~\ref{sec:toth}). The resolution is similar to the Brio-Wu test. MFM and grid/CT methods converge most rapidly to the exact solution (dotted), followed by MFV, which exhibits some residual noise in $\rho$ around $x\sim 2.1-2.4$ at this resolution. MFM/MFV both control $\divB$ errors well; the large $B_{y}$ discontinuity introduces a small (percents-level) offset in $B_{x}$ which converges away $\propto N^{-1}$ (nearly ideal), this is especially challenging for divergence-cleaning methods. SPH shows larger $\divB$ errors, and noise; with some systematic offset in the $B_{y}$ jump around $x\sim 2.2$ (as divergence cleaning is spread over several smoothing lengths), which converges slowly. The noise is reduced with larger neighbor number, but not eliminated. In all methods, the Powell scheme alone leads to systematically incorrect shock jumps in $B_{y}$ and $\rho$; as in Fig.~\ref{fig:briowu}, these do not converge. 
    \vspacerpostplot 
    \label{fig:toth}}
\end{figure}

\vspace{-0.5cm}
\subsection{Brio-Wu Shocktube: Capturing MHD Discontinuities \&\ Controlling Noise}
\label{sec:briowu}

Next we consider the \citet{brio:1988.shocktube} shocktube; this tests whether the code can accurately represent uniquely MHD shocks, rarefactions, and contact discontinuities. We initialize a 2D periodic box (size $0<x<4$, $0<y<0.25$, with $896\times56$ cells/particles) with left-state $(\rho,\,v_{x},\,v_{y},\,v_{z},\,B_{x},\,B_{y},\,B_{z},\,P) = (1,0,0,0,0.75,1,0,1)$, and right-state $(0.125,0,0,0,0.75,-1,0,0.1)$, with $\gamma=2$. Figs.~\ref{fig:briowu}-\ref{fig:briowu.bad} compare the results at time $t=0.2$. Note that our box is intentionally large and extends well beyond the ``active'' domain; the dynamically active region is only $\sim200$ elements across.\footnote{We have run this test with a number of different element configurations in our initial conditions, including: square, triangular, and hexagonal lattices, SPH and gravitational glasses, and random (Poisson) positions. We have also considered the case of equal particle masses (different particle spacing across the initial discontinuity) or unequal particle masses (same spacing). As expected from the derivation in \citet{gaburov:2011.meshless.dg.particle.method}, our MFM/MFV methods are only very weakly sensitive to these choices (slightly more noise appears in the ``worst case'' situation of random positions). Perhaps more surprising, our SPH results are also only weakly sensitive to this. We show results for a relaxed SPH glass but they are qualitatively identical to any of the other configurations. And we show explicitly in \paperone\ that the choice of equal/different particle masses makes a difference only in the magnitude of the errors just at the contact discontinuity.}

At this resolution, the CT-based grid code is well-converged, except for some post-shock ``ringing'' most visible in $v_{x}$. In MFM \&\ MFV methods, the agreement with the exact solution is good at this resolution. At $4x$ higher-resolution, we find that the MFM/MFV results are nearly indistinguishable from the exact solution. At lower resolution, there is some ``overshoot'' at the density discontinuity, with MFM -- this is discussed in detail in \paperone; it is mostly sensitive to the choice of slope-limiter (not the basic numerical method). The effects are much smaller in MFV, owing to mass fluxes allowing more sharply-captured density discontinuity. This also causes a small pressure ``blip'' in MFM at the contact discontinuity, but this converges away. Shock jumps and discontinuities are captured across $\sim2$ cells/particles in each direction; comparable to high-order grid methods. 

As an indicator of the $\divB$ errors, we plot the dimensionless magnitude of $h_{i}|\nabla\cdot {\bf B}|_{i} / |{\bf B}|_{i}$; our divergence cleaning keeps this generally at low values ($\ll 10^{-4}$), except at the magnetic shocks (the large discontinuities in $B_{y}$); but even there, the {\em maximum} value is still $< 10^{-2}$. The good divergence-cleaning is also manifest in $B_{x}$; at the same shocks, there are some jumps in $B_{x}$ generated, but this is returned to the correct value across $\sim2$ particles, and the magnitude of the deviations from the analytic $B_{x}=0.75$ is typically at the sub-percent level, at this resolution. 

With {\em no} divergence corrections whatsoever, the problem crashes. However, Fig.~\ref{fig:briowu.bad} shows that if we run with {\em only} the Powell terms in Eq.~\ref{eqn:source} (i.e.\ do not include the \citet{dedner:2002.divb.cleaning.scheme} divergence-cleaning and damping terms), this particular problem is not badly corrupted in most respects. As expected the divergence and deviation in $B_{x}$ are larger. More seriously, though, a systematically incorrect jump in $u$ appears, which {\em does not converge away} without divergence cleaning, even at $10x$ higher resolution in the $x$-direction. This problem occurs in MFM, MFV, and SPH.

SPH captures all the qualitative features. However, if we run with the same neighbor number (kernel size) as in our MFM/MFV methods, the noise is larger. This is shown in Fig.~\ref{fig:briowu.bad}. If we instead use the equivalent of a 3D neighbor number of $\sim 120$ (as opposed to the $\sim32$ we use for MFM/MFV), the noise is reduced, as expected. However, there is still much larger noise and post-shock ringing (compared to MFM/MFV), and significant overshoot in the velocities and $u$ at the rarefactions.\footnote{In extensive experiments, we found that the magnitude of these errors in SPH, at fixed resolution, is determined by the artificial viscosity and resistivity schemes. If we simply assume a constant (large) artificial viscosity (i.e.\ disable our normal ``shock detection'' switch) we are able to eliminate most of the noise, and reduce the overshoot, in SPH in Figs.~\ref{fig:briowu}-\ref{fig:briowu.bad}. However, such a choice would severely degrade the performance of our SPH implementation on almost every other problem we consider. As an attempted compromise, in Appendix~\ref{sec:spmhd.numerics}, we discuss modifications to the default \citet{cullen:2010.inviscid.sph} viscosity scheme and \citet{tricco:2013.sphmhd.methods} resistivity scheme, which we have used here. Using instead the default version of the viscosity scheme makes only small differences for large neighbor number (``SPH-hi''), but produces much larger (order-unity) noise levels at low-neighbor number (``SPH-lo'') on this test. In highly super-sonic tests, the difference is negligible.} Divergence-cleaning works in SPH, but is much less effective, especially around the contact discontinuity, where the divergence errors reach $\sim 10\%$. We stress though that these errors are resolution-dependent, and do converge away eventually. Better results (at a given resolution) can also be obtained on shocktube problems in SPH by increasing the strength of the artificial dissipation terms (viscosity/resistivity/conductivity); however this significantly degrades performance on other tests we consider. For an example of this test which demonstrates good agreement between SPH and the exact solution (by combining higher resolution and stronger dissipation), see \citet{tricco:2013.sphmhd.methods}.

We have also compared the \citet{ryu:1995.mhd.test.problems} MHD shocktube, which exhibits all seven MHD waves simultaneously. The qualitative results and differences between methods are the same as in the Brio-Wu test, so we do not show it here.

\vspace{-0.5cm}
\subsection{Toth Shocktube: The Critical Need for Divergence Cleaning Beyond Powell}
\label{sec:toth}

Next we consider the \citet{toth:2000.divB.constraint} shocktube; this tests super-sonic MHD shocks. It is particularly important because \citet{mignone:2010.ctu.mhd.divb.constraint} showed that a hyperbolic divergence-cleaning scheme such as the \citet{dedner:2002.divb.cleaning.scheme} method (or CT) is necessary to get the correct shock jump conditions at any resolution, in grid-based methods. 
We initialize a 2D periodic box with the same initial element configuration as the \citet{brio:1988.shocktube},\footnote{Again, we have considered a variety of initial element configurations. In the \citet{toth:2000.divB.constraint} shocktube we find these produce negligible differences.} with left-state $(\rho,\,v_{x},\,v_{y},\,v_{z},\,B_{x},\,B_{y},\,B_{z},\,P) = (1,10,0,0,5/\sqrt{4\pi},5/\sqrt{4\pi},0,20)$, and right-state $(1,-10,0,0,5/\sqrt{4\pi},5/\sqrt{4\pi},0,1)$ and $\gamma=5/3$. 

Fig.~\ref{fig:toth} compares the results at time $t=0.08$ (for clarity, we plot only a randomly-chosen subset of $500$ cells/particles for each method). Our MFM method does extremely well at this resolution; the only difference between it and high-resolution grid runs is a small ($\sim1\%$ level) deviation in $B_{x}$ introduced by $\divB$ errors at this resolution, and some noise at the shock in the small $v_{y}$. The $\divB$ errors are generally extremely small, $\sim1\%$ at super-sonic shocks and $\sim 10^{-4}$ elsewhere. MFV is similar, although there is substantially more noise at the same resolution in $\rho$ and $u$ in the post-shock region; this is seen in pure hydro in \paperone\ and in \citet{gaburov:2011.meshless.dg.particle.method}; the larger noise translates to slightly larger $\divB$ errors at shocks, hence slower convergence in $B_{x}$. Both MFM \&\ MFV are indistinguishable from the exact solution at $10x$ larger resolution in the $x$-direction. 

However, with the Powell-only (no divergence-cleaning) mode, we see that the shock is in the wrong place! The shock jump is systematically wrong in $\rho$, $P$ and $B_{y}$, and this leads to it being in the wrong place over time. Again, this appears even at infinite resolution. The noise, especially in $B_{x}$, is also much larger, and $\divB$ errors, as expected, are factor $\sim100$ larger. 

In SPH, the noise is much larger, even with much larger 3D-equivalent $\NNb=120$. This is especially noticeable in $v_{y}$, $B_{x}$, and $B_{y}$. In $B_{y}$, the shock jump is also systematically over-estimated by $\sim 2-5\%$, owing to the much larger ($\sim 10-30\%$) $\divB$ errors at the shock jump. As discussed below (\S~\ref{sec:discussion}), in SPH, divergence-cleaning cannot act over smaller scales than a few smoothing lengths, so the method has difficulty controlling the errors seen in the Powell-only case. The divergence errors are systematically larger by factors $\sim10-100$. With $\NNb=32$ as in our MFM/MFV methods, the noise is yet larger (order-unity fractional noise in $B_{x}$, $v_{y}$).

\begin{figure}
    \plotonesize{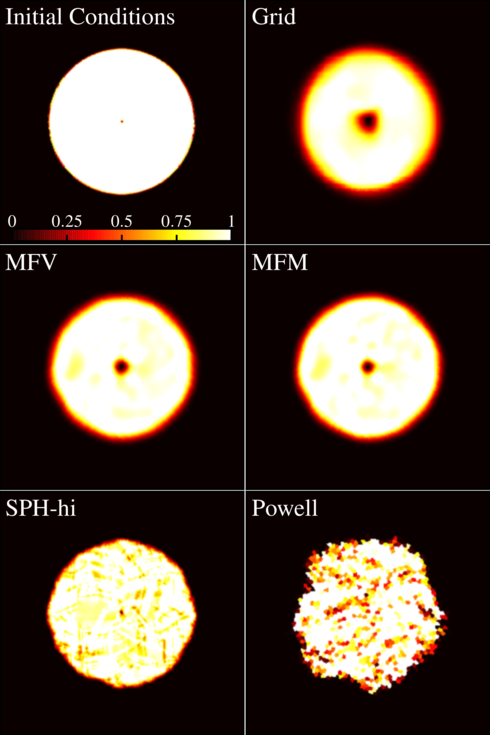}{0.98}
    \vspace{-0.25cm}
    \caption{Field loop advection test (\S~\ref{sec:field.loop}). An equilibrium, 2D field loop is advected uniformly across the domain; we plot $|{\bf B}|^{2}$ (in units of the initial loop value, as labeled) at time $t=20$, for tests with $256^{2}$ resolution. Here we use in initially Cartesian particle lattice with unequal-mass particles (the density within the loop is a factor $\sim2$ higher than the background). This is the natural configuration for fixed-mesh codes but represents a near ``worst case'' for Lagrangian codes. In all cases, the initial conditions (top left) should be reproduced identically. In non-moving grid methods, even at arbitrarily high order, advection errors diffuse the loop, while numerical resistivity reduces the central field strength. In MFM/MFV/SPH, the advection errors are eliminated, and numerical resistivity at the center is reduced; however divergence-cleaning and ``grid noise'' from different particle masses around the loop edge produce some diffusion and noise (the peak value of $h\,|\divB|/|{\bf B}|$ in any cell at any time remains $\ll 0.01$, however). The Powell-only scheme exhibits much more severe noise, because the $\divB$ errors are transported but not damped; this leads to non-linear corruption of the solution.
    \vspacerpostplot 
    \label{fig:field.loop.im}}
\end{figure}

\begin{figure}
    \plotonesize{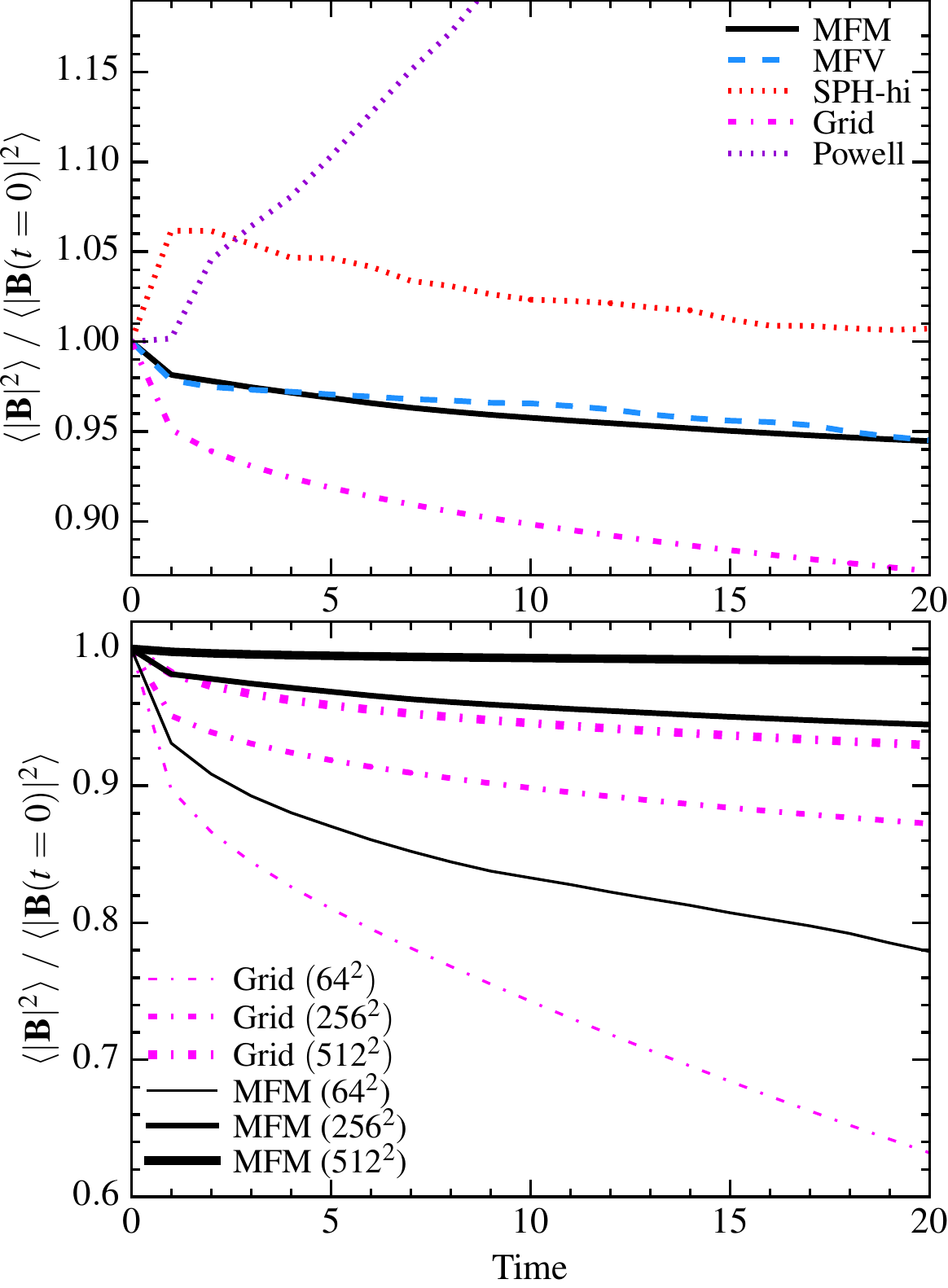}{0.9}
    \vspace{-0.25cm}
    \caption{Quantitative decay of the box-averaged magnetic energy in the field loop test (Fig.~\ref{fig:field.loop.im}), owing to numerical diffusion/resistivity. At infinite resolution, methods preserve the initial $\langle |{\bf B}|^{2} \rangle$. Because they are Lagrangian, MFM/MFV/SPH methods show much less dissipation than high-order grid methods at the same resolution (default runs here are $256^{2}$, but we compare $64^{2}-512^{2}$ MFM/grid runs for reference). 
    SPH shows some spurious initial growth of $\langle |{\bf B}|\rangle$ as divergence-cleaning acts on zeroth-order kernel errors, but the subsequent decay rate is close to MFM/MFV. Otherwise on this test SPH is not as sensitive to $\NNb$. Powell-only methods are unstable on this problem, and lead to artificial field amplification.
    \vspacerpostplot 
    \label{fig:field.loop.decay}}
\end{figure}

\begin{figure}
 \begin{tabular}{c}
  \vspace{0.1cm}
  \hspace{0.6cm}\includegraphics[width=0.84\columnwidth]{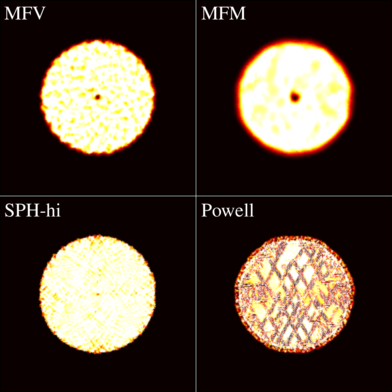} \\
  \hspace{-0.2cm}\includegraphics[width=0.95\columnwidth]{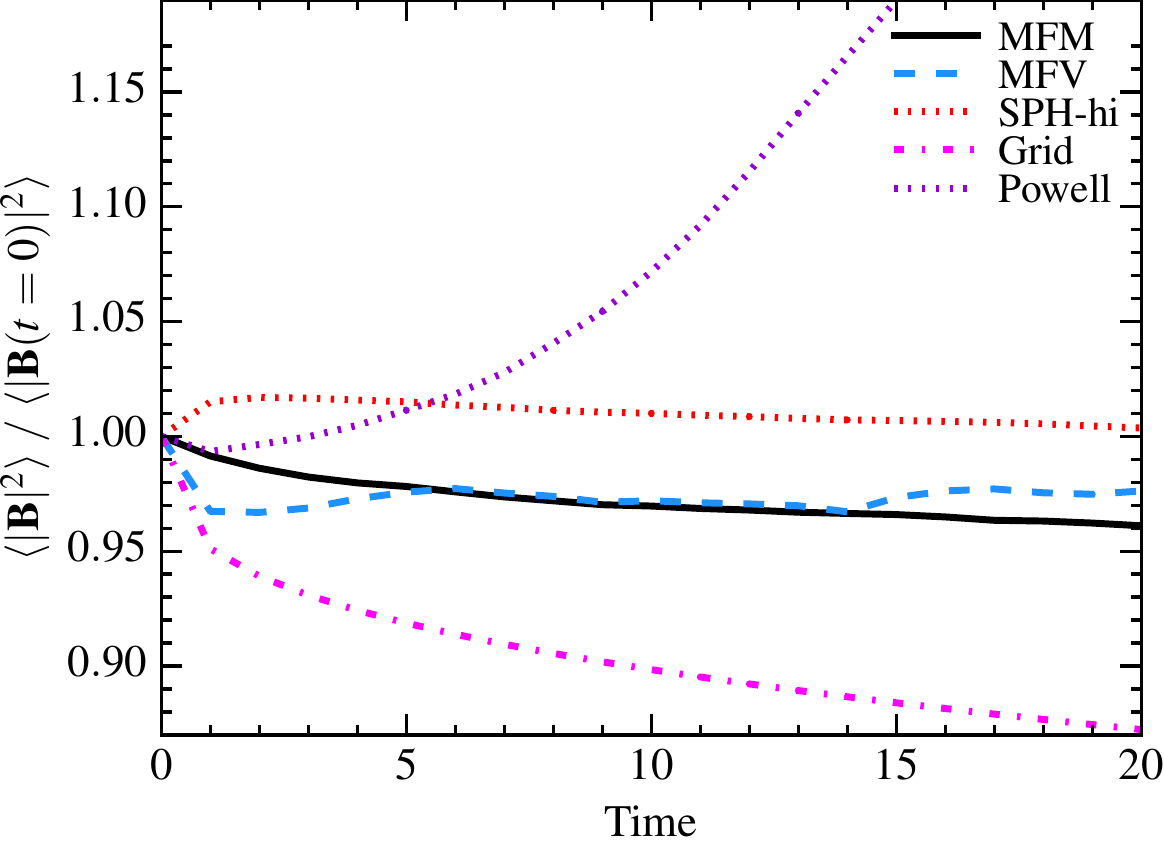}
 \end{tabular}
    \vspace{-0.25cm}
    \caption{As Figs.~\ref{fig:field.loop.im}-\ref{fig:field.loop.decay}, but with a different initial particle configuration: here a triangular lattice with constant particle mass (with fixed total particle number, so the resolution is a factor of $(\sqrt{2})^{2}$ higher within the magnetized loop). In this case, the errors are reduced in all our Lagrangian methods. In particular, the noise in SPH, which is sensitive to both the local particle arrangement and differences in particle masses, is greatly reduced. Qualitatively, the features in all cases are identical, however.
    \vspacerpostplot 
    \label{fig:field.loop.decay.triangles}}
\end{figure}

\vspace{-0.5cm}
\subsection{Advection of a Field Loop: Minimizing Numerical Diffusion}
\label{sec:field.loop}

The next test is a standard test of advection errors and numerical dissipation. We initialize a periodic 2D domain: inside a circle of $R=\sqrt{x^{2}+y^{2}} < R_{0}=0.3$ about the origin, we set $(\rho,\,B_{x},\,B_{y})=(2,\,B_{0}\,y/R,-B_{0}\,x/R)$ with $B_{0}=10^{-3}$. Outside the circle $(\rho,\,B_{x},\,B_{y})=(1,\,0,\,0)$, and everywhere $(P,\,v_{x},\,v_{y},v_{z},B_{z})=(1,2,1/2,0,0)$; this is an equilibrium configuration that should simply be advected. In Fig.~\ref{fig:field.loop.im}, we plot images of the magnetic energy density. In Fig.~\ref{fig:field.loop.decay}, we plot the total magnetic energy in the box as a function of time (which should remain constant at its initial value). For the sake of direct comparison between grid and mesh-free codes, in these plots we take the initial element configuration to be a Cartesian grid, with unequal-mass elements.

Advection of any configuration not perfectly aligned with the grid is challenging in grid codes; here the loop is continuously diffused away, at a rate that increases rapidly at lower resolution. In Lagrangian methods, on the other hand, stable configurations with bulk advection should be advected perfectly. In \paperone, we demonstrate that our MFM \&\ MFV methods can advect arbitrary pressure-equilibrium hydrodynamic configurations (including arbitrary scalar quantities) to within machine accuracy. However, here the introduction of the divergence-cleaning source terms leads to some initial diffusion of ${\bf B}$ in MFM/MFV. This is enhanced by the fact that the particle masses change discontinuously at the ``edge'' of the loop. But still, we clearly see the benefit of a Lagrangian method: convergence is much faster than in fixed-grid codes (even using CT); the dissipation in our $256^{2}$ simulation with MFM/MFV is approximately equivalent to that in {\small ATHENA} at $1024^{2}$. In the image, we see slightly more noise; this is the expected ``grid noise'' which is higher in meshless methods, but the diffusion is less (in particular, the ``hole'' which appears at the center owing to numerical resistivity is minimized).\footnote{Of course, in any of our Lagrangian methods, it is trivial to obtain perfect evolution of the field loop test for arbitrarily long times, if we simply disable any dissipation terms. In MFM/MFV this amounts to invoking the ``energy-entropy'' switch described in \paperone\ (setting it always to ``entropy,'' i.e.\ using purely adiabatic fluxes), and in SPH it amounts to disabling the artificial dissipation terms \citep[see examples in e.g.][]{rosswog:2007.sph.mhd,price:2012.sph.review}. Then, because the system is in uniform motion there is zero advection and the evolution is trivial. The same is true for moving-mesh codes. But these changes make the methods numerically unstable (and would produce disastrous errors at any shocks or discontinuities). We therefore will not consider such modifications further.}

If we use Powell-only divergence subtraction, the conservation errors associated with non-zero $\divB$ can actually lead to non-linear growth of ${\bf B}$, such that the total magnetic energy increases! The growth in this case is nearly resolution-independent, since it is sourced around the sharp discontinuity in the field at the edge of the loop. By the end of the simulation, the total magnetic energy in the Powell-only run has increased $\sim50\%$. The noise and asymmetry in the image have grown severely.

In SPH, there is an initial, brief but unphysical growth in $|{\bf B}|^{2}$; this comes from the divergence-cleaning being less effective than in MFM/MFV (so it behaves like the Powell case); however once enough diffusion and particle re-arrangement has occurred, the divergence-cleaning operator can work effectively, and the energy decays at approximately the same rate as our MFM/MFV calculations.

As noted above, the noise and dissipation in the mesh-free methods is enhanced by the artificial jump in element masses at the edge of the loop (this is not a ``natural'' configuration for our mesh-free methods). We therefore consider in Fig.~\ref{fig:field.loop.decay.triangles} the same test, with the same {\em total} number of elements in the box, with an initial triangular lattice configuration and equal-mass elements. The combination of reduced noise at the loop edge, and a factor $\sim 1.25^{2}$ higher resolution within the loop (since it is higher-density), does reduce the noise and errors significantly. However, the qualitative behavior in every case is identical.

\begin{figure}
    \plotonesize{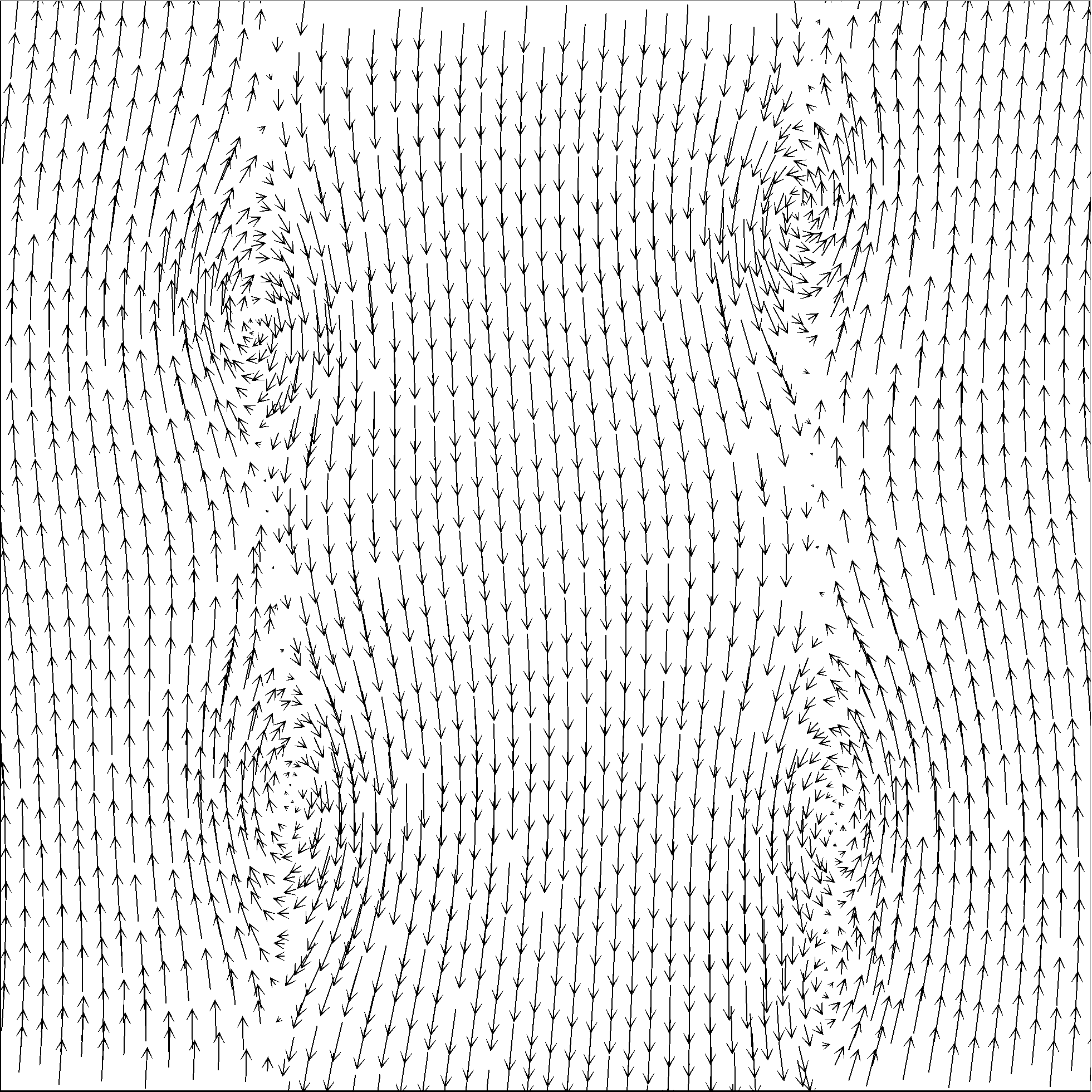}{0.9}
    \vspace{-0.25cm}
    \caption{Hawley-Stone current sheet (\S~\ref{sec:current.sheet}). We plot magnetic field lines (arrows indicate local field strength and direction) at time $t=5$ with $\beta=0.1$, $A=0.1$, in MFM. MFV, SPH, and grid-methods produce very similar results for these parameters. Reconnection along the current sheets leads to magnetic ``islands'' which grow and merge. Our new mesh-free methods are able to stably evolve the current sheet indefinitely, with $h\,|\divB|/|{\bf B}|\ll 10^{-2}$ for all cells at all times.
    \vspacerpostplot 
    \label{fig:current.sheet}}
\end{figure}

\begin{figure}
    \plotonesize{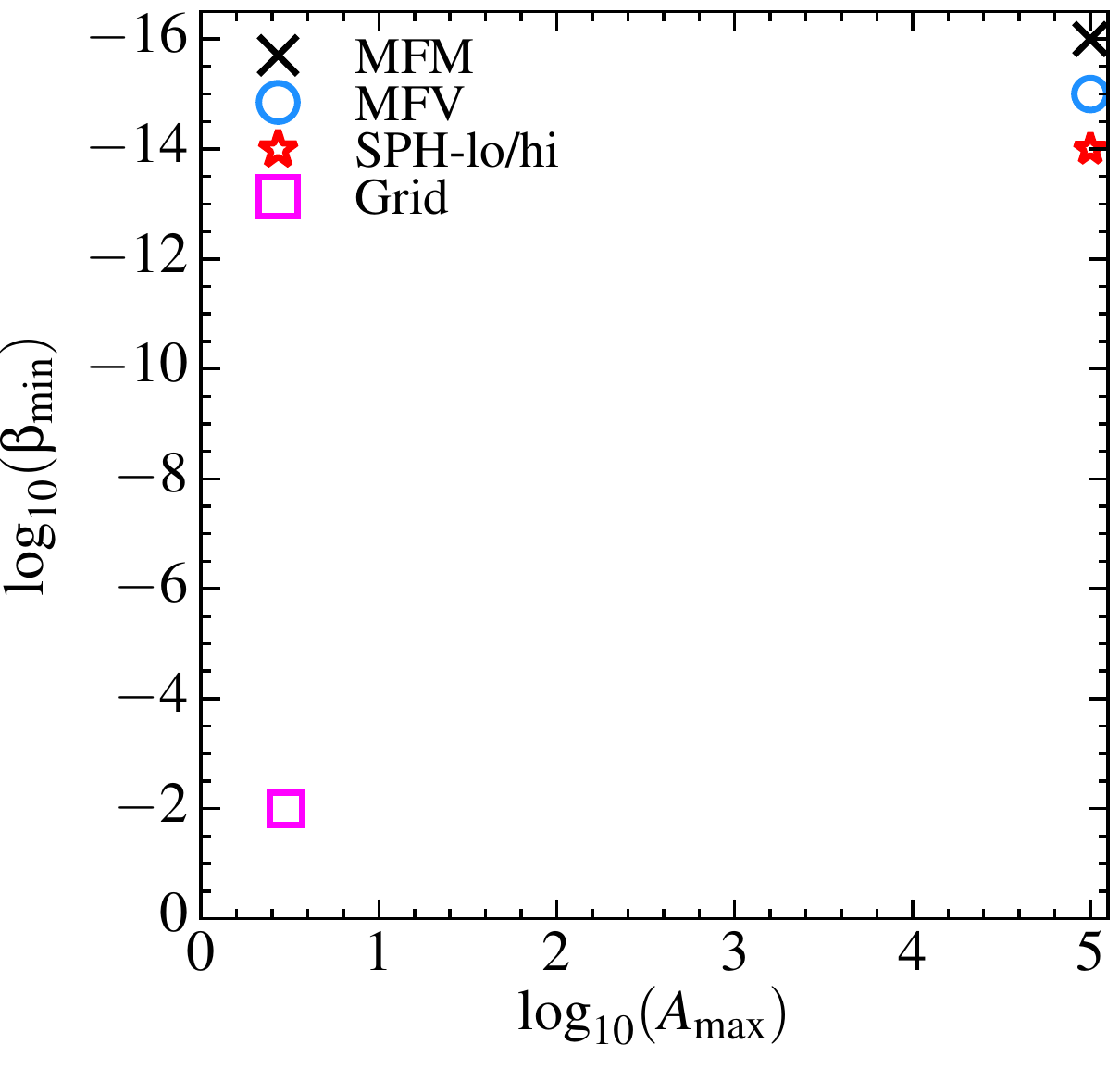}{0.9}
    \vspace{-0.25cm}
    \caption{Stability limits of the current sheet problem in Fig.~\ref{fig:current.sheet}. Given an initial pressure $P=\beta/2$, and velocity perturbation with amplitude $A$, we consider the minimum $\beta$ and maximum $A$ for which the problem can be evolved stably to time $t=10$ (further to the top-right is more-stable). In the grid-based CT method of {\small ATHENA}, the total-energy formulation of the code, coupled with high-accuracy subtraction needed for accurate CT, and advection errors when the fluid moves over the grid, mean that the method will crash (negative pressures result) for $\beta\le0.01$ or $A\ge3$. Combining the duel-energy formalism from \paperone, with a Lagrangian method that moves with the fluid, and using divergence-cleaning instead of CT, we are able to stably evolve the system until $\beta$ reaches machine-error levels $\sim 10^{-16}$, and arbitrarily large $A\gtrsim10^{5}$ (we have not considered larger $A$ only because the simulations become too expensive, not because they crash). 
    \vspacerpostplot 
    \label{fig:current.sheet.limits}}
\end{figure}

\vspace{-0.5cm}
\subsection{Hawley-Stone Current Sheet: Numerical Stability}
\label{sec:current.sheet}

This test follows \citet{hawley:1995.mhd.tests}. In a 2D periodic domain with $-0.5<x<0.5$, $-0.5<y<0.5$, we initialize $(\rho,\,P,\,v_{x},\,v_{y},\,v_{z},\,B_{x},\,B_{z})=(1,\,\beta/2,\,A\sin{(2\pi\,y)},\,0,\,0,\,0,\,0)$ and $\gamma=5/3$, with $B_{y}/(4\pi)^{1/2}=1$ for $|x|>0.25$ and $B_{y}/(4\pi)^{1/2}=-1$ otherwise. This is not a good test of algorithm accuracy, since the non-linear solution depends sensitively on the numerical dissipation in different methods. However, it is a powerful test of code robustness. Qualitatively, the solution should exhibit rapid reconnection along the initial current sheet, which will launch nonlinear polarized Alfven waves, that generate magnetosonic waves, while magnetic islands form, grow, and merge. For smaller $\beta$ and larger $A$, it becomes more difficult for algorithms to evolve without crashing or returning unphysical solutions (e.g.\ negative pressures in the Riemann problem).

Fig.~\ref{fig:current.sheet} shows the magnetic topology at time $t=5$ in a run with $\beta=0.1$, $A=0.1$. For these parameters, MFM, MFV, SPH, and grid methods (here, {\small ATHENA}) all look very similar.\footnote{For extensive description of this test problem in {\small ATHENA}, see \burl{http://www.astro.virginia.edu/VITA/ATHENA/cs.html}} The real test arises when we vary $\beta$ and $A$; in Fig.~\ref{fig:current.sheet.limits}, we plot the maximum $A$ and minimum $\beta$ which we are able to use in each algorithm before the code crashes or returns an unphysical result. {\small ATHENA} crashes after some small early-time evolution for $\beta \le 0.01$ or $A \ge 3$. The low-$\beta$ problem most likely owes to the fact that the method evolves total energy: when the magnetic energy dominates, this causes serious difficulty recovering the correct internal energy (since we must subtract two large numbers), eventually producing negative temperatures. Here we use a dual-energy formalism described in \paperone, which does not conserve total energy to machine error, but can handle essentially arbitrary ratios. So for our SPH, MFM, MFV implementations, $\beta_{\rm min}$ is limited by essentially machine error ($\beta_{\rm min}\sim 10^{-14}-10^{-16}$, depending on the formulation). For $A$, we find similar results; the increased stability owes both to the same dual-energy formalism above, but also to the Lagrangian method, which eliminates the advection errors that, in grid-based codes, become larger with the local fluid velocity. In non-moving grid codes, eventually, at any resolution, there is some bulk velocity which will wipe out the correct physical solution completely (necessitating still-higher resolution); this is avoided in Lagrangian methods. We explore only values of $A$ up to $\sim 10^{5}$ in this test because the timestep becomes so small that it is impractical to evolve the system to late non-linear times, but we suspect the robustness of the algorithms should hold to similar machine error levels (i.e.\ allowing $A\sim 10^{16}$).  

\begin{figure}
    \plotonesize{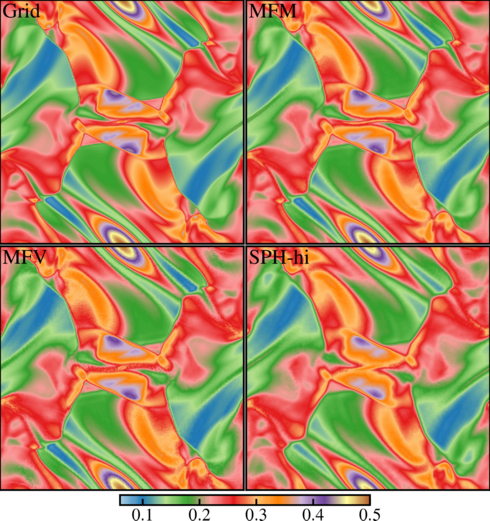}{0.98}
    \vspace{-0.25cm}
    \caption{The Orszag-Tang vortex (\S~\ref{sec:orszag.tang}). We show images of density $\rho$ (in code units, as labeled) at time $t=0.5$, in runs with $256^{2}$ elements (particles/cells). All methods develop the major qualitative features, though there is some additional smoothing in SPH. Note that the contact discontinuities and shocks are captured sharply in MFM/MFV methods. 
    \vspacerpostplot 
    \label{fig:orszag}}
\end{figure}

\begin{figure}
    \plotonesize{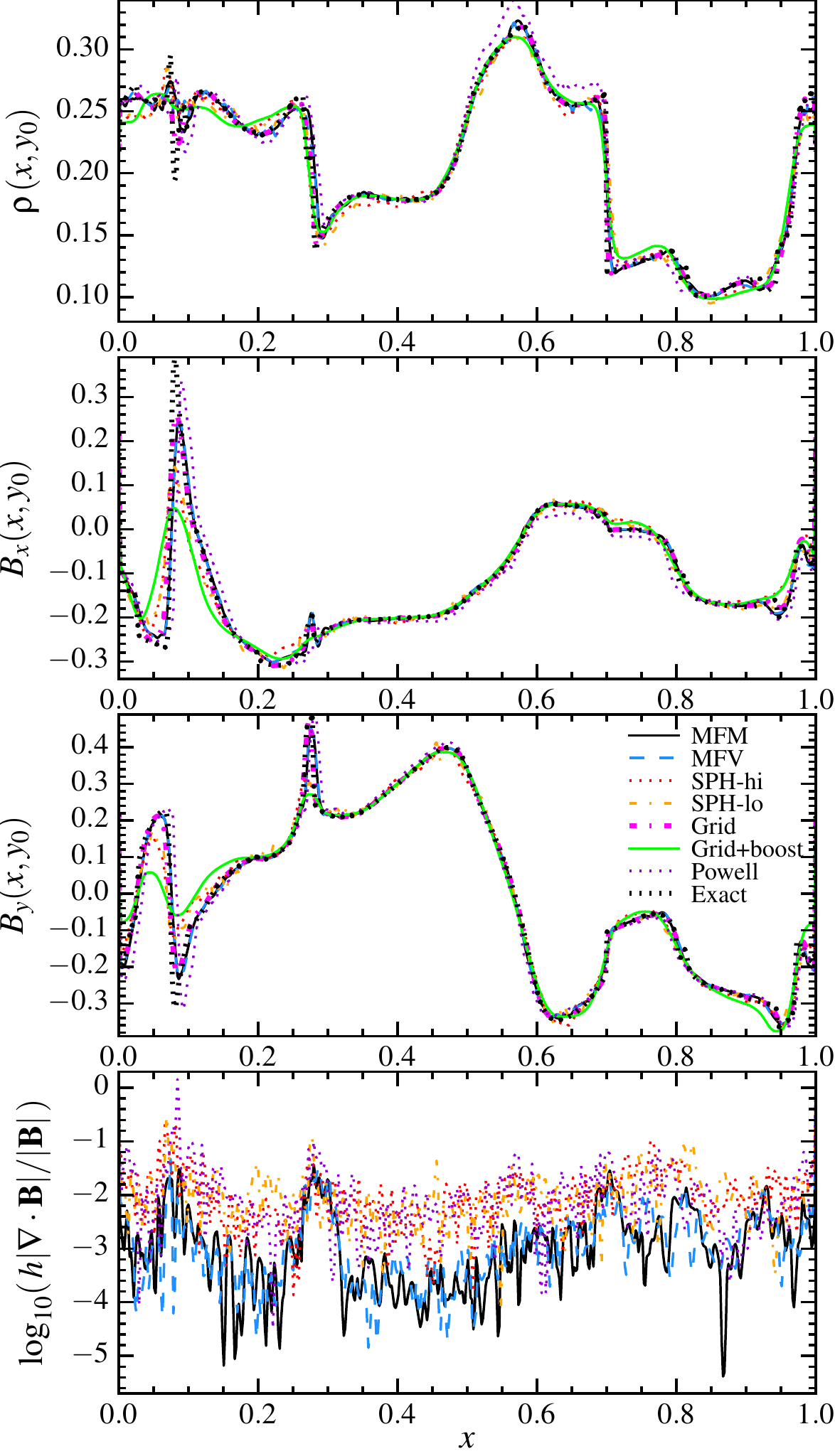}{0.98}
    \vspace{-0.25cm}
    \caption{Comparison of the Orszag-Tang problem from Fig.~\ref{fig:orszag}. We plot density $\rho$, magnetic field components $B_{x}$, $B_{y}$, and $\nabla\cdot{\bf B}$, in horizontal slices at $y_{0}=0.3125$. With $256^{2}$ cells, MFM, MFV, and grid methods have converged well to the exact solution (except for some small smoothing of the sharpest features). Here, the ``exact'' line is a $2048^{2}$ {\small ATHENA} PPM CT result; at resolution $>512^{2}$, MFM/MFV/unboosted-{\small ATHENA} results are indistinguishable. SPH performs well but shows further smoothing which converges more slowly. We also consider a grid simulation where the fluid is given an additional boost (a uniform $v_{x}=10$); the Lagrangian (MFM/MFV/SPH) methods are invariant to these boosts. But in grid codes the boost produces a smoothing of the ${\bf B}$ features at $x\sim0-0.2$; these require grid resolution $>1024^{2}$ to converge away. Using only Powell divergence subtraction leads to a systematic error in $\rho$ which is small, but does not converge away.
    \vspacerpostplot 
    \label{fig:orszag.numbers}}
\end{figure}

\begin{figure}
    \plotonesize{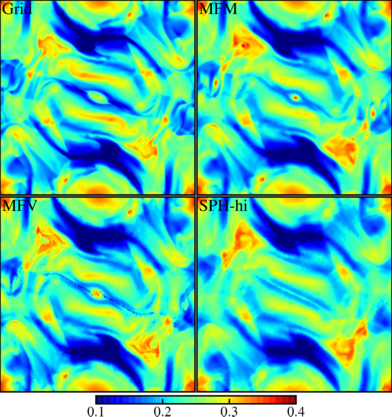}{0.98}
    \vspace{-0.25cm}
    \caption{Comparison of the Orszag-Tang problem as Fig.~\ref{fig:orszag}, but at time $t=1$, when parts of the flow have broken up into chaotic turbulence. Unsurprisingly the differences between methods have grown, but the results are still qualitatively similar. Note that the additional SPH smoothing in Fig.~\ref{fig:orszag} has here suppressed the formation of the central compact vortex; this feature is most sharply resolved in our MFM/MFV runs.
    \vspacerpostplot 
    \label{fig:orszag.image.t1}}
\end{figure}

\begin{figure}
    \plotonesize{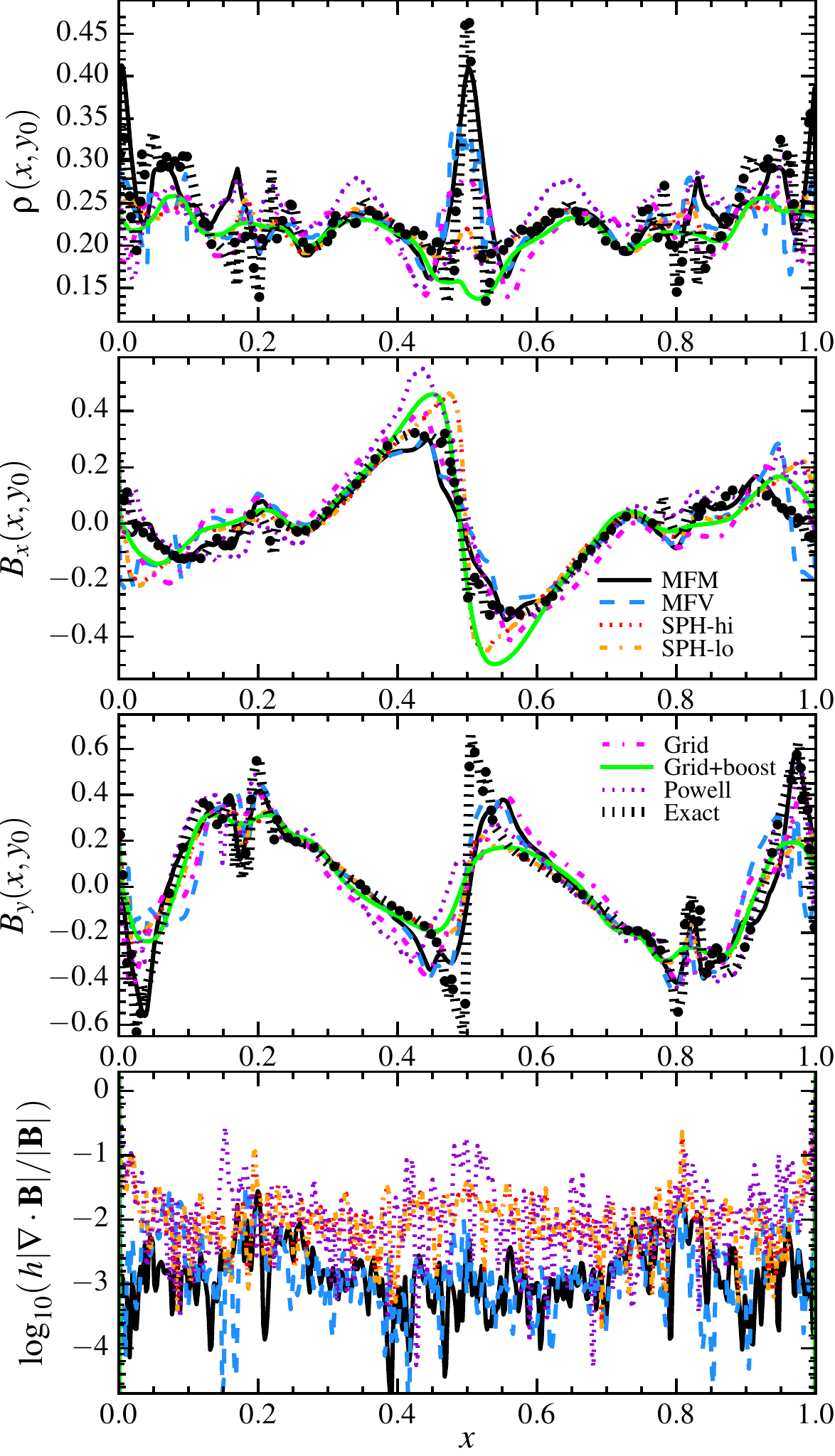}{0.98}
    \vspace{-0.25cm}
    \caption{Quantitative comparison of the Orszag-Tang problem as Fig.~\ref{fig:orszag.numbers}, but at time $t=1$. Here we take the slice through the center ($y=1/2$), where differences are maximized. The density plot demonstrates the presence of the central vortex in particular; it is suppressed by smoothing in the grid code and disappears in the boosted grid, SPH, and Powell results at this resolution.
    \vspacerpostplot 
    \label{fig:orszag.numbers.t1}}
\end{figure}

\begin{figure*}
 \begin{tabular}{lr}
  \includegraphics[width=0.95\columnwidth]{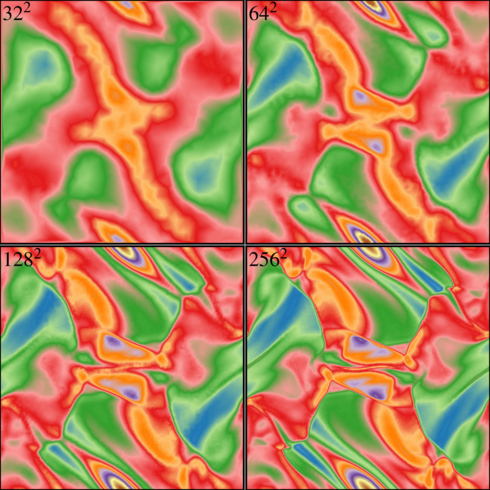} 
  \includegraphics[width=0.95\columnwidth]{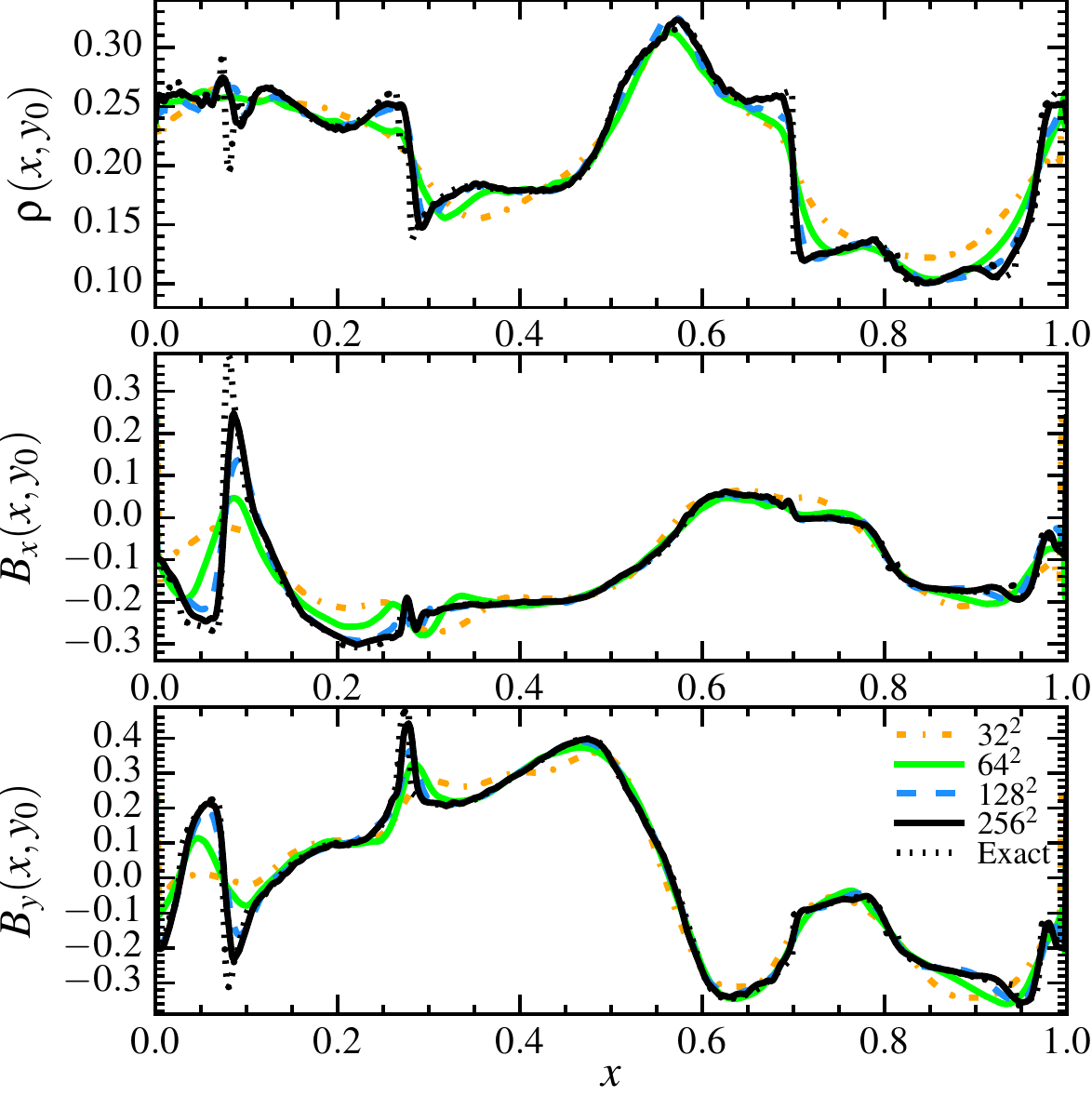}
 \end{tabular}
    \vspace{-0.25cm}
    \caption{Resolution study of the Orszag-Tang vortex in our MFM method (MFV is essentially identical). We show both images as Fig.~\ref{fig:orszag} and slices as Fig.~\ref{fig:orszag.numbers}, for several resolutions. Most major features are present even at low resolution; convergence with higher resolution is clear for all features. The formal $L_{1}$ and $L_{2}$ convergence accuracy is close to ideal scaling $\propto N^{-1}$ for this problem (given that it includes shocks and discontinuities). 
    \vspacerpostplot 
    \label{fig:orszag.resolution}}
\end{figure*}

\begin{figure}
    \plotonesize{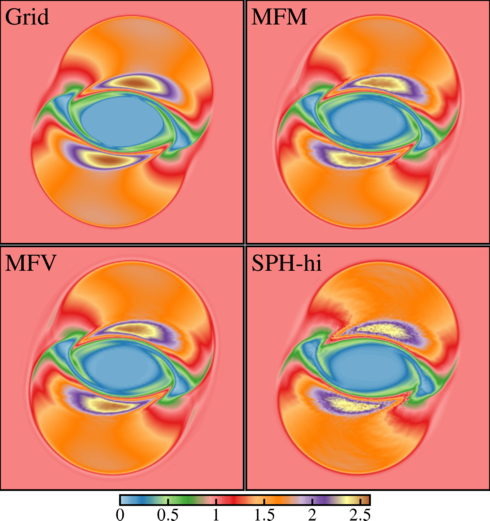}{0.9}
    \vspace{-0.25cm}
    \caption{Magnetic rotor (\S~\ref{sec:rotor}).  We show images as Fig.~\ref{fig:orszag} of the magnetic pressure $|{\bf B}|^{2}/2$ (in code units, as labeled), for runs with $256^{2}$ resolution. Most features appear identical; however note some difficulty in SPH capturing the pressure extrema at $y\approx 0.5\pm 0.15$, and additional noise in SPH (especially with low neighbor number). With Powell-only cleaning, similar errors and noise appear. 
    \vspacerpostplot 
    \label{fig:rotor}}
\end{figure}

\vspace{-0.5cm}
\subsection{Orszag-Tang Vortex: Shock-Capturing \&\ Super-Sonic MHD Turbulence}
\label{sec:orszag.tang}

The Orszag-Tang vortex is a standard MHD test which captures a variety of MHD discontinuities, and develops super-sonic MHD turbulence, which is particularly challenging for many methods. In a periodic 2D domain of unit size, we take $\gamma=5/3$ and set  $(\rho,\,P,\,v_{x},\,v_{y},\,v_{z},\,B_{x},\,B_{y},\,B_{z})$$=$$(25/(36\pi),\,5/(12\pi),\,-\sin{(2\pi\,y)},\\ \,\sin{(2\pi\,x)},\,-\sin{(2\pi\,y)}/(4\pi)^{1/2},\,\sin{(4\pi\,x)}/(4\pi)^{1/2},\,0)$.

Fig.~\ref{fig:orszag} compares images of the resulting density field at $t=0.5$; Fig.~\ref{fig:orszag.numbers} quantitatively compares these by plotting the density and $x$, $y$ components of ${\bf B}$ in a slice through $y=0.3125$ at the same time. At this (early) time, the flow contains a complicated set of shocks, but has not yet broken up into turbulence. Figs~\ref{fig:orszag.image.t1}-\ref{fig:orszag.numbers.t1} consider the same at $t=1$, when parts of the flow begin to evolve chaotically (amplifying the differences between methods). All runs here have $256^{2}$ resolution. In all the methods here, all of the key qualitative features are captured at $t=0.5$, and most at $t=1$, including several complicated shocks, discontinuities, and sharp features. At $t=0.5$, non-moving grid, MFM, and MFV solutions are essentially indistinguishable they have converged very well to the exact solution already, with only a slight smoothing of the sharpest features (as expected). Their ``effective resolution'' appears identical. At $t=1$, qualitative features are similar but quantitative differences appear; the MFM/MFV methods better resolve the central vortex/density peak and sharpest features in ${\bf B}$ (not surprising, since their Lagrangian nature automatically provides higher resolution in this region), but there are stronger oscillations in the low-density regions. However, if we boost the system by a constant velocity, at fixed resolution the stationary-grid solution degrades; for a $v_{x}=10$ boost it is more comparable to a $64^{2}$ run at both times. Obviously the Lagrangian methods avoid this source of error. Divergence errors are well-controlled at both times even in super-sonic shocks. 

SPH does nearly as well, although the features in ${\bf B}$ are more smoothed (similar to a $100^{2}$ MFM/MFV/grid result) at $t=0.5$, and this suppresses the appearance of the central density peak at $t=1$. On this problem, there is not a dramatic difference, interestingly, between SPH with normal versus large neighbor numbers. As we saw in the shocktube tests, the divergence errors are larger by a factor of several in SPH compared to MFM/MFV.

If we use a Powell-only cleaning scheme, the results are not too badly corrupted by $\divB$ errors; however we do see some systematic offsets, particularly in $\rho$ around $x\sim 0.6$, and the position of the ${\bf B}$ jumps around $x\approx0.1$ at $t=0.5$, and the central density peak and $B_{x}$ discontinuity at $t=1$, that do not appear to converge away.

Fig.~\ref{fig:orszag.resolution} demonstrates the convergence in our MFM method (MFV results are nearly identical here). As we increase the resolution, we clearly see good convergence towards the exact solution in {\em all} features here. Quantitatively, the convergence in the $L_{1}$ and $L_{2}$ norms of the plotted quantities is first-order, as expected due to the presence of shocks (the same is true in {\small ATHENA}). Compare this to older SPH MHD implementations, which only saw convergence in {\em some} features, while others converged slowly or not at all (a well-known issue in SPH; \citealt{stasyszyn:2013.divb.cleaning.mhd.gadget}).

We have also considered the pure driven-turbulent box problems from \paperone\ (rms Mach numbers $\sim0.3$ and $\sim8$), as well as the ABC dynamo \citep{Arnold:ABC.dynamo} with small initial seed fields; we confirm that the turbulent power spectra and growth rate of magnetic energy in the box agree well between MFM, MFV, and grid methods. This echoes our conclusions for the pure hydrodynamic case in \paperone. With Powell-only cleaning, however, the growth rate of the magnetic energy in the highly super-sonic case is artificially high (growing faster than the flow-crossing time, indicating a clear numerical artifact). In SPH, the results depend on neighbor number, in a manner that we demonstrate in detail for the MRI test problem below.

\begin{figure}
    \plotonesize{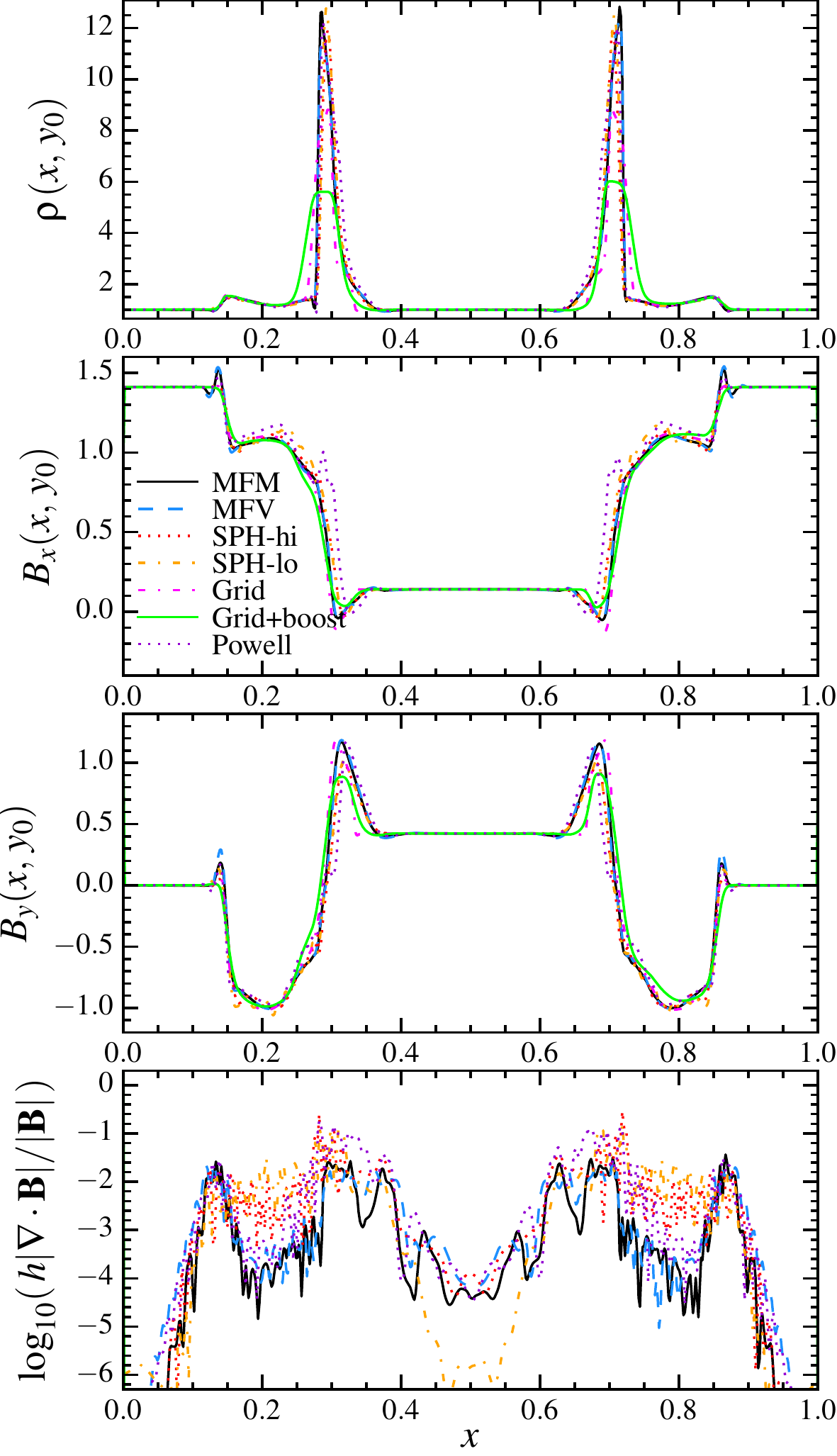}{0.9}
    \vspace{-0.25cm}
    \caption{Density and magnetic fields in a slice through $y_{0}=1/2$ in the MHD rotor problem in Fig.~\ref{fig:rotor}. The qualitative differences between methods are the same as in Fig.~\ref{fig:orszag.numbers} (the ``grid+boost'' run uses the same bulk $v_{x}=10$). MFM/MFV/grid methods agree well. Note that boosted grid methods smooth the pressure spikes significantly, and Powell-only cleaning leads to systematically incorrect shock positions. 
    \vspacerpostplot 
    \label{fig:rotor.numbers}}
\end{figure}

\vspace{-0.5cm}
\subsection{The Magnetic Rotor: Torsional MHD Waves}
\label{sec:rotor}

Next we consider the MHD rotor, a standard problem from \citet{balsara:1999.mhd.rotor.test} used to test strong torsional Alfven waves. A 2D domain of unit size and $\gamma=7/5$, is initialized with uniform $P=1$, ${\bf B}=(5/\sqrt{4\pi},\,0,\,0)$. Inside a circle of size $R_{0}=0.1$, $\rho=10$ and ${\bf v}=(-2\,y/R_{0},\,2\,x/R_{0},\,0)$; this is surrounded by a ring-shaped transition region from $R_{0}<R<R_{1}=0.115$ with $\rho = 1+9\,f(R)$, ${\bf v}=(-2\,y\,f(R)/R,\,2\,x\,f(R)/R,\,0)$ with $f(R)=(R_{1}-R)/(R_{1}-R_{0})$. At $R>R_{1}$, $\rho=1$ and ${\bf v}=(0,\,0,\,0)$. 

As with the Orszag-Tang vortex, Figs.~\ref{fig:rotor}-\ref{fig:rotor.numbers} plot images of the magnetic energy density and field values in horizontal slices, at time $t=0.15$. Again, all methods capture the key qualitative features. MFM/MFV and grid methods are very similar. Grid methods converge slightly more rapidly on the sharp ${\bf B}$ field ``spikes'' at $x=0.3,\,0.7$ if the bulk velocity is nil, but if the fluid is boosted appreciably, the density spikes are noticeably more smoothed. In either case, MFM, MFV, and grid methods exhibit convergence at the same order. Divergence errors are again well-controlled, even around large discontinuities in ${\bf B}$.

Here, for SPH, using the same neighbor number as MFM/MFV (3D-equivalent 32 neighbors) leads to significant noise, and some systematic errors in ${\bf B}$ around $x\sim 0.15-0.3,\,0.7-0.85$. These are resolved if a large neighbor number is used. However in both cases significant smoothing of the local extremum in the magnetic pressure (the brown patches at $y\approx 0.5\pm 0.15$ in Fig.~\ref{fig:rotor}) is evident, and $\divB$ is less accurately suppressed.

With only Powell-cleaning, significantly more noise is evident in the image; moreover, the systematic offset of the discontinuity positions in $B_{x}$ is clear, and this does not converge away.

\begin{figure}
    \plotonesize{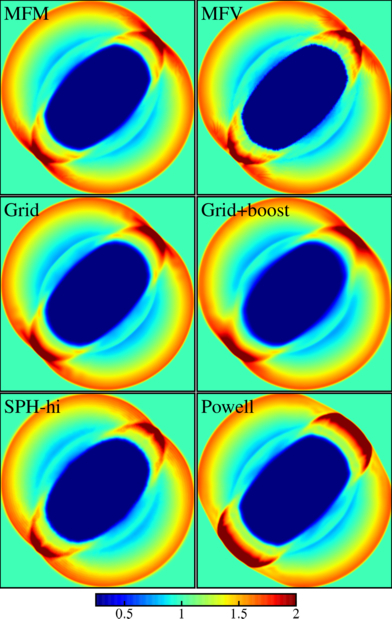}{0.98}
    \vspace{-0.25cm}
    \caption{Blastwave in a strongly magnetized medium (\S~\ref{sec:blastwave}). We plot density $\rho$ (as labeled) in $256^{2}$ simulations. The blastwave expands asymmetrically along field lines (initial ${\bf B}=(1/\sqrt{2},\,1/\sqrt{2},\,0)$), as expected. All methods capture the key behaviors; Lagrangian methods capture slightly more detail in the high-density regions, but with enhanced grid noise. Stationary-grid methods converge more slowly when the fluid is boosted (here, $v_{x}=10$); the errors break the symmetry of the solution noticeably at this resolution, but they do eventually converge away. Using only Powell/8-wave methods (no $\divB$-damping) leads to physically incorrect shapes of the high-density features in the top-right and bottom-left corners; these {\em do not} converge away with resolution. 
        \vspacerpostplot 
    \label{fig:blast}}
\end{figure}

\begin{figure}
    \plotonesize{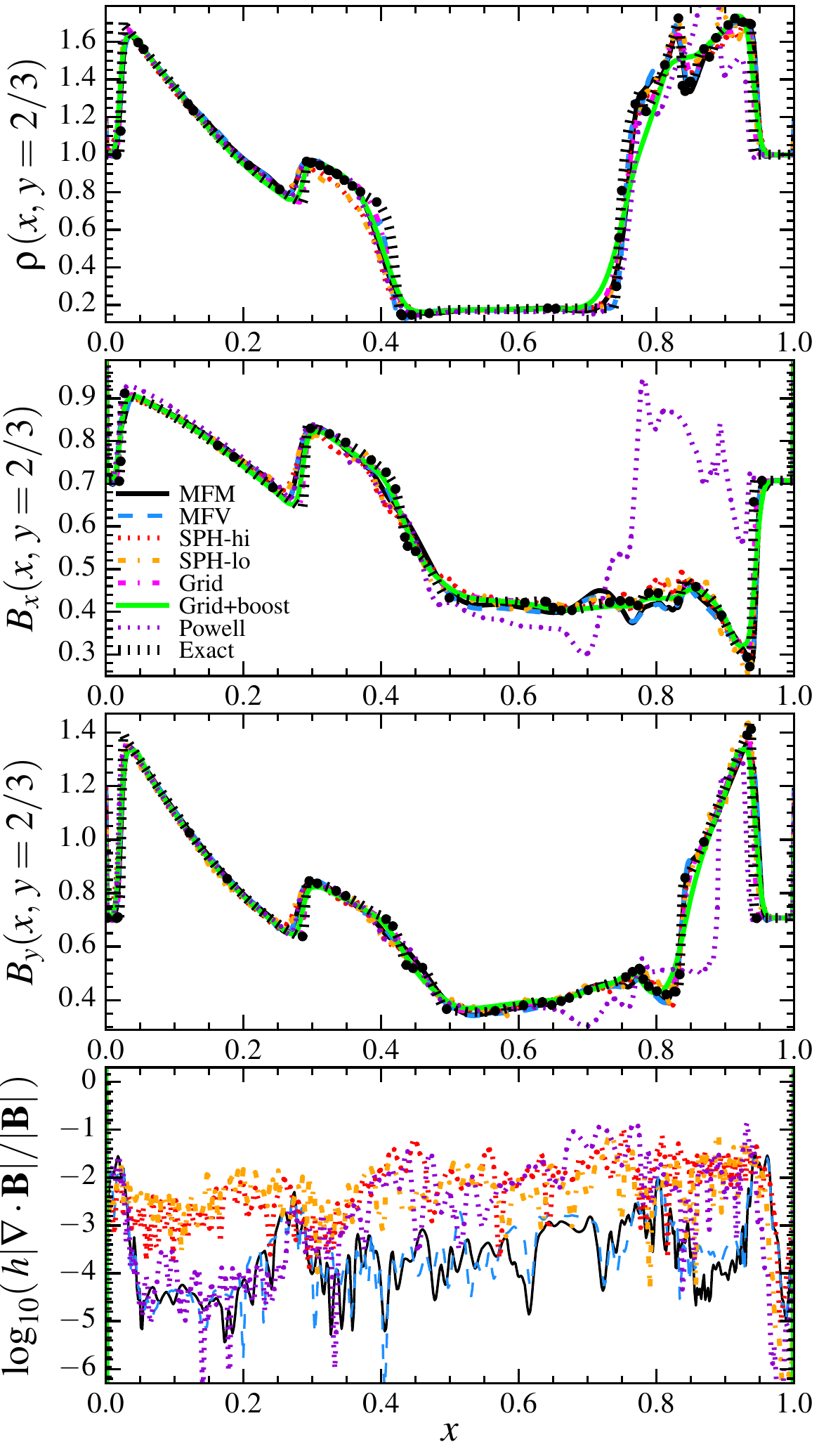}{0.9}
    \vspace{-0.25cm}
    \caption{Slices through the MHD blastwave in Fig.~\ref{fig:blast}. Most methods agree well; the convergence is good at this resolution except for SPH without a large neighbor number, or grids if the fluid is moving. The errors from the Powell-only (no-cleaning) method are dramatic: in $B_{x}$, we see factor $\sim2-3$ systematically incorrect results. 
    \vspacerpostplot 
    \label{fig:blast.numbers}}
\end{figure}

\begin{figure*}
    \plotsidesize{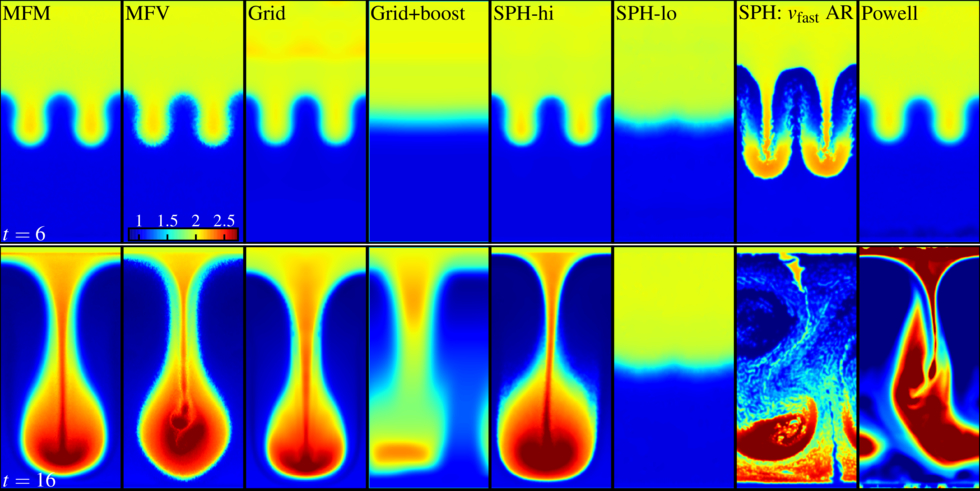}{0.9}
    \vspace{-0.25cm}
    \caption{Magnetic Rayleigh-Taylor instability (\S~\ref{sec:mixing}). We show the density $\rho$ (as labeled), in both early ({\em top}, $t=6$) and late-time ({\em bottom}, $t=16$) results at $128\times256$ resolution. MFM, MFV, and grid methods (with no bulk motion) all converge to the correct solution. In the early stages they are nearly identical; even in late stages there is little difference (small differences appear at late times in MFV owing to growth of the slightly-larger grid noise at early times). As in the pure-hydro case, in non-moving grid methods, ``boosting'' the fluid grid by a uniform velocity slows down mode growth and breaks symmetry, unless significantly higher resolution is used. In SPH, reasonable results can be obtained if large neighbor numbers are used (although some large-scale asymmetry appears in the non-linear solution because of imperfect $\divB$-cleaning); with smaller neighbor number (the same as used in MFM/MFV methods), low-order errors in SPH dominate and no mode grows. However, the SPH results are also very sensitive to the form of the artificial dissipation (artificial resistivity or AR) employed: we compare the results if we use the fast magnetosonic velocity $v_{\rm fast}$ in AR, as advocated by \citet{tricco:2013.sphmhd.methods} (which reduces noise in super-sonic problems), instead of our default choice (the Alfven velocity). For details, see Appendix~\ref{sec:spmhd.numerics}. This produces too much dissipation around the contact discontinuity, which damps the ${\bf B}$-field to sub-critical values; so the instability behaves (incorrectly) like the pure-hydro case.
    \vspacerpostplot 
    \label{fig:rt}}
\end{figure*}

\begin{figure}
    \plotonesize{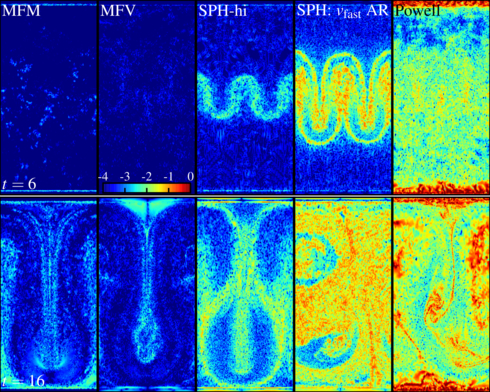}{0.98}
    \vspace{-0.25cm}
    \caption{Divergence errors in the MHD RT instability from Fig.~\ref{fig:rt}. We plot $\log_{10}(h\,|\divB|/|{\bf B}|)$ (as labeled), for the same times as Fig.~\ref{fig:rt}. The grid methods here use CT so $\divB=0$ to machine precision. MFM/MFV methods maintain $\log_{10}(h\,|\divB|/|{\bf B}|) < -2$ (for {\em every} particle/cell) even into the late non-linear evolution (most of the values at $>-4$ are boundary effects, in fact). In SPH, the zeroth-order errors around a contact discontinuity lead to less-accurate cleaning in these regions. In the $v_{\rm fast}$ artificial resistivity (AR) case, $h\,|\divB|/|{\bf B}|\sim1$ around these discontinuities; this is not because $\divB$ is large, but because the AR over-damps ${\bf B}$ at the discontinuity. In the Powell case, order-unity errors appear at late times.
       \vspacerpostplot 
    \label{fig:rt.divB}}
\end{figure}

\begin{figure}
    \plotonesize{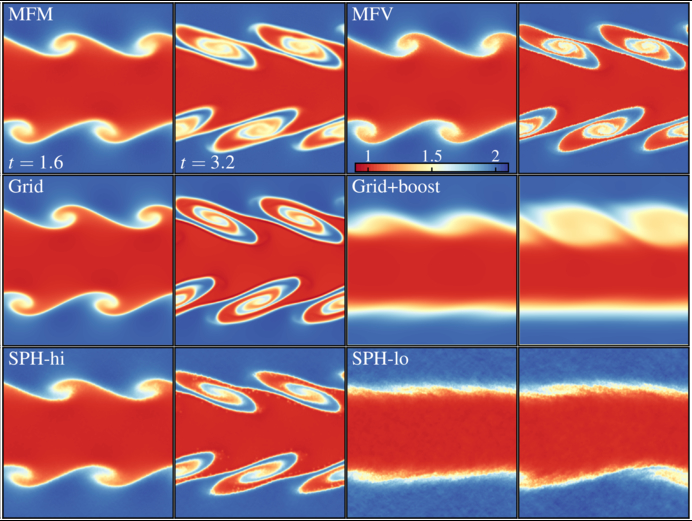}{0.98}
    \vspace{-0.25cm}
    \caption{Magnetic Kelvin-Helmholtz (KH) instability (\S~\ref{sec:mixing}). We show density $\rho$ (as labeled) in the non-linear phases $t=1.6$ and $t=3.2$ (KH growth timescale $\approx0.7$), at $256^{2}$ resolution. Differences between methods resemble the RT test (Fig.~\ref{fig:rt}). MFM, MFV, and non-boosted grid results agree well into the non-linear growth phase (where different grid noise effects lead to small departures). Boosting the fluid in the grid case or using typical neighbor numbers for SPH break symmetry and suppress the instability at this resolution. Divergence errors in each case closely resemble those in Fig.~\ref{fig:rt.divB} around the contact discontinuities.
       \vspacerpostplot 
    \label{fig:kh}}
\end{figure}

\begin{figure}
    \plotonesize{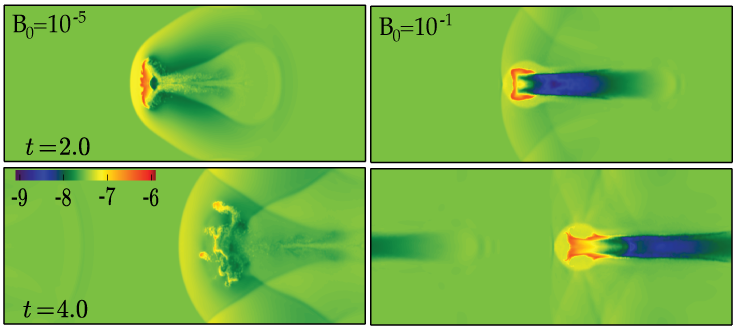}{0.98}
    \vspace{-0.25cm}
    \caption{Magnetic ``blob'' test (\S~\ref{sec:mixing}). We show density $\log_{10}(\rho)$ (arbitrary units) at $t=2$ and $t=4$, with initial vertical field $B_{0}=10^{-5}$ ({\em top}) and $10^{-1}$ ({\em bottom}) in code units. The weak-field case is essentially identical to the pure hydrodynamic case studied in \paperone, as expected; a combination of non-linear Rayleigh-Taylor and Kelvin-Helmholtz instabilities ``shred'' the cloud. The strong-field case suppresses cloud destruction. MFV is shown, but differences between methods are qualitatively identical to the RT and KH tests above.
       \vspacerpostplot 
    \label{fig:blob}}
\end{figure}

\vspace{-0.5cm}
\subsection{Magnetized Blastwave: Strong Shocks \&\ Grid-Alignment Effects}
\label{sec:blastwave}

Next we consider a strong blastwave in a magnetic field; this is another standard problem, which tests the ability of the code to handle strong shocks and preserve symmetry. In the hydrodynamic version of this test, it is also very challenging for grid-based codes to avoid strong grid-alignment and preferential propagation of shocks along the grid (see \paperone); however these effects are reduced by the asymmetric forces in the MHD problem. 

We initialize a 2D periodic domain of unit size with $\gamma=5/3$, zero velocity, $\rho=1$, ${\bf B}=(1/\sqrt{2},\,1/\sqrt{2},\,0)$, and pressure $P=10$ within a circle at the center of initial size $R<R_{0}=0.1$ and $P=0.1$ outside. Figs.~\ref{fig:blast}-\ref{fig:blast.numbers} show images and slices through the solution at time $t=0.2$, for $256^{2}$ runs. Visually, the MFM/MFV, SPH (with high neighbor number), and grid (with zero boost to the fluid) solutions look similar. The blast expands correctly, with the cavity growing more rapidly in the direction along the field lines. There is some additional detail visible in the high-density regions in the Lagrangian solutions (the ``dimpling'' in the upper-right and lower-left); this is real and appears at slightly higher resolution in the grid calculation as well. The SPH solution is slightly more smoothed along the shock fronts. Here we also show the Powell-only and boosted-grid results; unlike the Orszag-Tang and rotor problems, here the differences are plainly visible by-eye. The Powell case develops incorrect features; the boosted-grid case loses symmetry.

Quantitatively, we see these effects in Fig.~\ref{fig:blast.numbers}. Note that MFM/MFV methods exhibit very small $\divB$ values; SPH exhibits larger $\divB$, but still quite small. In general, different methods agree fairly well quantitatively. However, for the Powell-only case, the value of $B_{x}$, in particular, is seriously wrong in the upper-right portion of the solution. We emphasize that the error is {\em worse} if we re-simulate this with Powell-only cleaning but a resolution of $1024^{2}$. The failure of this method to guarantee the correct shock jumps corrupts the entire late-time solution. Note that, in SPH (using the full \citet{dedner:2002.divb.cleaning.scheme} cleaning), the magnitude of the $\divB$ errors is not much smaller than our Powell-only run, but this incorrect behavior does not appear. This emphasizes that the {\em magnitude} of $h\,|\divB|/|{\bf B}|$ is not, by itself, the whole story; non-linear errors are damped by the \citet{dedner:2002.divb.cleaning.scheme} approach (at the cost of some additional local diffusion) which would otherwise build up coherently at later times. Even quite large values of $\divB\sim0.1-1$ can be tolerated, if the algorithm includes a proper damping formulation (as our MFM/MFV/SPH implementations do).

\begin{figure*}
    \plotsidesize{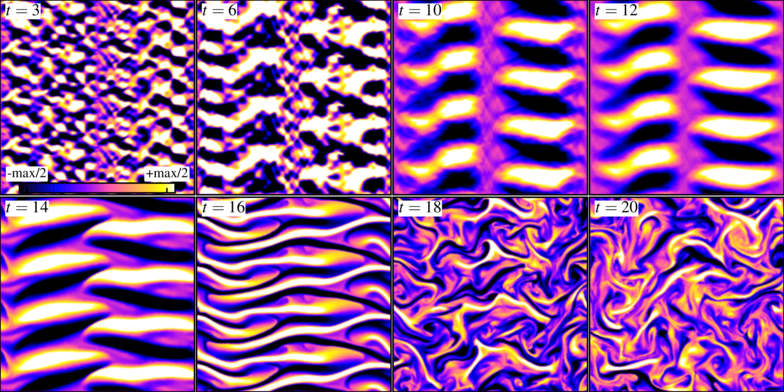}{0.98}
    \vspace{-0.25cm}
    \caption{Magneto-rotational instability (MRI; \S~\ref{sec:mri}) test problem. We show the results from a 2D axisymmetric shearing box (horizontal/vertical axes correspond to radial/vertical coordinates), with vertical $B_{z}$ set so the fastest-growing modenumber is $m=4$. We plot the azimuthal/toroidal component $B_{y}$ of the magnetic field (the behavior of the radial component is similar, but with reversed sign), in an MFM calculation at $256^{2}$ resolution (units are scaled to the maximum/minimum values in each frame as labeled, since the absolute value of $B_{y}$ grows exponentially). The initial seed noise is amplified on the correct timescale, and from times $t\sim 5-16$, the dominant $m=4$ mode pattern is clearly visible. At late times, the non-linear modes break up into turbulence, as expected. MFM, MFV and high-order CT-based grid methods (see \citealt{guan:2008.shearing.box.mri}) produce similar results. 
    \vspacerpostplot 
    \label{fig:mri}}
\end{figure*}

\begin{figure}
    \plotonesize{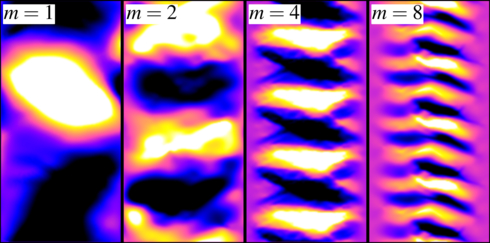}{0.98}
    \vspace{-0.25cm}
    \caption{MRI, as Fig.~\ref{fig:mri} (same colorscale; just showing the $0<x<0.5$ half of the box), at times when the linear mode dominates, for runs with initial $B_{z}$ set so that the fastest-growing mode corresponds to $m=1,\,2,\,4,\,8$, as labeled (MFM shown, MFV is nearly identical). In each case, the correct mode is clearly visible. We find approximately $4$ cells/particles are required across each ``node'' in the linear direction to resolve the correct mode growth, the same as the number of cells in PPM grid methods.
    \vspacerpostplot 
    \label{fig:mri.modes}}
\end{figure}

\begin{figure}
    \plotonesize{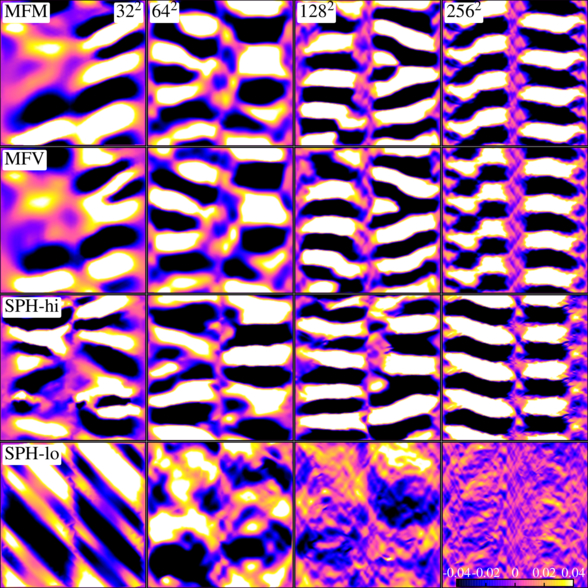}{0.98}
    \vspace{-0.25cm}
    \caption{MRI (with mode number $m=4$), as Fig.~\ref{fig:mri}, as a function of resolution and numerical method. MFM/MFV methods are very similar, and resolve the MRI with $m=4$ with as few as $32^{2}$ elements. Modern SPH MHD is, in fact, also able to capture the MRI modes, if a larger neighbor number is used. If we use a smaller neighbor number in SPH, the low-order errors completely swamp the correct solution.
    \vspacerpostplot 
    \label{fig:mri.res}}
\end{figure}

\vspace{-0.5cm}
\subsection{MHD Rayleigh-Taylor \&\ Kelvin-Helmholtz Instabilities: Fluid Mixing in MHD}
\label{sec:mixing}

Fluid mixing instabilities are astrophysically important and historically challenging for SPH methods; the hydrodynamic forms of these are discussed at length in \paperone. In MHD, non-zero $|{\bf B}|$ suppresses the growth of small-scale modes. If magnetic fields are too strong, no interesting modes grow. If fields are too weak, the case is essentially hydrodynamic. But there is an interesting, MHD-specific regime when the fields strengths are near-marginal; we consider this in a Rayleigh-Taylor (RT) problem. 

We take initial conditions from \paperone\ and \citet{abel:2011.sph.pressure.gradient.est}: in a 2D domain with $0<x<1/2$ (periodic boundaries) and $0<y<1$ (reflecting boundaries), we take $\gamma=1.4$, ${\bf B}/\sqrt{4\pi}=(0.07,\,0,\,0)$, and $\rho(y)=\rho_{1}+(\rho_{2}-\rho_{1})/(1+\exp{[-(y-0.5)/\Delta]})$ with $\rho_{1}=1$, $\rho_{2}=2$, $\Delta=0.025$. Initial pressures are assigned to produce a gradient in hydrostatic equilibrium with a uniform gravitational acceleration $g=-1/2$ in the $y$ direction (at the interface, $P=\rho_{2}/\gamma=10/7$ so $c_{s}=1$). An initial $y$-velocity perturbation $v_{y} = \delta v_{y}\,(1+\cos{(8\pi\,(x+1/4))})\,(1+\cos{(5\pi\,(y-1/2))})$ with $\delta v_{y}=0.025$ is applied in the range $0.3<y<0.7$ (otherwise the velocities are zero).\footnote{Following the method in \paperone, we use the routines generously provided by R.\ O'Leary to construct the mesh-free initial conditions, then re-interpolate this onto the {\small ATHENA} grid. This is critical to ensure that the same ``seed'' grid noise is present in both methods, which in turn is necessary to see similar behavior in the late-time, non-linear phase of evolution.}

Fig.~\ref{fig:rt} shows the resulting density field at intermediate and late times, in a $128\times256$ simulation. In MFM, MFV, grid/AMR with non-moving fluid, and SPH (with sufficiently large neighbor number), the linear growth of the field is essentially identical; this is consistent with our pure-hydro results in \paperone. Even the non-linear, late-time results agree reasonably well (although there is some symmetry-breaking in SPH which owes to less-accurate $\divB$-cleaning, even at large neighbor number). There is slightly more ``grid noise'' in MFV at this resolution (which leads to small differences in the late-time evolution). Just like in the pure hydrodynamic case, in SPH, the results are very sensitive to neighbor number: if a smaller neighbor number is used (say, the same as we use in our MFM/MFV methods), then the initial seed mode is too weak -- it is overwhelmed by the zeroth-order errors in the method, and no modes grow. Similarly, if we boost the fluid by a constant velocity in stationary-grid methods, advection errors break symmetry and dramatically suppress the growth of the mode at this resolution (much higher resolution is required to match the accuracy of the Lagrangian methods). All of this is consistent with our results from the hydrodynamic case in \paperone. 

If we consider the Powell-only case, the {\em linear} mode evolution appears reasonable -- the growth rates are only slightly suppressed. However, in the {\em non-linear} phase, the solution is totally corrupted! The non-linear errors accumulate, if only Powell-type schemes are used (once again, because in this method, divergence errors are only transported, not suppressed), until they overwhelm the real solution. Clearly, tests restricted to the linear regime are not sufficient to validate divergence-cleaning schemes. 

In SPH, we also find another difficulty unique to MHD. Numerical stability in SPH requires somewhat ad-hoc artificial dissipation terms for ${\bf B}$, the ``artificial resistivity.'' As discussed in Appendix~\ref{sec:spmhd.numerics}, and \citet{rosswog:2007.sph.mhd,tricco:2013.sphmhd.methods}, this carries ambiguities that are not present in hydrodynamics: the appropriate ``signal velocity'' for the resistivity could be the sound speed, Alfven velocity, or magnetosonic speed, and in some cases resistivity should be applied in rarefactions. The correct answer depends on the type of MHD discontinuity. In Godunov methods such as MFM/MFV/grid codes, the correct form of the dissipation is provided by the appropriate solution to the Riemann problem. But in SPH, even with the state-of-the-art ``switches'' used here, this is difficult to assign correctly. By default, we use the Alfven velocity in this switch. But \citet{tricco:2013.sphmhd.methods} show that this can lead to too-low a resistivity in super-sonic MHD turbulence, in turn producing shock breakup and serious noise (see their Fig.~7). They recommend increasing the dissipation by using the fast magnetosonic speed $v_{\rm fast}$ instead. However, if we do that for this problem, it produces excessive dissipation of the magnetic field around the contact discontinuity.\footnote{For the test in Fig.~\ref{fig:rt} labeled ``$v_{\rm fast}$ AR,'' we keep everything about our SPH MHD method identical (including the ``switch'' and maximum viscosity $\alpha_{B}=0.1$), except replace the Alfven velocity in Eq.~\ref{eqn:cijB} with the fast magnetosonic speed.} This makes the problem behave (incorrectly) like the pure-hydrodynamic case.

In Fig.~\ref{fig:rt.divB}, we show full 2D maps of the divergence $h\,|\divB|/|{\bf B}|$. in MFM/MFV, these are extremely well-controlled, with median values $<10^{-4}$ and maxima $<10^{-2}$. In our default SPH implementation, they are larger by a factor of a few. In the ``$v_{\rm fast}$ AR'' SPH run, we see much larger values along the contact discontinuity. These are not, however, caused by poor $\divB$-control; rather, excessive dissipation of ${\bf B}$ around the discontinuity leads to a local sharp depression of $|{\bf B}|$ (the denominator), as opposed to $|\divB|$. In the Powell runs the divergence errors are (as expected) much larger.

We have also compared the magnetized Kelvin-Helmholtz instability, shown in Fig.~\ref{fig:kh}. The initial conditions follow \citet{mcnally:2012.kh.test.comparison} (see also \paperone), a 2D setup with $256^{2}$ resolution elements following two streams with $P=5/2$, $\gamma=5/3$, and $(\rho,\,v_{x})=(1,-0.5)$ and $(2,\,0.5)$, with a $1\%$ amplitude initial seed mode and small interface region between the streams. We add a uniform ${\bf B}=(0.1,\,0,\,0)$, about a factor $\approx2$ below the critical value which suppresses the instability, and show results at $t=1.6$ and $t=3.2$ (where the KH growth time is $\approx 0.7$). Here, we obtain qualitatively identical conclusions to our RT test. In the linear and even non-linear stages, MFM/MFV and non-boosted grid results agree well, with more small-scale structure in the non-linear stages in the meshless methods (owing to increased grid noise). Quantitatively, the total magnetic energy in the box grows, with excellent agreement between these methods and converged solutions until $t\approx 2.5$ (well into the non-linear phase). However, the $v_{x}=10$ boost completely suppresses mode growth in stationary grid methods at this resolution. In SPH, a reasonable, answer can be obtained with sufficiently high neighbor number {\em and} an appropriate choice for the artificial resistivity, but the instability is totally suppressed if we use typical neighbor numbers and behaves as if the field were much weaker in the ``$v_{\rm fast}$ AR'' case. Comparing $\divB$, we see the same behavior around the contact discontinuity as in the RT test; the relative performance of different methods is essentially identical, although in all cases the $\divB$ errors are systematically smaller by a factor of a couple.

Finally, in Fig.~\ref{fig:blob}, we also compare an MHD version of the ``blob'' test from \citet{agertz:2007.sph.grid.mixing}. The setup is identical to our detailed study of different methods on this problem in \paperone\ (featuring a cold cloud in pressure equilibrium in a hot wind tunnel), but with an initially constant field in the vertical direction. As expected the field strongly suppresses the non-linear RT and KH instabilities that tend to disrupt the cloud in the hydrodynamic case \citep[for a detailed study, see][]{shinstone:3d.mhd.shock.cloud}. Our qualitative conclusions (comparing different methods) are identical to the tests above.

\begin{figure}
    \plotonesize{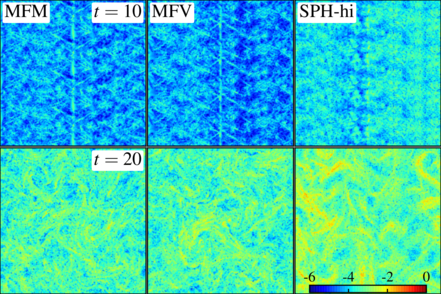}{0.98}
    \vspace{-0.25cm}
    \caption{Divergence errors in the MRI test in Fig.~\ref{fig:mri.res} at $256^{2}$ resolution. We plot $\log_{10}(h\,|\divB|/|{\bf B}|)$ (as labeled), for two times, $t=10$ when the fastest-growing mode dominates, and $t=20$ when the system has broken up into turbulence (Fig.~\ref{fig:mri}). In both cases, the errors are well-controlled in MFM/MFV; in SPH some regions reach larger $h\,|\divB|/|{\bf B}| > 0.01$ in the turbulent phase; however these almost entirely correspond to regions of nearly-vanishing $|{\bf B}|$ (i.e.\ $\langle h\,|\divB| \rangle / \langle |{\bf B}| \rangle \ll 0.01$), so the errors are not dynamically significant.
     \vspacerpostplot 
    \label{fig:mri.divB}}
\end{figure}

\begin{figure*}
    \plotsidesize{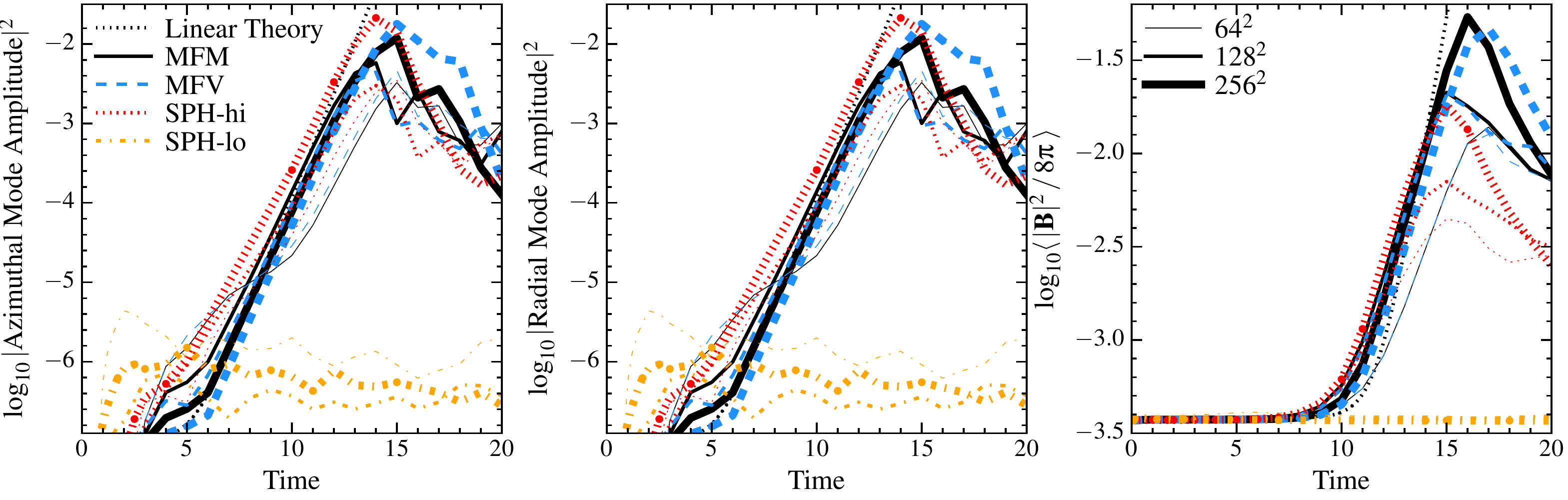}{0.98}
    \vspace{-0.25cm}
    \caption{Growth of the $m=4$ MRI  modes in Figs.~\ref{fig:mri}-\ref{fig:mri.res}. We measure the $m=4$ Fourier amplitude in the azimuthal ($B_{y}$; {\em left}) and radial ($B_{x}$; {\em center}) magnetic fields, as well as the average magnetic energy ($\langle |{\bf B}^{2}/8\pi| \rangle$ in the box ({\em right}). We compare each to the expectations of linear theory (dotted black line). We compare three different resolutions ($64^{2}$, $128^{2}$, $256^{2}$, in progressively thicker lines). MFM/MFV methods give similar results; in both cases the simulations converge to the correct linear growth rate quickly at resolutions above $\sim 64^{2}$. This agrees well with state-of-the-art grid CT codes. The peak mode amplitude also increases with resolution; once the MRI breaks into turbulence, the field amplitude declines, as it should. For SPH, we see no MRI at {\em any} resolution unless we use a large neighbor number. With sufficiently large neighbor number, we obtain growth, but convergence to the correct growth rate is slower (even at $256^{2}$, the growth rate is suppressed by $\sim 30\%$). 
    \vspacerpostplot 
    \label{fig:mri.growth}}
\end{figure*}

\begin{figure}
    \plotonesize{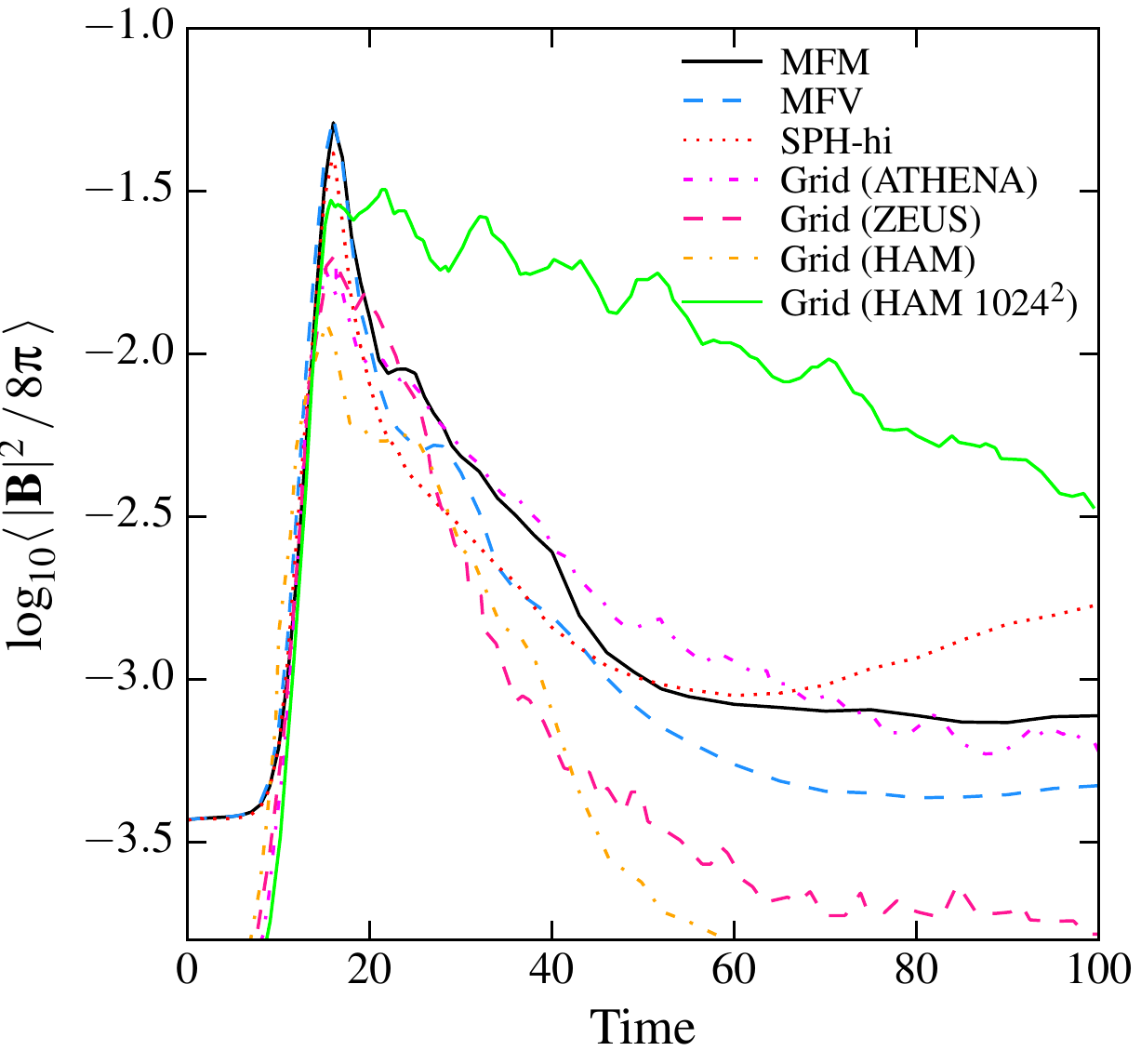}{0.9}
    \vspace{-0.25cm}
    \caption{Late-time evolution of the magnetic energy, for the $m=4$ case, at $256^{2}$ resolution. We compare results from three different unsplit, CT-based grid codes in \citet{guan:2008.shearing.box.mri}. The 2nd-order {\small HAM} code is most diffusive, while the 3rd-order PPM {\small ATHENA} is least-diffusive ({\small ZEUS} is intermediate). In all cases, the linear growth rates and peak amplitudes are similar and agree well with our meshless methods; however the decay rate is sensitive to the details of numerical dissipation. Our meshless results at fixed resolution are most similar to {\small ATHENA}, the least-diffusive of the grid-based codes considered. 
    \vspacerpostplot 
    \label{fig:mri.late}}
\end{figure}

\vspace{-0.5cm}
\subsection{Magneto-Rotational Instability: Can Meshless Methods Capture the MRI?}
\label{sec:mri}

We next consider the magneto-rotational instability (MRI), one of the most important and historically challenging MHD phenomena. We consider a 2D axisymmetric shearing box defined as in \citet{guan:2008.shearing.box.mri}; this is a locally-Cartesian box where the $x$ coordinate represents the radial direction $R$, and the other coordinate is the vertical direction $z$ (azimuthal axisymmetry is assumed). Boundary conditions are periodic in $z$ and shear-periodic in $x$: $f(x,\,z)=f(x+n_{x}\,L_{x},\, z+n_{z}\,L_{z})$ for all values $f$ except the azimuthal velocity $v_{\phi} = v_{y}(x,\,z) = v_{y}(x+n_{x}\,L_{x},\,z+n_{z}\,L_{z}) + n_{x}\,q\,\Omega\,L_{x}$, where $q=-(1/2)\,d\ln{\Omega}^{2}/d\ln{R}=3/2$ for a Keplerian disk, $n_{x}$ and $n_{y}$ are integers representing the box periodicity, and $\Omega$ is the mid-plane orbital frequency. In this approximation, the momentum equations are also modified with the source terms $D(\rho\,{\bf v})/Dt = -2\,(\Omega\,\hat{z})\times (\rho\,{\bf v}) + 2\,\rho\,q\,\Omega^{2}\,x\,\hat{x}$. We initialize a box of unit size ($-0.5<x<0.5$, $-0.5<z<0.5$) and $\Omega=1$, with $(\rho,\,P,\,v_{x},\,v_{y},\,v_{z},\,B_{x},\,B_{y},\,B_{z})=(1,\,1,\,0,\,\delta v,\,-q\,x,\,0,\,0,\, B_{0}\,\sin{(2\pi\,x)})$, where $B_{0} = \sqrt{15}/(8\pi\,m)$ is set so that the most unstable wavelength $\lambda_{\rm MRI}=1/m$ corresponds to a mode number $m$. Here $\delta v$ is set to a uniform random number in the range $\pm0.005$ for each cell, to seed the instability. 

Fig.~\ref{fig:mri} shows the results at various times from a $256^{2}$ MFM calculation with $m=4$. We see the expected behavior; the random ${\rm B}$-field fluctuations seeded by the velocity perturbation grow, and quickly are dominated by the fastest-growing ($m=4$) mode, until at late times the non-linear modes break up into MRI turbulence. The same behavior appears in MFV, grid, and even SPH methods. In Fig.~\ref{fig:mri.modes}, we compare results when the linear mode dominates, for runs with initial $B_{z}$ set such that $m=1,\,2,\,4,\,8$, with MFM at the same resolution. In each case, we see the correct mode grows. Fig.~\ref{fig:mri.res} compares different methods and resolutions (for $m=4$). We see that MFM and MFV produce nearly identical results. Even at $32^{2}$ resolution, a reasonable mode structure emerges. In general, with MFM/MFV methods, we find about $\sim 4$ resolution elements (particles/cells) in linear dimension across each ``node'' are needed to see reasonable modes (for $m=2$, we see growth for $16^{2}$ resolution); this is the same as in higher-order (PPM) grid-based codes using CT \citep[see][]{guan:2008.shearing.box.mri}.

For SPH, we see -- perhaps surprisingly -- reasonable behavior, if we use large enough neighbor numbers. Compared to previous SPH MHD implementations, the fact that the new methods combine many basic hydrodynamic improvements (larger kernels to reduce noise, evolution of conserved variables, Lagrangian-derived magnetic forces, and properly-derived divergence-cleaning) leads to the ability to follow the MRI. A very similar implementation of SPH MHD by \citet{gradsphmhd.methods.paper} and preliminary results from \citet{tricco:thesis} have also demonstrated this. However the noise level in SPH is higher, even with this larger neighbor number, at lower resolutions. And if we use a smaller neighbor number in SPH, comparable to what is used in our MFM/MFV methods, we see the SPH gradient errors totally dominate the solution, and prevent any mode growth \citep[see also][]{tricco:thesis}.

Fig.~\ref{fig:mri.divB} compares $h\,|\divB|/|{\bf B}|$. Even in the late non-linear turbulence stage, these are maintained at $<10^{-2}$ (median $\sim 10^{-4}$) in MFM/MFV, and $<0.1$ (median $\sim 10^{-3}$) in SPH. 

Figs.~\ref{fig:mri.growth}-\ref{fig:mri.late} compare the growth of the MRI modes quantitatively. We measure the amplitude of the $m=4$ Fourier mode in radial ($B_{x}$) or azimuthal ($B_{y}$) field components (the maximum of the vertical $m=4$ mode amplitude in the 2D Fourier transform), at each time, for the simulations above. We also compare the volume-averaged magnetic energy $\langle E_{B} \rangle = \langle |{\bf B}|^{2} / 2 \rangle$ in the box. The linear-theory prediction is that the fastest-growing $B_{x}$ and $B_{y}$ modes should grow $\propto \exp{(0.75\,t)}$, so the magnetic energy should scale $\langle E_{B} \rangle = E_{0} + \delta E\,\exp{(1.5\,t)}$, where $E_{0}$ is the initial energy and $\delta E$ is a seed perturbation amplitude. For $B_{x}$, $B_{y}$, and $E_{B}$, we see good convergence with MFM/MFV to the correct linear growth rate at resolutions above $\sim 64^{2}$ for $m=4$; this agrees well with state-of-the-art CT grid methods. At late times, the mode saturates and then the energy must decay according to the anti-dynamo theorem; this is expected. In SPH, we see no MRI growth unless we go to large neighbor numbers; we then do obtain growth, but the convergence in growth rates is slower. 

In Fig.~\ref{fig:mri.late}, we compare the late-time evolution of the magnetic energy to grid methods. The linear growth rate and peak $|{\bf B}|$ amplitude agree well with high-order grid-CT calculations. The late time decay rate of the magnetic energy is known to be sensitive to the numerical diffusivity of the method \citep[see][and references therein]{guan:2008.shearing.box.mri}: we therefore compare three different grid codes at the same resolution (and one much higher): {\small HAM} (most diffusive), {\small ZEUS}, and {\small ATHENA} (in PPM, CT, CTU mode; least diffusive). The late-time behavior in our meshless methods agrees remarkably well with {\small ATHENA}.

\begin{figure*}
    \plotsidesize{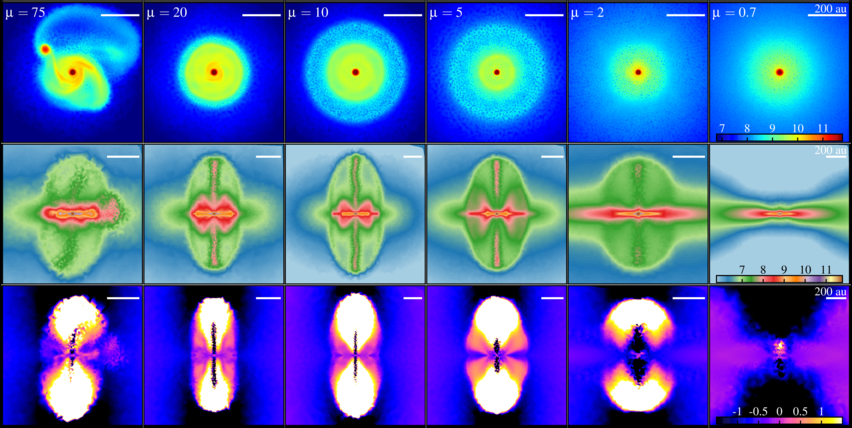}{0.98}
    \vspace{-0.25cm}
    \caption{Proto-stellar jet test (\S~\ref{sec:core}). A magnetized, rotating proto-stellar core with initial  mass-to-flux ratio $\mu \propto |{\bf B}|^{-1}$ collapses under self-gravity until it forms an accretion disk, which winds up the magnetic field and launches a jet. We show results for MFM calculations with $50^{3}$ total particles/cells in the core, and various $\mu$ (columns labeled), just after one free-fall time of the initial core. Scale bar in each panel shows $200$\,au. {\em Top:} Density ($\log_{10}(n/{\rm cm^{-3}})$), as labeled) in a slice through the collapsed disk midplane (face-on). The central ``protostar'' has collapsed by a factor of $>10^{4}$ in density; the disk is more extended and lower-density for stronger ${\bf B}$. For $\mu\lesssim1$, the disk is quasi-stable, and should not form a jet. For $\mu \gtrsim 20$, the disk should fragment into multiple protostars. {\em Middle:} Density ($\log_{10}(n/{\rm cm^{-3}})$) in a slice through the rotation axis. Here the jets are plainly visible. {\em Bottom:} Radial velocity ($v_{r}/{\rm km\,s^{-1}}$) in a slice through the rotation axis. Even at low resolution, and very weak ${\bf B}$, our new meshless methods are able to capture jet-launching. 
    \vspacerpostplot 
    \label{fig:core.bfield}}
\end{figure*}

\begin{figure}
    \plotonesize{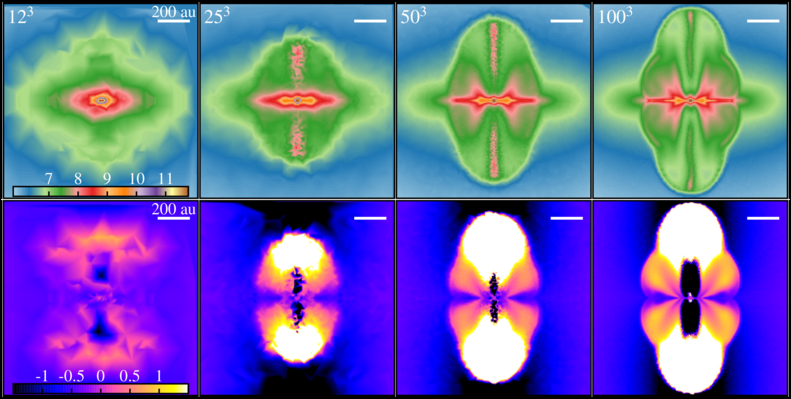}{0.98}
    \vspace{-0.25cm}
    \caption{Resolution study of the collapsing core as in Fig.~\ref{fig:core.bfield}, using MFM (MFV is similar), with initial $\mu=10$. We compare at fixed time ($t=1.05\,t_{\rm ff}$). The resolution quoted is the total number of cells/particles in the collapsing spherical core. At higher resolution, additional fine-structure in the disk and outflow continue to appear; the jet also forms slightly earlier, so it has propagated further. However, there is already some (albeit weak) jet/polar outflow structure at $12^{3}$; by $\sim25^{3}$, the existence of a jet and much of the global structure is already resolved, and the jet momentum/mass are converged within tens of percent of the highest-resolution run. This is remarkably low resolution; usually, $>128^{3}$ resolution is needed for similar convergence in grid calculations.
    \vspacerpostplot 
    \label{fig:core.res}}
\end{figure}

\begin{figure}
    \plotonesize{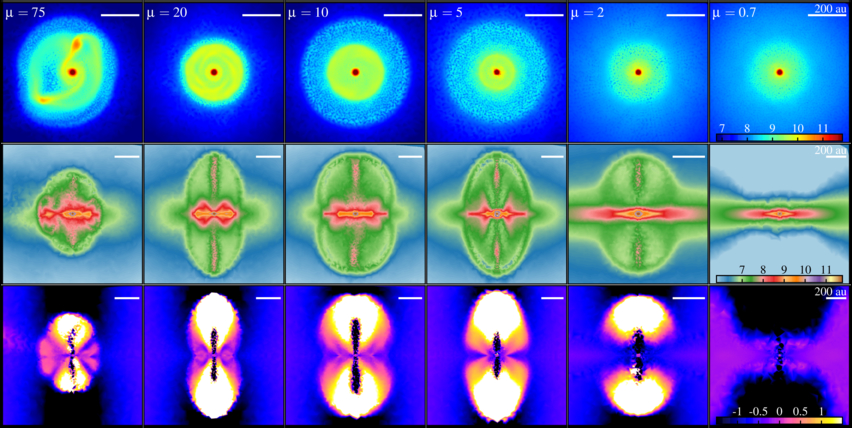}{0.98}
    \vspace{-0.25cm}
    \caption{As Fig.~\ref{fig:core.bfield}, for MFV (shown at identical times to Fig.~\ref{fig:core.bfield}). The results are similar up to  non-linear details. Note that the $\mu=20$ case is marginally unstable to fragmentation; so small changes to the particle splitting/merging algorithm in MFV (which can seed grid noise) can lead to modest perturbations that induce earlier fragmentation (and weaker jets); the ``default'' case here allows no splitting/merging (most similar to MFM).
    \vspacerpostplot 
    \label{fig:core.mfv}}
\end{figure}

\begin{figure}
    \plotonesize{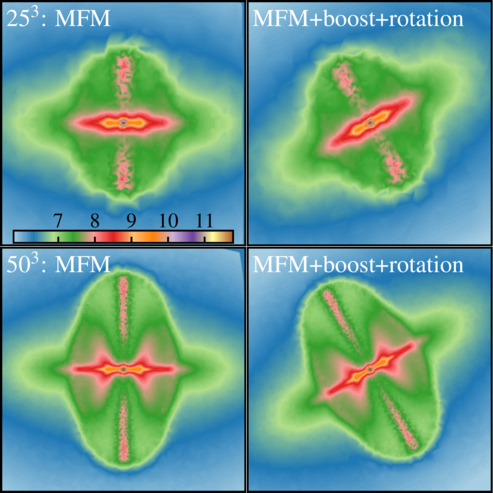}{0.98}
    \vspace{-0.25cm}
    \caption{Effects of rotations and boosts on our meshless methods in the protostellar jet test. {\em Left:} Our ``default'' MFM run from Fig.~\ref{fig:core.bfield} with $\mu=10$, resolution $50^{3}$, zero net bulk motion and an initial angular momentum/magnetic field axis aligned with the $\hat{z}$ direction. {\em Right:} Same, but with the rotation/field axis rotated $30^{\circ}$ in the $x-z$ plane, and the entire box boosted by a uniform $\delta v_{x}=10\,{\rm km\,s^{-1}}$, $\delta v_{y}=\delta v_{z}=2\,{\rm km\,s^{-1}}$. Our mesh-free methods are invariant to boosts (bulk motion) of the box and rotations/grid mis-alignments at all resolutions; however these pose serious challenges for grid-based calculations.
    \vspacerpostplot 
    \label{fig:core.rotation}}
\end{figure}

\begin{figure}
    \plotonesize{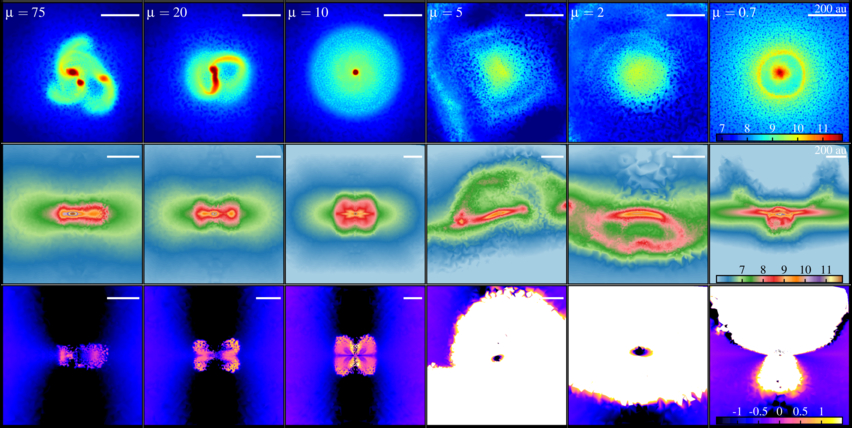}{0.98}
    \vspace{-0.25cm}
    \caption{As Fig.~\ref{fig:core.bfield}, for SPH-hi. SPH has much greater difficulty capturing the important behaviors at low resolution. At low-$|{\bf B}|$, the SPH ``artificial resistivity'' over-damps ${\bf B}$ and suppresses jets. At high-$|{\bf B}|$ ($\mu\ll 10$), less-accurate $\divB$-cleaning leads to unstable errors that disrupt the disk.
    \vspacerpostplot 
    \label{fig:core.sph}}
\end{figure}

\begin{figure}
    \plotonesize{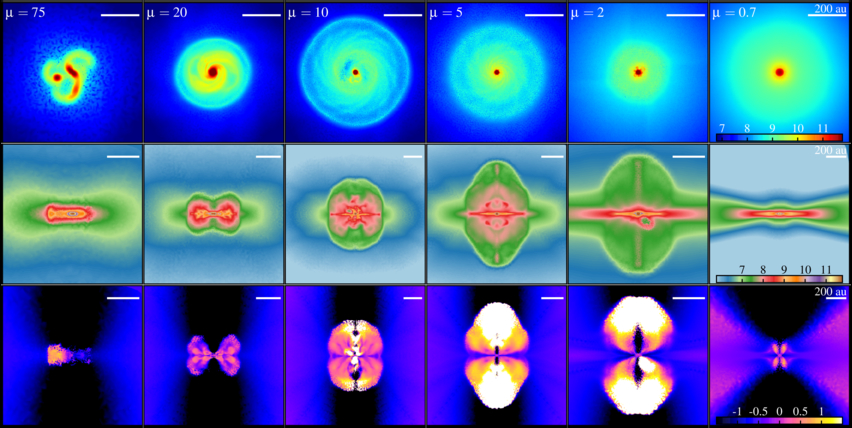}{0.98}
    \vspace{-0.25cm}
    \caption{As Fig.~\ref{fig:core.sph}, for SPH-hi but at higher resolution ($100^{3}$). At higher resolution, SPH is better able to control $\divB$ errors and jets emerge at high-${\bf B}$ ($\mu<10$); however the SPH numerical resistivity still over-damps the weak-field case.
    \vspacerpostplot 
    \label{fig:core.sph.hires}}
\end{figure}

\begin{figure}
    \plotonesize{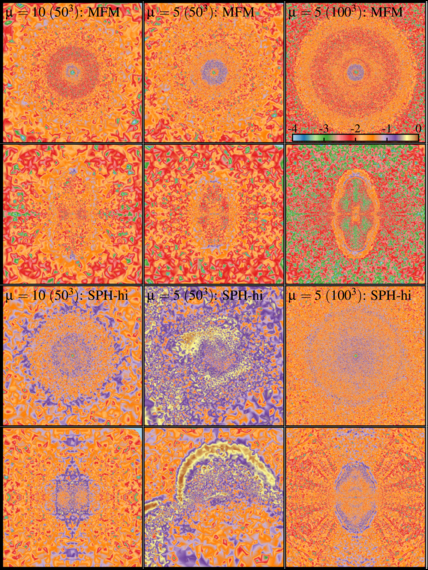}{0.98}
    \vspace{-0.25cm}
    \caption{Divergence errors for some of the protostellar core tests. We show face-on ({\em upper}) and edge-on ({\em lower}) slices as Figs.~\ref{fig:core.bfield}-\ref{fig:core.sph.hires}, for MFM ({\em top}; MFV is similar) and SPH-hi ({\em lower panel}). The unstable behavior in SPH at low resolution is clearly related to poor control of the $\divB$-errors. We show two $\mu$ values at $50^{3}$ resolution and compare $\mu=5$ at higher ($100^{3}$) resolution. In all cases $\divB$ errors decrease with resolution.
    \vspacerpostplot 
    \label{fig:core.divB}}
\end{figure}

\vspace{-0.5cm}
\subsection{Collapse of a Magnetized Core: Preserving Symmetry \&\ Launching MHD Jets}
\label{sec:core}

Next, we consider collapse of a magnetized proto-stellar core and launching of a proto-stellar jet. This is a less quantitative problem; however, there are several key qualitative phenomena. A rotating, weakly-magnetized, self-gravitating gas sphere is initialized. This collapses under self-gravity quasi-spherically to much higher densities, testing the ability to follow the fluid in compressions and collapse over many orders of magnitude, and whether the gravity-hydro coupling is conservative. The collapse is arrested by the formation of a disk (requiring good angular momentum conservation in the code), which slowly contracts via magnetic braking. The braking and field amplification in collapse and the subsequent disk-driven dynamo test the ability of the code to follow initially weak fields and MHD instabilities. Finally, a jet is launched: following this requires the code have good symmetry preservation, and most critically maintain low values of $\divB$, or else the protostar will be ejected from the disk by accumulated errors \citep[see][]{price:2012.sph.mhd.jets}. 

Following \citet{hennebelle:2008.protostellar.jets}, we initialize a 3D box of size-length $0.15$\,pc, in which the hydrodynamic forces are periodic but gravity is not. We initialize a constant-density sphere of radius $R_{0}=0.015$\,pc and mass $1\,M_{\sun}$, in a uniform, non-moving background (filling the box) with factor $\approx360$ lower density. The sphere is set in rigid-body rotation such that the orbital time is $4.7\times10^{5}$\,yr; this corresponds to a ratio of kinetic-to-potential energy ${\rm KE}/|{\rm PE}|\approx 0.045 \ll 1$. The magnetic field is initialized to a constant value $B_{0}$, aligned with the angular momentum vector of the sphere. At all times, the system is forced to obey the barotropic equation of state $P = (0.2\,{\rm km\,s^{-1}})^{2}\,\rho\,\sqrt{1 + (\rho/\rho_{0})^{4/3}}$, with $\rho_{0}=10^{-14}\,{\rm g\,cm^{-3}}$. The value of the field $B_{0}$ is chosen to correspond to different mass-to-magnetic flux ratios, defined in the traditional fashion relative to the ``critical'' value quasi-stability of a spherical cloud, $\mu = (M/\Phi)/(M/\Phi)_{\rm crit}$ where $(M/\Phi)\equiv M/(\pi\,R_{0}^{2}\,B_{0})$ and (in our units) $(M/\Phi)_{\rm crit}\equiv 0.126\,G^{-1/2}$. With $M$ fixed, $\mu\propto B_{0}^{-1}$, and for our choices $\mu=10$ corresponds to $B_{0} = 61\,\mu\,G$. 

Fig.~\ref{fig:core.bfield} shows the results from our MFM method, for various values of $\mu$ ($B_{0}$), using $60^{3}$ {\em total} cells/particles in the box, or $\approx 50^{3}$ cells in the collapsing sphere (we use equal-mass particles, so the initial packing is denser in the sphere). The times are chosen to be some (short) time after a jet forms, in each case close to $t \sim 4\times10^{4}\,{\rm yr} \sim t_{\rm ff}$, where $t_{\rm ff}\equiv \sqrt{3/(2\pi\,G\,\rho)}$ is the gravitational free-fall time. Fig.~\ref{fig:core.mfv} shows the same for MFV.

Even at this (modest) resolution, we are able to see all of the key behaviors above over a wide range of $\mu$. As expected, for $\mu\lesssim 1$, a stable, thick disk forms, which does not spin up the field sufficiently to launch a jet. But for larger $\mu>1$ (where the disk collapses rapidly), a jet is launched and propagates well into the diffuse medium as accretion onto the protostar continues. The collimation and strength of the jets vary with $\mu$ as expected, and are consistent with AMR simulations using CT and run with much higher resolution \citep[see][]{hennebelle:2008.protostellar.jets}. A more quantitative exploration of this is an interesting scientific question, but outside the scope of our study here. We stress that as the fields become weaker, this problem becomes more challenging, since any excess numerical dissipation will tend to suppress field amplification and jet-launching. This is well-known from grid-based studies using Riemann solvers with different diffusivity. What is remarkable is that we are able to see jets even at $\mu\gtrsim10$ at this resolution. For weaker fields than about $\mu\sim20$, various authors have noted that the magnetic field becomes insufficient to stabilize the disk, which then fragments; we see this in our weak-field case. However, even here, we see that each fragment/protostar is launching its own ``mini-jet,'' which then precess owing to the orbital motion of the fragments! 

To see these behaviors, particularly in the weak-field case, usually requires quite high resolution ($>128^{3}$) in grid-based codes \citep{hennebelle:2008.protostellar.jets,pakmor:2011.arepo.mhd}. In Fig.~\ref{fig:core.res}, we show a resolution study at fixed time in our fiducial MFM, $\mu=10$ case. While the disk continues to show more fine structure, and the jets are launched slightly earlier (so have propagated further) at higher resolution, the presence of a jet is clearly resolved, and the jet outflow rate and momentum are converged to within tens of percent, at $25^{3}$ resolution; some jet is even visible at $12^{3}$ resolution! 

We can see these behaviors even at very low resolution because this test combines many advantages of our new methods. It is clearly Lagrangian, with high-dynamic range collapse; the fact that $N$-body codes naturally couple to our methods with fully adaptive gravitational softening providing second-order accuracy (\paperone) means that we can follow the entire collapse  without any special ``switches'' required in our gravity solver. The smooth, continuous adaptive resolution provided by our method avoids the low-order errors which tend to artificially damp magnetic fields, inherent to AMR refinement boundaries. Once the disk forms, it is critical that the method be able to accurately conserve angular momentum and not degrade orbits; this is especially challenging in AMR codes (assuming $\sim 50-100^{3}$ cells across the central disk, even high-order AMR methods will degrade the orbits within a couple of periods; see \paperone\ \&\ \citealt{de-val-borro:2006.disk.planet.interaction.comparison,hahn:2010.disk.gal.orientations.ramses,duffell:2012.disco.method.protoplanetary.disk}). 

Fig.~\ref{fig:core.rotation} demonstrates that, because they are Lagrangian and mesh-free, our methods here are trivially invariant to both (a) boosting the entire fluid (so the core is moving super-sonically through the box), and/or (b) rotating the core at an arbitrary angle to the coordinate axes. This is {\em not} true in non-moving grid methods: either of these changes in grid codes implies a very significant loss of ``effective resolution'' or accuracy at fixed resolution. Based on comparison experiments with {\small RAMSES}, we estimate that for the $\mu=10$, rotated+boosted case in Fig.~\ref{fig:core.rotation}, achieving comparable accuracy in AMR to our $\sim 50^{3}$ ($10^{5}$-element) MFM run requires $\sim 10^{8}$ cells.

SPH, however, has greater difficulty with this problem. Fig.~\ref{fig:core.sph} shows the same survey of $\mu$ at $50^{3}$ resolution in SPH; in no case does a stable jet develop. For small initial fields ($\mu \gtrsim 10$), artificial resistivity too-efficiently dissipates the field, so the case resembles pure hydro (either weak or no jets appear). For large fields, the method has difficulty maintaining stability: the central disk breaks up, and the proto-star is ejected! These errors are resolution-dependent, however. At higher resolution, we do begin to recover the correct behavior in SPH; Fig.~\ref{fig:core.sph.hires} shows the same at $100^{3}$ resolution. Here the high-${\bf B}$ cases begin to resemble the MFM/MFV results. The low-${\bf B}$ cases still exhibit too much dissipation (too-weak jets), but the field values are (very slowly) converging towards the MFM/MFV result.\footnote{We have explicitly checked, for the proto-stellar jet test in SPH, whether using the Alfven or magnetosonic velocity in the artificial resistivity (as in \S~\ref{sec:mixing}) makes a large difference; it does not. Likewise the choice of ``pressure''-SPH (PSPH) versus traditional ``density''-SPH (TSPH), discussed in \paperone, is not especially significant for the core/jet problem. We have also experimented with a wide range of different artificial viscosity and conductivity parameters and find the qualitative results are similar. However, as noted in \citet{tricco:2013.sphmhd.methods}, it is important that the artificial resistivity be limited to a small maximum value; our default here is $0.1$ (see Appendix~\ref{sec:spmhd.numerics}), if we raise this to $1$, the ${\bf B}$ fields are artificially damped away to negligible values during collapse.}

Fig.~\ref{fig:core.divB} compares the divergence errors; this illustrates the source of the instability in SPH. In MFM/MFV (essentially identical here), the errors are well-controlled at the $10^{-3}-10^{-2}$ level (a few particles in the disk midplane reach larger values, but these are in the regime where the thin disk is completely unresolved vertically so the change in ${\bf B}$ must occur across a single particle). We also clearly see that the errors decrease with resolution (the median $h\,|\divB|/|{\bf B}|$ decreases by a factor of $\sim 5$ from $50^{3}$ to $100^{3}$ resolution). In SPH, we see much larger $\divB$, as in many cases above -- most critically, we see that where the disks have gone unstable and ``kicked out'' the protostars corresponds (in every case) to $h\,|\divB|/|{\bf B}|\gg1$. This is consistent with previous studies \citep{burzle:2011.protostellar.outflows,price:2012.sph.mhd.jets}. Because the $\divB$ errors decrease with resolution, going to higher-resolution suppresses this and allows stable evolution of the jets.

\begin{figure*}
    \plotsidesize{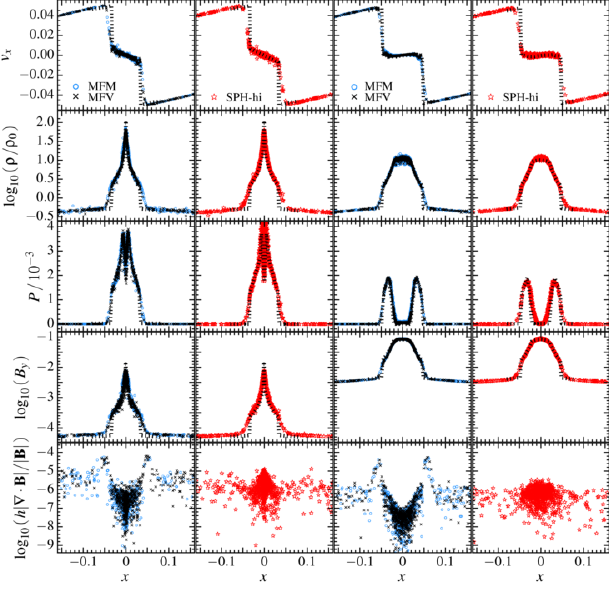}{0.75}
    \vspace{-0.25cm}
    \caption{Magnetic Zeldovich pancake (\S~\ref{sec:zeldovich}) at $z=0$. An initially small (linear) sinusoidal density perturbation collapses along one dimension ($\hat{x}$) in a 3D cosmological expansion, forming a caustic at $z=1$, with a small initial perpendicular magnetic field ${\bf B}=B_{0}\,\hat{y}$. We plot velocity in the collapse direction ($v_{x}$), density $\rho$ (relative to the box-mean, $\rho_{0}$), pressure $P$, and field $B_{y}$. All cells/particles are plotted; the resolution is $\sim 100^{3}$. {\em Left:} Weak-initial field case. Here the results are similar to the pure-hydro case (see Fig.~30) in \paperone; $B_{y}$ grows via compression in the center and declines adiabatically with the Hubble flow outside the shock. {\em Right:} Strong-initial field case. Here, the shock center is magnetically dominated, flattening the density peak and generating a pressure cavity. 
    In both cases, MFM/MFV converge well to the exact solution (here, a $2048^{3}$-equivalent, one-dimensional grid-based PPM, CT calculation from {\small ENZO}) -- a 1D, $512$-element MFM calculation is indistinguishable from the exact solution. At lower ($100^{3}$) resolution, the outer, lower-density shock at $x=\pm 0.03$ is broadened by $\sim 2$ particles in the linear direction (about the same as obtained in second-order grid methods). The inner region, including the detailed shape of the central density and ${\bf B}$-field, are already well-converged. The shape and boundaries of the low pressure cavity in the strong-field case are well-resolved, but the lowest value of the {\em thermal} pressure converges slowly in the central region where $P_{\rm thermal} \ll |{\bf B}|^{2}/2$ (i.e.\ where it is dynamically insignificant). 
    In SPH, shocks are broadened by about twice as many particles in the linear direction, and the density peak shape converges slightly more slowly. The shape of the boundaries of the central pressure cavity in the strong-field case converge significantly more slowly than in MFM/MFV/grid codes. The $\divB$ errors are negligible in this problem.
    \vspacerpostplot 
    \label{fig:zeldovich}}
\end{figure*}

\vspace{-0.5cm}
\subsection{MHD Zeldovich Pancake: Testing Cosmological MHD Integration \&\ Extremely High Mach Number Shocks}
\label{sec:zeldovich}

We now consider an MHD Zeldovich pancake, following a single Fourier mode density perturbation in an Einstein-de Sitter space, to test the code's ability to deal with cosmological integrations, small-amplitude perturbations, large Mach numbers, and anisotropic cell arrangements. We initialize a linear perturbation following \citet{zeldovich:1970.pancakes}, with parameters chosen to match \citet{li:2008.cosmo.grid.mhd}, in a periodic box with side-length $L=1$ and $\gamma=5/3$. Assume the unperturbed fluid elements have uniform density with co-moving position $q$ along $\hat{x}$ as $z\rightarrow \infty$, then co-moving positions and fluid quantities at the initial redshift $z_{i}=20$ of the simulation are $x = q - A\,\sin(k\,q)/k$, $\rho = \rho_{0} / (1-A\,\cos(k\,q))$, $v_{\rm pec} = -H_{0}\,\sin(k\,q)/k$, where $A = (1+z_{c})/(1+z_{i})$, $k=2\pi/L$, $\rho_{0}=3\,H_{0}^{2}/(8\pi\,G)$ the critical density, and $z_{c}=1$ is the redshift of caustic formation. In our particular (arbitrary) code units, $H_{0} \approx 0.18$, and the initial $u=9.3\times10^{-8}$ and ${\bf B} = B_{0}\,\hat{y}$ (comoving). We consider a weak and strong field case, with $B_{0} = 1.25\times10^{-4}$ and $7.7\times10^{-3}$, respectively. These are rather unconventional units, but we note that the solution can trivially be rescaled; it depends only on the dimensionless parameters $z_{c}$ and the initial $\beta = P_{\rm thermal}/P_{\rm magnetic}$. The perturbation should grow linearly along the $\hat{x}$ direction under self-gravity, until it goes non-linear and eventually collapses into a shock (caustic) at $z=1$. 

At early times, the flow is smooth and obeys a known linear solution; we confirm that our MFM/MFV methods reproduce this with second-order convergence (even in cosmological integration with self-gravity coupled to MHD). In Fig.~\ref{fig:zeldovich} we plot the results at $z=0$. There are now clear strong shocks (pressure jumps of factors $\sim10^{8}$) and factor $\sim1000$ compressions. At various locations, the kinetic, thermal, and magnetic energy form an extremely disparate hierarchy: in the weak-field case $\rho\,|{\bf v}_{\rm ram}|^{2}\sim 10^{10}\,\rho\,c_{s}^{2} \sim 10^{14}\,|{\bf B}|^{2}$, or in the strong-field case, $\rho\,|{\bf v}_{\rm ram}|^{2} \sim 10^{9}\,|{\bf B}|^{2} \sim 10^{15}\,\rho\,c_{s}^{2}$. As a result, Eulerian codes which evolve only the total energy almost invariably crash or produce unphysical results (e.g.\ subtraction of very large numbers giving negative pressures). The Lagrangian nature of our method greatly reduces these errors, and the dual-energy formalism (\paperone; Appendix~D) allow the method to deal smoothly with disparate hierarchies. 

Another key aspect of the 3D test is that it involves highly anisotropic compressions: the gas collapses by a factor of $\sim1000$ along $\hat{x}$, but there is no collapse in the other directions. We discuss this at length in \paperone: for AMR methods, this makes it very expensive to achieve the same resolution as our mesh-free methods, because (since grid cells are cubical), refinement must take place along the $\hat{y}$ and $\hat{z}$ dimensions with the collapse; for moving-mesh methods, it requires very careful and accurate cell refinement and regularization schemes (which can introduce other errors). But it is handled continuously and naturally in our mesh-free Lagrangian schemes. For the same total number of resolution elements, this allows our Lagrangian methods to achieve a factor $\sim 10$ better spatial resolution along $\hat{x}$ in the central regions, compared to AMR (in 3D).

In the weak-field case, we confirm the results from \paperone: the method can handle large energy hierarchies; arbitrarily strong shocks are resolved across $\sim 2$ linear cells/particles for MFM/MFV, and smeared across a factor $\sim2-3$ larger range in SPH; co-moving integration with self-gravity is accurate; and because of their ability to handle asymmetric cell/particle distributions, the Lagrangian methods converge more rapidly in high-density regions compared to AMR. In the strong-field case, we also confirm that trace ${\bf B}$-fields are correctly amplified, and recover the correct jumps and rarefactions.

On this particular test, we obtain similar results with the \citet{powell:1999.8wave.cleaning}-only divergence cleaning. The reason is that there is negligible field in the $\hat{x}$ direction, and the growth of the perpendicular $B_{y}$ component is driven by simple, effectively one-dimensional compression/expansion.

\begin{figure}
    \plotonesize{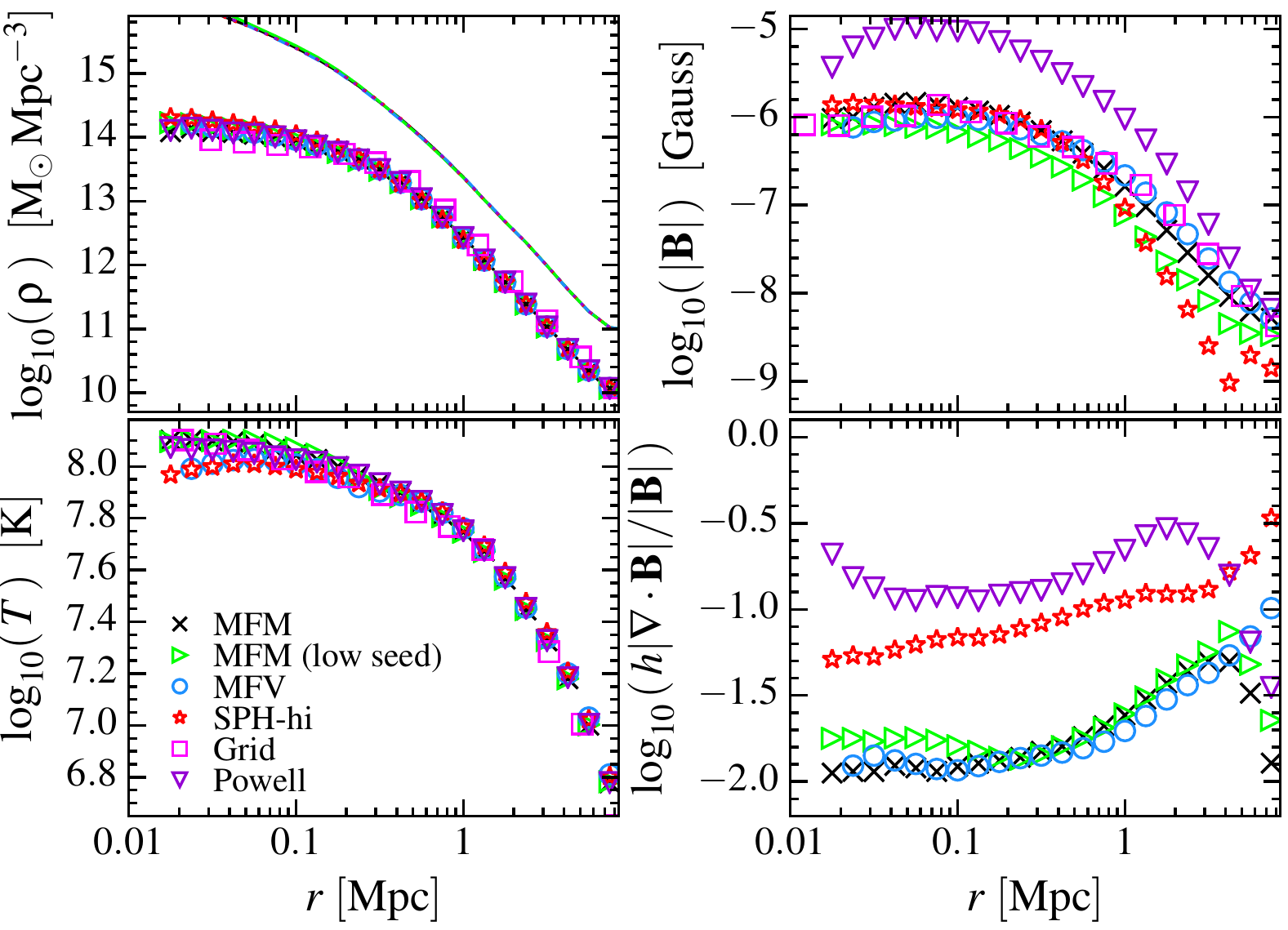}{0.99}
    \vspace{-0.25cm}
    \caption{Magnetic Santa Barbara cluster (\S~\ref{sec:sb}). A high-resolution sub-volume of a cosmological simulation is followed; it forms a cluster-mass dark matter halo; we show radially averaged profiles at $z=0$. 
    {\em Top Left:} Gas density (symbols) and dark matter density (lines). {\em Bottom Left:} Temperature. SPH shows slightly lower central-$T$ (see \paperone). {\em Top Right:} Magnetic field. The central rms $|{\bf B}|\sim \mu G$, independent of the numerical method. With Powell-only cleaning, however, the magnetic energy is artificially amplified (as in the field loop problem) to order-of-magnitude excessive values. In SPH, excess artificial resistivity leads to some excessive damping at large radii. The ``low seed'' run features a seed field a factor $\sim 10^{5}$ weaker than the default run: the final ${\bf B}$ is nearly independent of the seed. {\em Bottom Right:} Divergence errors. Absolute values in $|\nabla\cdot {\bf B}|$ are taken {\em before} averaging: the mean $|\langle h \nabla \cdot {\bf B} / |{\bf B}| \rangle | \sim 10^{-8}$ is approximately 6 orders of magnitude smaller. In MFM/MFV, the well-resolved ($<1\,$Mpc) region maintains small divergence errors. In SPH the errors are factor $\sim 5$ larger. 
    \vspacerpostplot 
    \label{fig:sb.cluster}}
\end{figure}

\vspace{-0.5cm}
\subsection{The MHD Santa Barbara Cluster: Cosmological MHD Integration in Turbulent Flows \&\ Divergence-Control}
\label{sec:sb}

Next we consider the MHD ``Santa Barbara Cluster'' from \citep{frenk:1999.sb.cluster}; the hydrodynamic case is again in \paperone. The test is a ``zoom-in'' where we initialize a high-resolution Lagrangian region in a low-resolution Einstein-de Sitter cosmological background, which collapses to form an object of a rich galaxy cluster mass at $z=0$. The cluster ICs are in \citet{frenk:1999.sb.cluster}; a periodic box of side-length $64\,h^{-1}\,{\rm Mpc}$ is initialized at redshift $z=49$,  in a flat Universe with dark matter density $\Omega_{\rm DM}=0.9$, baryonic $\Omega_{\rm b}=0.1$, Hubble constant $H_{0}=50\,{\rm km\,s^{-1}\,Mpc^{-1}}$. The gas is non-radiative with $\gamma=5/3$ and initial $T=100\,$K. We add to this a trace initial seed field ${\bf B}=B_{0}\,\hat{z}$. The initial ${\bf B}$ direction and magnitude should be unimportant, provided it is small.

Fig.~\ref{fig:sb.cluster} shows the resulting profiles of density, temperature, magnetic field, and $\divB$ at $z=0$, across simulations using different methods (the average $|{\bf B}|$ and $h\,|\divB|/|{\bf B}|$ values plotted in Fig.~\ref{fig:sb.cluster} are magnetic energy weighted averages). The grid result here is taken from high-order (PPM unsplit CTU) AMR simulations using CT in \citet{miniati:2011.mhd.charm.santabarbara}. The dark matter density profile (essentially determined by the $N$-body solver) is nearly identical in all runs, as expected (as is the final pressure profile, given by hydrostatic equilibrium). There are some very small differences in the central gas density, temperature, and entropy; these are discussed extensively in \paperone. 

In all methods, for a wide range of $B_{0}$, ${\bf B}$ is amplified to $\sim\mu$G in the cluster core. In our ``default'' runs, we seed $B_{0}=10^{-8}$\,G ($4\times10^{-12}$\,G co-moving). This gives plasma $\beta \equiv P_{\rm gas}/P_{\rm mag} \ge 50$ everywhere in the ICs, so the magnetic pressure is unimportant. As long as this is true, the final ${\bf B}$ profile is nearly independent of $B_{0}$. We have verified this in all methods: for $B_{0}\gtrsim 10^{-7}$\,G, the initial $\beta\sim1$ and the hydrodynamic properties ($\rho$, $T$) as well as maximum $|{\bf B}|$ begin to change. For sufficiently small $B_{0}$, in practice, numerical resistivity and truncation errors will swamp the field and suppress growth: we obtain similar final ${\bf B}$ profiles for minimum $B_{0} \ge (10^{-14},\ 10^{-11},\ 10^{-9},\ 10^{-12})$\,G in (MFM, MFV, SPH, AMR), respectively. The fact that we can use such small minimum $B_{0}$ (comparable to roundoff errors) in MFM reflects the extremely low numerical dissipation of advected quantities inherent to the method. The large value in SPH reflects the artificial resistivity errors discussed above, whereby fields with $\beta \gg 100$ tend to be artificially over-damped. Even for $B_{0}=10^{-8}$\,G, we already see some SPH over-damping in the cluster outskirts.

In MFM/MFV methods, $\divB$ is well-controlled, with mean {\em absolute} values of $h\,|\divB|/|{\bf B}|\approx 0.01$ in the resolved region of the cluster. The errors rise towards the outskirts owing to (1) decreasing resolution there, and (2) boundary effects (the setup of this ``zoom in'' involves no gas outside the initial Lagrangian region, so there are vacuum boundaries on one side of many particles). The median $h\,|\divB|/|{\bf B}|$ is $\sim 10^{-4}-10^{-3}$. In SPH, the errors are less well-controlled, reaching $\sim 0.1$; however this appears to have little or no effect on the solution. 

With Powell-only cleaning, however, the field is artificially amplified to order-of-magnitude larger values in the cluster core; this is essentially the compounded version of the erroneous growth seen in the field loop test. This is qualitatively different from the behavior seen in any runs with cleaning, despite the fact that $h\,|\divB|/|{\bf B}|$ (while large) is not much larger than our with-cleaning SPH runs. Once again, this emphasizes that in Powell-only cleaning, the {\em non-linear} error terms can integrate unstably (i.e.\ build coherently). In contrast, with a properly-applied \citet{dedner:2002.divb.cleaning.scheme} divergence-damping, these terms are stabilized and \citet{tricco:2012.sphmhd.methods} show that the ${\bf B}$ field cannot artificially self-amplify (rather, the sense of the errors will be to produce additional numerical dissipation, just like most other sources of error). So even if $\divB$ is nominally the {\em same} over the course of a simulation, runs with \citet{dedner:2002.divb.cleaning.scheme} cleaning will avoid many of the more serious instabilities of Powell-only cleaning.

\begin{figure}
    \plotonesize{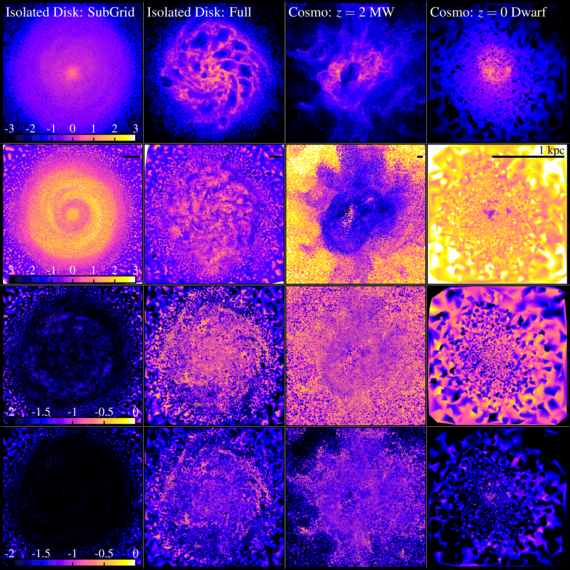}{0.99}
    \vspace{-0.25cm}
    \caption{Galactic disk test (\S~\ref{sec:galaxies}). We simulate a galaxy disk (gas, stars, and dark matter) with full self-gravity, star formation, cooling \&\ gas chemistry, and stellar feedback/mass return in SNe, stellar winds, and radiation (photo-heating and radiation pressure). We consider four galaxy models: an isolated (non-cosmological) starburst disk, with either the sub-grid \citet{springel:multiphase} ``effective equation of state'' model for the ISM ({\em left}; this does not explicitly treat the small-scale ISM turbulence and multiphase structure), or the ``full'' FIRE physics modules from \citet{hopkins:2013.fire} ({\em second from left}; these explicitly treat the multi-phase, supersonically turbulent ISM), as well as a cosmological zoom-in simulation of a Milky-Way mass galaxy run to $z=2$ with the full FIRE physics ({\em second from right}; undergoing a major merger at this time), and a cosmological zoom-in of a dwarf with the FIRE physics ({\em right}; at $z=0$). 
    {\em Top:} Gas density $\log_{10}(n/{\rm cm^{-3}})$, in a slice through the galaxy midplane, as labeled. {\em Second Row:} Plasma $\log_{10}(\beta)$ ($\beta\equiv P_{\rm thermal}/P_{\rm magnetic}$), in the same slice. The ${\bf B}$ values reach approximate equipartition with the thermal+turbulent energy, producing a wide range in $\beta$. {\em Third Row:} Divergence error $\log_{10}(\,h\,|\divB|/|{\bf B}|\,)$. These are reasonably controlled but can reach local values $\gtrsim 10\%$, because particles are being re-arranged by gravity and feedback on timescales much faster than the fast magnetosonic ``response time'' for divergence-cleaning. {\em Bottom:} $\log_{10}(\,h\,|\divB|/\sqrt{|{\bf B}|^{2} + 8\pi\,P_{\rm thermal}}\,)$. This shows the divergence error relative to the total hydrodynamic pressure: here typical values are $\lesssim 10^{-2}$ -- the large nominal values of $|\divB|/|{\bf B}|$ generally appear only in regions where the magnetic fields are dynamically irrelevant.
    \vspacerpostplot 
    \label{fig:iso.disk}}
\end{figure}

\begin{figure}
    \plotonesize{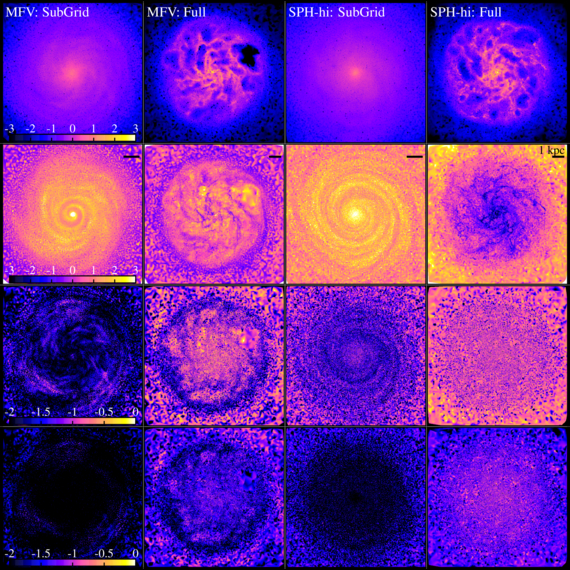}{0.99}
    \vspace{-0.25cm}
    \caption{As Fig.~\ref{fig:iso.disk}, but for our MFV and SPH-hi methods. Here we just compare the isolated disk, with the simplified sub-grid (``smoothed ISM'') physics or ``full ISM'' (multiphase, turbulent) physics, as labeled. The gas density distributions, star formation rates, and galactic outflow rates are similar (for the same physics) in each method, although SPH predicts somewhat stronger/weaker fields in the inner/outer regions of the galaxy. SPH has larger divergence errors by a factor of $\sim 3-10$, consistent with our previous tests.
    \vspacerpostplot 
    \label{fig:iso.disk.method}}
\end{figure}

\vspace{-0.5cm}
\subsection{Simulated Galaxies: Testing Code Robustness in Multi-Physics Applications}
\label{sec:galaxies}

We now consider a ``stress test'' of the methods here, adding magnetic fields to simulations of galaxies using state-of-the-art physics models. Specifically, we consider three initial conditions: (1) an isolated (non-cosmological) galaxy (with a pre-existing stellar disk and bulge, gas disk, and dark matter halo), designed to represent a starburst/M82-like system (model {\bf Sbc} in \citealt{hopkins:fb.ism.prop}, with a $1\,\mu$G seed field); (2) a cosmological ``zoom-in'' simulation of a dwarf galaxy (with $z=0$ halo mass $10^{10}\,\msun$; model {\bf m10} in \citealt{hopkins:2013.fire}, with a $10^{-10}\,\mu$G seed field); and (3) a ``zoom-in'' of a Milky Way-like system, run to $z=2$ (model {\bf m12i} in \citealt{hopkins:2013.fire}; $10^{-10}\,\mu$G seed field).  We intentionally study low-resolution versions of each -- using a factor of $\sim 64$ fewer particles than the ``production runs'' in those papers -- in order to maximize the numerical challenge. For each galaxy, we activate the full suite of physics from the FIRE (Feedback In Realistic Environments) simulations described in \citet{hopkins:rad.pressure.sf.fb,hopkins:stellar.fb.winds,hopkins:stellar.fb.mergers} and \citet{faucher-giguere:2014.fire.neutral.hydrogen.absorption}. This includes: self-gravity; gas cooling and chemistry; star formation; cosmological expansion; the interaction of gas, stars, and dark matter; energy, mass, momentum, and metal injection from supernovae and stellar winds; and radiation-matter interactions in the form of photo-ionization, photo-electric heating, and radiation pressure.

We emphasize that there is no simple ``correct'' solution for the ${\bf B}$-field evolution in these tests, and our concern here is not whether this particular model of the physics is correct or complete (we know, in fact, that these examples are under-resolved). Rather, we test (1) whether or not the algorithms we have developed can run (at all), without crashing or returning unphysical solutions; and (2) how well they control the divergence errors. This is extremely challenging: essentially every numerically difficult situation our test problems have considered above will occur here (and often be poorly-resolved). In addition, gas is dis-continuously added \&\ removed from the system (by stellar mass loss and star formation), and non-MHD forces (gravity and stellar feedback) are constantly re-arranging the particles/cells on timescales much faster than the local magnetosonic crossing times. 

Fig.~\ref{fig:iso.disk} shows the results for our MFM runs, and Fig.~\ref{fig:iso.disk.method} compares MFV and SPH for a subset of the initial conditions. With the ``full'' FIRE physics, all the runs develop a multi-phase, super-sonically turbulent medium, with strong galactic outflows \citep[for details, see][]{hopkins:clumpy.disk.evol,hopkins:2013.merger.sb.fb.winds,hopkins:virial.sf}. The ${\bf B}$ fields are amplified to values in very rough equipartition with the thermal+turbulent energy of the disk: in cold molecular clouds ($T=100$\,K, $v_{\rm turb}\sim 10\,{\rm km\,s^{-1}}$, $n\sim 10-10^{3}\,{\rm cm^{-3}}$), this implies $|{\bf B}|\sim 10-100\,\mu$G and small plasma $\beta \sim 10^{-2}$, while in SNe-heated bubbles with $T\sim10^{7}$\,K and $n\sim 0.01$ we find $\beta$ as large as $\sim 10-100$. The algorithms are stable under arbitrarily long integration.

In general, we find that $\divB$ is reasonable well-controlled, with mean values of $h\,|\divB|/|{\bf B}|\sim 0.03-0.1$ in MFM/MFV, and somewhat larger $\sim0.1-0.2$ in SPH. Although still within this range, the errors are clearly larger in our cosmological zoom-in runs. This owes to two facts: they are less well-resolved, and they are less dynamically relaxed systems (being perturbed by mergers, accretion, etc). However, we stress that these relatively high values of $h\,|\divB|/|{\bf B}|$ are totally dominated by the regions where the magnetic fields are dynamically irrelevant (i.e.\ where $|{\bf B}|$ is small). In these regions, the fields are passive, and our divergence cleaning scheme has essentially no time to ``respond'' to the constant, super-sonic, turbulent particle rearrangement. We therefore instead plot $h\,|\divB|/\sqrt{|{\bf B}|^{2} + 8\pi\,P_{\rm thermal}} = h\,|\divB| / |{\bf B}|\,\sqrt{1+\beta}$, which compares the $\divB$ error to the total MHD pressure (the relevant term for the MHD forces); here we see that the errors are actually well-controlled at the $\lesssim 10^{-2}$ level. It is also worth noting that, like in \cite{pakmor:2011.arepo.mhd}, the errors are ``locally offsetting'' -- if we smooth/average the fields over a few neighboring particles, $|\divB|$ rapidly vanishes. 

We also compare a more simplified ISM/star formation model; this is the sub-grid \citet{springel:multiphase} ``effective equation of state'' model. Here, the turbulence and phase structure of the ISM is not resolved, but replaced with a prescription which forms stars, kicks gas out of the galaxy in a ``wind,'' and pressurizes the gas such that certain large-scale properties can be recovered (this is the model used in popular large-volume simulations such as {\small Illustris} or {\small EAGLE}; \citealt{vogelsberger:2013.illustris.model}). By design, in these ``sub-grid'' models the small-scale phase structure and turbulence are smoothed over in the ISM (although the simulation still includes super-sonic motion, self-gravity, star formation, and galactic winds). With a smoother gas distribution, our divergence-control scheme does an excellent job maintaining errors $<10^{-2}$. Compare this to Figs~13-15 in \citealt{pakmor:2012.mag.field.disk.evol.weak.fx}, who apply the Powell-only scheme to essentially the same problem in {\small AREPO}, and find mean errors $h\,|\divB|/|{\bf B}|\approx 1$ (nearly independent of their resolution).

\vspace{-0.5cm}
\section{Performance}
\label{sec:performance}

In \paperone, we compare performance in terms of speed and memory useage, across MFM/MFV, our modern SPH-hi, and ``bare-bones'' SPH (the fastest but least accurate form of SPH, using no higher-order switches for diffusion, the simplest SPH forms of the hydrodynamic equations, and small neighbor numbers), and both moving-mesh and grid/AMR codes. Although performance is always problem-dependent, we found in general that speed (at fixed number of resolution elements) in MFM/MFV was comparable to (or slightly faster than) ``bare-bones'' SPH, and a factor of $\sim 1.5-2.5$ faster than SPH-hi, which is itself comparable to the speed of moving-mesh algorithms. The memory requirements are very similar across MFM/MFV/SPH-hi/SPH-lo, and substantially (factor $\sim2$) lower than in moving-mesh or AMR methods. 

In terms of zone-cycles per second, the MHD versions of the MFM/MFV algorithms are slower than the hydro-only algorithm by about $\approx 30\%$ (run time a factor of $1.3$ larger), owing to the additional fluid quantities (and their gradients) which must be evolved and reconstructed (and extra terms in the Riemann solver/equation of motion).\footnote{To compare performance of the MHD algorithms, consider a simple test which essentially counts cycles per second. We initialize a 3D box with a standing linear wave of negligible fractional amplitude ($10^{-6}$; just so the values of gradients, and solution of the Riemann problem, are not trivially vanishing), and evolve it for a short time. We consider a pure hydro case, and then add a trace magnetic field that does not alter the dynamics.} The fractional difference in SPH is about $15\%$. Compared to many other MHD implementations -- particularly constrained-transport methods -- this is extremely efficient: in {\small ATHENA} using CT the speed difference is a factor $\approx 2.3$ (and in moving-mesh methods can be much larger; see \citealt{mocz:2014.constrained.transport.mhd}). The memory requirements of MFM/MFV/SPH are also somewhat higher in MHD compared to pure hydro, but only by the amount needed to carry the additional MHD quantities (${\bf B}$, $\psi$), their gradients, and time derivatives. 

Comparing performance across different MHD methods in our tests is non-trivial, because the non-linear dynamics are slightly different. But on all our tests, MFV and MFM are nearly identical in cost (MFV is systematically $\approx 5\%$ slower, owing to the need to calculate mass fluxes). If we consider an idealized, pure-MHD test and force identical timesteps (i.e.\ compare cycles per second), we find that MFM is only $\approx 5-10\%$ more expensive than SPH-lo (which requires no reconstruction or Riemann problem and uses the same neighbor number); however as in the hydro case, larger timesteps are allowed in MFM, so we actually find in our real runs that MFM is slightly {\em faster} than even SPH-lo and ``bare-bones'' SPH MHD with the same neighbor number. SPH-hi (with $N_{\rm NGB}=120$ in 3D), on the other hand, is a factor $\approx 2 - 2.5$ more expensive than MFM (because $\sim 4x$ as many neighbors are needed). 

Even comparing complicated, highly non-linear problems with gravity, like the MHD Zeldovich pancake or Santa Barbara cluster, we find similar results. On these problems MFV is $\approx 10\%$ more expensive than MFM (the additional difference owes to variable particle masses in the gravity tree), and SPH-hi is a factor $\approx 2-2.5$ slower (for the same particle number). Furthermore, since we have shown that achieving the same accuracy in SPH requires significantly larger number of resolution elements (and convergence in SPH is slow), we conclude that, at fixed {\em accuracy}, our MFM/MFV methods are less expensive than SPH by factors of $\sim2-10$ (depending strongly on the problem).

\vspace{-0.5cm}
\section{Discussion}
\label{sec:discussion}

We have extended the mesh-free MFM and MFV finite-volume, arbitrary Lagrangian-Eulerian Godunov-type hydrodynamics methods from \paperone\ to include ideal MHD. We have implemented a second-order accurate, conservative formulation of these methods into our code {\small GIZMO}, together with state-of-the-art implementations of MHD in SPH. We systematically compared these methods to the results from grid codes and analytic solutions on a wide range of test problems, and find that the MFM and MFV methods are at least competitive with state-of-the-art grid MHD codes, and in many cases may have some advantages. 

Critically, we find that our meshless methods can indeed capture phenomena like the MRI (with the correct growth rates and mode structure), formation of magnetically-driven jets in collapsing cores, MHD fluid-mixing instabilities (the Rayleigh-Taylor and Kelvin-Helmholtz instabilities), and both sub and super-sonic MHD turbulence. In fact we find convergence on these problems in our MFM/MFV methods is comparable to, and in some cases even faster than, AMR codes using constrained transport. This supports the similar conclusions found by \citet{gaburov:2011.meshless.dg.particle.method}, studying the MFV method on a smaller set of problems.\footnote{We note that our major extension of the work in \citet{gaburov:2011.meshless.dg.particle.method} is to extend the methods to include (and test) the MFM method, as well as to conduct a systematic comparison with SPH and grid codes, on a wide variety of tests not considered in that paper. We have also made many subtle improvements of the MFV algorithm (all described here and in \paperone); these do not change the qualitative behavior on any tests, but do tend to decrease numerical noise.} Historically, these problems have been difficult for SPH; our new methods, however, do not suffer from the low-order errors that typically cause problems for SPH MHD. But we also show that even SPH, with the most current implementation of MHD, is able to capture most of these phenomena, albeit at the cost of larger kernels and some ``by hand'' adjustment of artificial dissipation parameters. 

\vspace{-0.5cm}
\subsection{The Divergence Constraint \&\ Conservation}

Most importantly, we find that, using a state-of-the-art implementation of the \citet{dedner:2002.divb.cleaning.scheme} divergence-cleaning scheme (re-discretized appropriately for our new methods), we are able to maintain $\nabla\cdot {\bf B}\approx 0$ to sufficient accuracy that divergence errors do not corrupt the solutions to any of our test problems. Typically, this amounts to a ``worst-case'' $h\,|\nabla\cdot{\bf B}| / |{\bf B}| \sim 0.01$ even in highly non-linear problems evolved for long times. In smooth flows and/or highly-resolved problems, more typical values are $h\,|\nabla\cdot{\bf B}| / |{\bf B}| \sim 10^{-4}$. 

This is important: without any $\nabla\cdot{\bf B}$ correction, the MHD equations are numerically unstable, and most problems will either crash or converge to unphysical solutions. The simplest ``fix'' in the literature is to just subtract the unstable terms, the so-called \citet{powell:1999.8wave.cleaning} or ``8-wave'' cleaning. However, we show that {\em the Powell cleaning alone converges to the wrong solution on most test problems}. 

The problem is, in certain types of MHD discontinuities, Powell-cleaning alone produces the wrong jump conditions, even in the limit of {\em infinite} resolution, because the errors occur across a single resolution element and are zeroth-order. This has been shown before for a limited range of problems in fixed-grid codes; here we show the same applies to a wide range of problems in {\em all} the methods considered here. In the Brio-Wu and Toth shocktubes, the shock jump conditions are wrong, the same problem leads to {\em qualitatively} incorrect features appearing in MHD blastwaves (less dramatic versions of these errors appear in both the Orszag-Tang vortex and MHD rotor). In advection of a magnetic field loop, the field strength can grow unstably -- the same errors disastrously corrupt the non-linear growth of the MHD Rayleigh-Taylor and Kelvin-Helmholtz instabilities, and lead to {\em orders-of-magnitude} incorrect growth of seed fields in cosmological MHD turbulence (e.g.\ the Santa Barbara cluster test). In the protostellar jet test, associated momentum errors can ``kick'' the core out of its disk.

With a good implementation of divergence-cleaning, we find that all of these errors are eliminated, provided that converged solutions are considered. This is clearly critical to almost any interesting astrophysical problem. Unfortunately, it means that many previous MHD studies \citep[see, for example][]{dolag:2009.mhd.gadget,burzle:2011.protostellar.outflows,pakmor:2012.mag.field.disk.evol.weak.fx,zhu:magnetized.wd.merger}, which relied only on the simpler Powell-type schemes, may need to be revisited.

It is worth noting that, of the tests we explore here, a combination of the MRI, protostellar jet launching, magnetic Santa Barbara cluster, and non-linear magnetic Rayleigh-Taylor instability, appear to be the most challenging to {\em simultaneously} capture accurately. We encourage authors of future MHD methods papers to include these as opposed to only focusing on a subset of problems like the MHD rotor, Orszag-Tang vortex, and shocktubes, which we find comparatively ``easy'' and not as useful.

\vspace{-0.5cm}
\subsection{MFM vs.\ MFV vs.\ Moving-Mesh Methods}

In all of our tests, we find small differences between our meshless finite-volume (MFV) and meshless finite-mass (MFM) methods; those (minimal) differences are similar to what we saw in pure hydrodynamics tests in \paperone. MFV, with mass fluxes, is able to more sharply capture contact discontinuities and minimize overshoot in the density/velocity fields around them. However, the additional fluxes lead to enhanced ``grid noise,'' so the method is slightly more noisy. 

In practice, the differences are sufficiently small that the ``better'' method will depend on the problem. For some purposes (e.g.\ cosmological simulations), it is extremely useful to maintain approximately constant particle/cell masses (because the dynamics are dominated by gravity in an $N$-body solver); this is accomplished more elegantly and significantly more accurately with MFM than with MFV plus cell splitting/merging (which is more analogous to an AMR-type code). But in other cases, high resolution might be desired in low-density regions of the flow, in which case MFV (possibly used in the mode where cells do not move exactly with the fluid velocity) is more natural. 

We have not presented a detailed comparison with moving-mesh codes, because a public moving-mesh MHD code capable of running the tests here is not available; however, a few of the test problems here have been considered in other studies with the moving-mesh codes {\small AREPO}, {\small TESS}, and {\small FVMHD3D} \citep{duffell:2011.TESS,pakmor:2011.arepo.mhd,gaburov:2012.public.moving.mesh.code}. In each of these cases the results are very similar to ours here (especially similar to our MFV method). This is consistent with our extensive hydrodynamic comparison in \paperone, and expected, given that the methods are closely related.

Most importantly, on all tests both MFM/MFV methods exhibit good convergence properties and capture all of the key qualitative phenomena, even at relatively poor resolution.

\vspace{-0.5cm}
\subsection{Comparison to SPH MHD}
\label{sec:discussion:sph}

Historically, it has been very difficult to capture non-trivial MHD phenomena with SPH. However, in the last few years there have been tremendous improvements to almost every aspect of the basic hydrodynamic algorithms in SPH, as well as the specific discretization of MHD (see references in \S~\ref{sec:intro}). As a result, we find that state-of-the-art SPMHD is, in fact, able to capture most of the important MHD phenomena studied here, including non-linear MRI, dynamo effects, magnetic jet launching, and fluid mixing instabilities. 

However, convergence in SPH is still very slow; in almost every case, SPH is still significantly more noisy, less accurate, and more diffusive at fixed resolution compared to our MFM/MFV methods, and requires some ``by hand'' tweaking of numerical parameters to give good results on all tests. There are two fundamental problems: first, SPH requires ``artificial diffusion'' (viscosity, conductivity, resistivity) terms, which are somewhat ad hoc. The resistivity term in particular is challenging, as discussed in \S~\ref{sec:mixing}: the correct ``signal velocity'' and question of whether resistivity should be applied at all is much less clear than, say, artificial viscosity, as it depends on the type of MHD discontinuity (not just whether one is present). We are unable to find a single ``switch'' the works best for all cases, and we show that using an even slightly less-than-ideal choice (e.g.\ using the magnetosonic versus Alfven speeds for the resistivity signal velocity) can catastrophically corrupt certain problems (such a fluid mixing instabilities and/or jet launching). Similar conclusions were reached in \citet{tricco:2013.sphmhd.methods}. A potential solution to this is the replacement of the artificial resistivity with the full solution of a Riemann problem between particles, as in ``Godunov SPH'' schemes \citep[see][]{iwasaki.2011:sphmhd.reimann.solver}. Second, and more fundamental, SPH has low-order errors which can only be suppressed by increasing the neighbor number in the kernel. This leads to an effective loss of resolution and higher diffusivity. However not increasing the neighbor number to some very large value ($\gg 100$ in 3D) leads to disastrously large errors and noise in most of our test problems. Similar problems are well-known in the pure-hydro case (see \paperone), but they are much more problematic in MHD, because of how they interact with the artificial resistivity and divergence cleaning terms. Kernel-scale noise seeded by the low-order errors produces magnetic divergences, which are then subtracted off and damped away, potentially corrupting the real solution (and preventing the algorithm from identifying ``real'' divergence errors). For this reason, the typical $\divB$ errors and numerical diffusion are $\sim 2$ orders of magnitude larger in SPH (even with $>100$ neighbors) compared to MFM/MFV methods (with just $32$).\footnote{Note that some authors have attempted to control the noise in SPH MHD by performing operations only on ``re-smoothed'' quantities \citep[see][]{dolag:2009.mhd.gadget,stasyszyn:2013.divb.cleaning.mhd.gadget}. This is similar in spirit to increasing the kernel size, and similarly leads to a loss of resolution and increase in numerical diffusion. But it is not clear whether the discretized equations after re-smoothing actually consistently represent the true hydrodynamic equations (they are not, mathematically, the Lagrangian-derived SPH equations), so it remains unclear whether such methods can actually converge (at any resolution) to the correct solution.}

Still, provided sufficiently high resolution and large kernel neighbor number are used, together with care in choosing the artificial diffusion parameters specific to the problem, we conclude that ``modern'' SPH MHD can produce accurate solutions. And SPH MHD may still have some limited advantages in specific contexts. The artificial diffusion operators are wholly operator-split from the hydrodynamic operators; when there are extreme energy hierarchies between kinetic, magnetic, and thermal energies, it ensures that small errors in any of the three terms do not directly appear in the others. It can handle free surfaces trivially and maintain numerical stability with vacuum boundaries; however the MHD equations will not be correct at these boundaries (the zeroth-order errors become order-unity, although they are numerically stable). And it remains the most computationally simple method we study.

\vspace{-0.5cm}
\subsection{Comparison to AMR}
\label{sec:discussion:amr}

In all cases, our new mesh-free methods (MFM/MFV) appear competitive with state-of-the-art grid-based codes (e.g.\ third-order PPM methods, with constrained transport, and CTU-unsplit integration, as in {\small ATHENA}). We find no examples where there are qualitative phenomena that either method cannot capture, nor any examples where we cannot converge to similarly accurate solutions. Of course, there are quantitative differences in the convergence rates, and errors at fixed resolution, which depend on the method. Despite the fact that Eulerian codes can use CT-methods to maintain the divergence constraint, we identify several problem classes in which convergence is faster in our Lagrangian methods than in AMR. 

Not surprisingly, these tend to be problems where advection, angular momentum conservation, self-gravity and/or following large compressions are important -- these are the obvious areas where Lagrangian methods have an advantage. For example, we see significantly faster convergence on the field-loop advection problem (our MFM/MFV methods produce about the same numerical dissipation as grid methods at $4^{D}$-higher resolution, where $D$ is the number of dimensions.) The mesh-free algorithms are robust to arbitrary ``boosts,'' which degrade the non-moving grid solutions on problems with mixing and/or contact discontinuities (e.g.\ the Rayleigh-Taylor and Kelvin-Helmholtz instabilities, Orszag-Tang vortex, MHD rotor, and blastwave/explosion problems). The errors caused by these boosts are, of course, resolution-dependent (and will converge away in grid codes), but this means that convergence to a desired accuracy on these problems is usually faster in our MFM/MFV methods than in grid-based methods, if the fluid is being advected at super-sonic velocities. This difference also means our new methods are robust to arbitrary velocities in the current sheet test (while some stationary-grid methods will crash for modestly super-sonic motion around the sheet). Perhaps most dramatically, the protostellar core collapse/MHD jet problem combines high-dynamic range collapse, self-gravity, and evolution of a global thin disk (angular momentum conservation) -- as a result, convergence is much faster in MFM/MFV than in AMR methods. Qualitative phenomena (e.g.\ the jet momentum/mass) start to converge at resolutions as low as $10^{4}$ cells/particles, compared to at least $0.3-1\times10^{7}$ cells in AMR (``effective'' resolutions of $\gtrsim 20,000^{3}$).\footnote{It is a common mistake to refer to ``kernels'' in mesh-free methods as ``resolution elements'' the same way single-cells are referred to in grid codes. This is wrong. In our MFM/MFV methods, the correct identification is to think of each particle as equivalent to a cell in a grid code (with about the same ``effective resolution per cell/particle''). The kernel represents the number of neighbors in causal contact for hydrodynamics; so the correct analogy is to the {\em stencil} used in a grid code (number of neighbors with adjacent faces, plus those needed for gradient calculations). Our default choice for MFM/MFV, then, of $\approx 32$ in 3D, is actually quite similar to what is obtained in moving Voronoi-mesh, AMR, and higher-order (PPM) Cartesian grid codes.} And the resolution demands become more severe in AMR if the disk is rotated and/or moving with respect to the coordinate axes: like in \paperone\ with a simple Keplerian disk problem, this requires $\gg 512^{3}$ resolution in AMR in the disk for good behavior, plus a comparable number of elements along the jet, to prevent it from numerically grid-aligning (artificially bending) and being destroyed. In contrast, our new methods are trivially invariant to such rotations and boosts, at any resolution.

Of course, on other problems, grid methods converge more rapidly. In smooth, pressure-dominated flows, the ``grid noise'' is minimized in truly fixed-grid (non-AMR) methods, so convergence in the highly sub-sonic regime (Mach numbers $\lesssim 0.01$) is usually faster. On shock-tube type problems (e.g.\ the Brio-Wu \&\ Toth problems above), where our errors are dominated by the noise introduced by divergence-cleaning and non-zero $\divB$ errors, we see significantly faster convergence in grid-based codes that can use constrained transport to eliminate these errors entirely (a similar factor $\sim 4^{D}$ as above). And of course, by virtue of {\em not} being Lagrangian, in high-dynamic range problems Eulerian codes will better-resolve {\em low}-density regions of the flow.

In short, we see no ``inherently'' superior method between AMR and our new mesh-free methods, simply differences in the accuracy achievable at fixed resolution or computational cost, which are highly problem-dependent.

\vspace{-0.5cm}
\subsection{Areas for Improvement \&\ Future Work}
\label{sec:discussion:future}

This is a first study, and there are many potential areas for improvement. Several possibilities discussed in \paperone\ (more accurate quadrature rules, generalizing to higher-order fluid reconstructions, better-optimized kernel functions) apply as well to the MHD case. 

The most dramatic improvement to the meshless methods here, however, would come from incorporating constrained transport. Recently, \citet{mocz:2014.constrained.transport.mhd} demonstrated that constrained transport could be successfully incorporated into moving-mesh algorithms; there is no conceptual reason why the algorithm described there cannot be applied to our MFM/MFV methods, since they are conservative finite-volume schemes with a well-defined set of effective ``faces'' and a partition of unity. In contrast, there is no clear way to generalize this to SPH (given the inherent zeroth-order inconsistency in SPH derivative operators, it is not clear whether it is possible under any circumstances to derive a CT-SPH method). However, the method in \citet{mocz:2014.constrained.transport.mhd} has not yet been extended to three dimensions and to adaptive timesteps, in a efficient manner which can run in competitive time. Therefore we have not considered it here, but it is certainly worthy of more detailed exploration in future work.

Absent a complete CT implementation, some progress might be made using locally divergence-free gradient representations, or (similarly) vector potentials. Previous efforts have been made in this area in both SPH MHD and discontinuous Galerkin methods \citep[see e.g.][]{miyoshi:2011.divergence.cleaning.comparison,mocz:2014.galerkin.arepo,stasyszyn:2015.vector.potential.sph}. These can offer some improvements; however, they usually sacrifice consistency and/or conservation, and by only providing locally divergence-free terms, it is by no means clear that they actually reduce the relevant errors driving numerical instability \citep{price:vector.potential}. But again, further study is needed.

In SPH, some errors (e.g.\ the zeroth-order errors) are inherent to the method. Others, however, could be decreased. There has been considerable work on improved switches for the artificial viscosity; similar work is needed for the artificial resistivity \citep[following the work of][]{tricco:2013.sphmhd.methods}. In particular, it would greatly expand the flexibility of the method if a switch were devised which could correctly interpolate between the relevant propagation speeds of the resistivity (which depends on the type of MHD discontinuity).

\vspace{-0.5cm}
\acknowledgments 
We thank Paul Duffell, Jim Stone, Evghenii Gaburov, Ryan O'Leary, Romain Teyssier, Colin McNally, our referee Daniel Price, and many others for enlightening discussions and the initial studies motivating this paper. Support for PFH was provided by the Gordon and Betty Moore Foundation through Grant \#776 to the Caltech Moore Center for Theoretical Cosmology and Physics, an Alfred P. Sloan Research Fellowship, NASA ATP Grant NNX14AH35G, and NSF Collaborative Research Grant \#1411920. Numerical calculations were run on the Caltech compute cluster ``Zwicky'' (NSF MRI award \#PHY-0960291) and allocation TG-AST130039 granted by the Extreme Science and Engineering Discovery Environment (XSEDE) supported by the NSF. 
\\

\vspace{-0.2cm}
\bibliography{/Users/phopkins/Dropbox/Public/ms}

\begin{appendix}


\section{The Smoothed-Particle MHD Implementation in GIZMO}
\label{sec:spmhd.numerics}

As noted in the text, the magnetic terms in our implementation of SPMHD follows that in the series of papers by \cite{tricco:2012.sphmhd.methods,tricco:2013.sphmhd.methods}. The full SPH hydrodynamics algorithms in {\small GIZMO} are given explicitly in \paperone. This includes both our implementation of ``traditional'' SPH and ``modern'' PSPH. We note that the additions for MHD are independent of whether TSPH or PSPH is used, but we will adopt the modern PSPH formulation as our ``default.'' 

With the hydrodynamics in place, the additions for MHD in SPH are as follows. First, note that we directly evolve the conservative quantities $V{\bf B}$ and $m\psi$, as in the MFM and MFV methods. As noted by \citet{price:sph.mhd.lagrangian,bate:2014.core.collapse.rad.mhd.sph}, this has several advantages over explicitly evolving ${\bf B}$ and $\psi$ -- namely, it improves the overall conservation properties of the code, eliminates significant errors associated with compression/expansion of the particles/fluid elements, reduces noise from poor particle order, allows us to write several of the key equations in manifestly anti-symmetric form (allowing conservation to be maintained even under individual time-stepping), and greatly simplifies some of the hydrodynamic calculations. Overall we find a net improvement in accuracy with the $V{\bf B}$ approach as opposed to directly evolving ${\bf B}$ in SPH; however there can be some advantages to the latter (for example, slightly reduced storage and conservation of an initial $B_{x}=0$ in 1D MHD; see \citealt{price:sph.mhd.lagrangian}). With this in mind, the primitive variables ${\bf B}$ and $\psi$ are constructed in SPH from the conserved variables as follows: 
\begin{align}
{\bf B}_{i} \equiv& \frac{(V{\bf B})_{i}}{V_{i}} \equiv \frac{\bar{\rho}_{i}}{m_{i}}\,{(V{\bf B})_{i}}\\
{\psi}_{i} \equiv& \frac{(m{\psi})_{i}}{m_{i}} \\
\bar{\rho}_{i} \equiv& \sum_{j} m_{j}\,W({\bf x}_{i}-{\bf x}_{j},\,h_{i}) 
\end{align}
where $\bar{\rho}_{i}$ is the usual SPH density estimator (constructed from neighbors, hence distinct from the actual density field evaluated at the position of particle $i$ in our MFM/MFV methods).

The conserved variable $(V{\bf B})$ is evolved according to: 
\begin{align}
\frac{d(V{\bf B})_{i}}{d t} =& \sum_{j}\,\frac{m_{i}\,m_{j}}{\Omega_{i}\,\bar{\rho}_{i}^{2}}\,\left({\bf v}_{j}-{\bf v}_{i} \right)\,\left( {\bf B}_{i}\,\cdot\,\nabla_{i}W_{ij}(h_{i})\right) \\ 
\nonumber +& \sum_{j}\,\left({\bf B}_{i}-{\bf B}_{j} \right)\,\frac{m_{i}\,m_{j}\,(\alpha^{B}_{i}+\alpha^{B}_{j})\,c^{B}_{ij}}{(\bar{\rho}_{i} + \bar{\rho}_{j})^{2}}\,\delta W_{ij} \\
\nonumber -& \sum_{j}\,m_{i}\,m_{j}\,\left[\frac{\psi_{i}}{\Omega_{i}\,\bar{\rho}_{i}^{2}}\,\nabla_{i}W_{ij}(h_{i})
+\frac{\psi_{j}}{\Omega_{j}\,\bar{\rho}_{j}^{2}}\,\nabla_{i}W_{ij}(h_{j})\right] \\
\Omega_{i} \equiv& 1 -\sum_{j}\,\frac{m_{j}}{\bar{\rho}_{i}}\,\left(W_{ij}(h_{i}) + \frac{|{\bf x}|_{ij}}{\Ddim}\,\frac{\partial W(|{\bf x}|_{ij}/h_{i})}{\partial |{\bf x}|_{ij}} \right) \\ 
\alpha^{B}_{i} \equiv& {\rm MIN}\left[ \alpha^{B}_{\rm max}\, , \,{\rm MAX}\left(  \frac{h_{i}\,|\nabla\otimes{\bf B}_{i}|}{|{\bf B}_{i}|} \, , \, \alpha^{B}_{\rm min} \right) \right] \\
|\nabla \otimes {\bf B}_{i}| \equiv& \left[ \sum_{j}\,\sum_{k}\,\left| \frac{\partial B_{i}^{(k)}}{\partial x_{j}} \right|^{2} \right]^{1/2} \\ 
\delta W_{ij}  \equiv& \,\left( \nabla_{i}W_{ij}(h_{i}) + \nabla_{i}W_{ij}(h_{j}) \right) \cdot \hat{{\bf x}}_{ij} \\
\label{eqn:cijB} c^{B}_{ij} \equiv& {\rm MAX}\left[ \frac{\alpha^{B}_{c}}{2}\left(v_{A,\,i}+v_{A,\,j} \right)\, , \,\frac{1-\alpha^{B}_{c}}{2}\,\left({v^{\rm fast}_{ij}+ v_{ji}^{\rm fast}}\right) \right]\\ 
(v^{\rm fast}_{ij})^{2} \equiv& \frac{1}{2}\,\left[{c_{s,\,i}^{2} + v_{A,\,i}^{2}} + 
\sqrt{\left( c_{s,\,i}^{2} + v_{A,\,i}^{2} \right)^{2} - {4\,c_{s,\,i}^{2}\,v_{A,\,i}^{2}\,(\hat{{\bf B}}_{i}\cdot \hat{{\bf x}}_{ij})^{2}}}\right] 
\end{align}
Here ${\bf x}_{ij} \equiv {\bf x}_{i}-{\bf x}_{j}$, $\Ddim$ is the number of spatial dimensions, $\otimes$ denotes the outer product, $\nabla \otimes {\bf B}$ is the $\Ddim\times\Ddim$ gradient matrix of ${\bf B}_{i}$ computed using our second-order consistent matrix gradient method (and $|\nabla\otimes{\bf B}|$ is the Frobenius norm of the matrix), $\hat{\bf x}$ is the unit vector ${\bf x}/|{\bf x}|$, $v_{ij}^{\rm fast}$ is the fast magnetosonic wave speed between particles, $c_{s,\,i}$ is the particle sound speed (usually computed as $c_{s,\,i} \equiv (\gamma\,P_{i}/\bar{\rho}_{i})^{1/2}$), and $v_{A}$ is the usual Alfven speed.

The first term in ${d(V{\bf B})_{i}}/{d t}$ is the induction equation. The precise form of this is derived {\em exactly} from the SPMHD Lagrangian; any other form will introduce errors in conservation and potential numerical instabilities \citep[see][]{tricco:2012.sphmhd.methods}. The $\Omega$ terms here and throughout are derived from the same Lagrangian approach following \citet{springel:entropy}, and account for variations in the ``smoothing length'' $h$ between particles.\footnote{The functional form of the $\Omega$ terms is slightly different here versus in \citet{tricco:2012.sphmhd.methods}, because we use the particle number density $n_{i}$, rather than $\bar{\rho}_{i}$, to determine the SPH smoothing length, but this has almost no effect on our results in any test problem. The two formulations are identical if particle masses are equal, which is also usually the case.} 

The second term in ${d(V{\bf B})_{i}}/{d t}$ is the artificial resistivity \citep{price:2005.sph.mhd.resistivity}. Just like artificial viscosity and conductivity in the pure hydrodynamic case (still present here), artificial dissipation terms are necessary in SPH for all hydrodynamic quantities to account for discontinuities. However unlike artificial viscosity, artificial resistivity is still needed in rarefactions to prevent numerical instability, so this term is always ``active'' between neighbors (independent of whether they are approaching or receding). The form here is motivated by (although significantly different from) the dissipation in a Reimann problem; the important aspect is the ``switch'' $\alpha^{B}$, which one would like to have a large value when there is a sharp discontinuity in ${\bf B}$, and a vanishing value in smooth flows. This is approximately accomplished by using the switch proposed in \citet{tricco:2013.sphmhd.methods}: $\alpha^{B} \propto h\,|\nabla \otimes {\bf B}| / |{\bf B}$. In \citet{tricco:2013.sphmhd.methods}, the authors show this is considerably more accurate, and less diffusive away from shocks, than the ``standard'' (constant-$\alpha^{B}$) approach.\footnote{We actually further improve on this formalism, by using our matrix-based gradients to determine $\nabla \otimes{\bf B}$. Just like with the higher-order artificial viscosity switches proposed in \citet{cullen:2010.inviscid.sph}, the use of second-order consistent gradients (as opposed to the zeroth-order inconsistent SPH gradient estimator) greatly improves the accuracy of the switch (helping it trigger in the ``correct'' locations).} 

Note that, in the artificial resistivity term, the appropriate ``signal velocity,'' $c_{ij}^{B}$, is ambiguous (the physically correct value depends on actually solving the relevant Reimann problem to determine the type of MHD shock). \citet{tricco:2013.sphmhd.methods} adopt the mean fast magnetosonic speed, \citet{price:2005.sph.mhd.resistivity} adopt the RMS Alfven speed; here we adopt a compromise. When $v_{A} \gg c_{s}$, we find the \citet{tricco:2013.sphmhd.methods} speed usually gives better results (the same conclusion they reached in their test problems). However, when $c_{s}\gg v_{A}$, and there is particle disorder (either because of motion induced by external forces or near discontinuities), the problem is that the zeroth-order SPH errors always seed non-trivial ($\sim 1\%$-level) $\alpha_{i}^{B}$, so even if there is a smooth, continuous gradient in ${\bf B}$, the resistivity is triggered and the magnetic fields are damped on a sound-crossing time. For some of the problems in this paper, for example the MHD RT and KH instabilities and the SB cluster, this suppresses the {\em mean} field by an order of magnitude, and leads to a {qualitatively} incorrect solution. This is remedied if a wavespeed which vanishes with $|{\bf B}|$ (e.g.\ a multiple of the Alfven speed) is used. We therefore allow for the use of either wavespeed in principle, with the parameter $\alpha_{c}^{B}$ in Eq.~\ref{eqn:cijB}, but adopt $\alpha_{c}^{B}=1$ as our ``default'' (i.e.\ simply set $c_{ij}^{B}$ to the mean Alfven speed). However in the tests described in \S~\ref{sec:mixing}, we consider $\alpha_{c}^{B}=0$, i.e.\ setting $c_{ij}^{B}$ to the mean fast magnetosonic speed.

The third term in ${d(V{\bf B})_{i}}/{d t}$ is the divergence-cleaning term, $\propto \nabla \psi$. The particular functional form is again Lagrangian-derived; as pointed out in \citet{tricco:2012.sphmhd.methods}, this is especially important, since not just any form of the gradient estimator can be used. Rather, it is necessary to use one which operates in appropriate conjugate pairs with the gradients used for the $\nabla \cdot {\bf B}$ estimation and pressure-gradient (hydrodynamic force) operations, or else the resulting cleaning scheme will be numerically unstable, and simply fail to clean the ``correct'' divergences.

The conserved variable $(m\psi)$ is evolved according to: 
\begin{align}
\frac{d(m\psi)_{i}}{d t} =& v_{\rm sig,\,i}^{2}\,\sigma_{h}\,\frac{m_{i}}{\Omega_{i}\,\bar{\rho}_{i}} \, \sum_{j}\,m_{j}\,\left( {\bf B}_{i}-{\bf B}_{j} \right)\cdot \nabla_{i}W_{ij}(h_{j}) \\
\nonumber &- (m\psi)_{i}\,\frac{\sigma_{p}\,v_{\rm sig,\,i}}{f_{\rm kern}\,h_{i}}
\end{align}
The first term in ${d(m\psi)_{i}}/{d t}$ is the corresponding source term for the divergence-cleaning field (the hyperbolic term). Again, the functional form inside the summation is strictly tied to the functional form of the cleaning in ${d(V{\bf B})_{i}}/{d t}$.\footnote{Here we follow the ``difference'' formulation from \citet{tricco:2012.sphmhd.methods}, which they show provides the greatest stability and minimizes errors among the formulations they compare.} The second term in ${d(m\psi)_{i}}/{d t}$ is the parabolic damping.\footnote{Note that, in closer analogy to grid-based methods, we evolve $(m\psi)$, and not, for example, $(V\psi)$ or $\psi$. This produces an essentially identical set of equations for the $\psi$ evolution as in \citet{tricco:2012.sphmhd.methods}. We do not include their additional advection term 
\begin{align}
-\frac{\psi_{i}}{2\,\Omega_{i}\,\bar{\rho}_{i}}\sum_{j}\,m_{i}\,m_{j}\,({\bf v}_{i}-{\bf v}_{j})\cdot \nabla_{i}W_{ij}(h_{i})
\end{align}
in the evolution equation for $\psi$ in \citet{tricco:2012.sphmhd.methods}. Like them, we found that this term does nothing to improve behavior on our tests, and can de-stabilize the cleaning procedure in simulations where the velocity divergence is large (e.g.\ cosmological runs), without additional timestep restrictions. However we have run almost every test in this paper with the term active and find (provided proper care is used in timestepping) that the differences are negligible. Also following \citet{tricco:2012.sphmhd.methods}, we have experimented with an artificial dissipation term for $\psi$. However, because in SPMHD there is no $\psi$ flux, we find (as these authors did) that this produces no improvement in performance on any test problems here, and only increases the numerical diffusion.} Here $f_{\rm kern}$ is a constant defined for convenience that depends on the kernel shape ($=1/2,\,1/3$ for the cubic/quartic splines), for $h$ defined as the kernel radius of compact support.

In both of these, $v_{\rm sig,\,i}$ is the {maximum} signal velocity calculated between all neighbors, as described in the text. This signal velocity is modified compared to the standard hydrodynamic case by the replacement of the sound speed $c_{s,\,i}$ with the fast magnetosonic speed $v_{ij}^{\rm fast}$ between particles. That replacement applies for all places where the signal velocity appears -- for example, in the artificial viscosity terms, and calculation of the CFL condition/timesteps.

The only remaining equation is the magnetic force: 
\begin{align}
\nonumber \frac{d(m\,{\bf v})_{i}}{d t}{\Bigr|}_{\rm B} \equiv& 
\sum_{j}\,m_{i}\,m_{j}\,\left[ \frac{{\bf M}_{i}}{\Omega_{i}\,\bar{\rho}_{i}^{2}}\cdot \nabla_{i}W_{ij}(h_{i})
+ \frac{{\bf M}_{j}}{\Omega_{j}\,\bar{\rho}_{j}^{2}}\cdot \nabla_{i}W_{ij}(h_{j}) \right] \\
-& {\bf B}_{i}\,\sum_{j}\,m_{i}\,m_{j}\,\left[ \frac{{\bf B}_{i}}{\Omega_{i}\,\bar{\rho}_{i}^{2}}\cdot \nabla_{i}W_{ij}(h_{i})
+ \frac{{\bf B}_{j}}{\Omega_{j}\,\bar{\rho}_{j}^{2}}\cdot \nabla_{i}W_{ij}(h_{j}) \right] \\
{\bf M}_{i} \equiv& \frac{{\bf B}_{i}\otimes {\bf B}_{i}}{\mu_{0}} - \frac{|{\bf B}_{i}|^{2}}{2\,\mu_{0}}\,{\bf I}
\end{align}
where ${\bf M}$ is the Maxwell stress tensor and ${\bf I}$ is the identity matrix. The first term here is the usual MHD acceleration $d{\bf v}/dt \propto \rho^{-1}\,\nabla \cdot {\bf M}$; again, the particular functional form of the gradient operator derives necessarily from the SPH Lagrangian (see \citealt{price:2005.sph.mhd.resistivity}; note it is essentially identical to the form of the Lagrangian-derived SPH hydrodynamic force in \citealt{springel:entropy}, with ${\bf M}$ replacing the pressure $P$). The second term is the \citet{borve:2001.sph.mhd.regularization} implementation of Powell 8-wave cleaning -- namely, subtracting the unphysical part of the equation of motion proportional to $\nabla \cdot {\bf B}$. As in the main text, this is necessary to prevent catastrophic numerical instability (here in the form of the tensile instability). 

There are now four numerical parameters that must be set: the artificial resistivity terms $\alpha_{\rm min}^{B}$, $\alpha_{\rm max}^{B}$, and the divergence-cleaning terms $\sigma_{p}$, $\sigma_{h}$. The divergence-cleaning parameters are discussed extensively in the text; we find a best compromise on all problems in this paper using the ``default'' values $\sigma_{h}=1$, $\sigma_{p}=0.1$.\footnote{While $\sigma_{p}\sim0.1$ is typical for mesh-based codes and appears optimal for our MFM/MFV methods, \citet{tricco:2012.sphmhd.methods} favor $\sigma_{p}=0.1$ in 2D but $\sigma_{p}=0.8-1$ (i.e.\ more rapid damping of $\psi$) in 3D for SPH. However this is based just two tests; and in at least one $\sigma_{p}\sim0.1$ actually minimizes the {maximum} value of $h\,|\nabla\cdot{\bf B}|/|{\bf B}|$. Moreover different definitions of $c_{h}$ make direct comparison difficult. While we certainly agree that larger $\sigma_{p}$ is beneficial on some tests, we find it can lead to substantially larger divergence errors on others; hence we adopt the more ``conservative'' cleaning parameter ($\sigma_{p}\sim0.1$).} As noted there, $\sigma_{h}$ is a ``nuisance'' parameter that can be folded into the definition of the cleaning speed -- we include it here only for completeness. For the artificial resistivity terms, unless otherwise specified we take $\alpha_{\rm min}^{B}=0.005$, $\alpha_{\rm max}^{B}=0.1$. We have experimented extensively with these, and find these are best compromise values. The $\alpha_{\rm max}^{B}=0.1$ choice follows \citet{tricco:2013.sphmhd.methods}; a much larger value (e.g.\ $\alpha_{\rm max}^{B}\sim1$) completely diffuses away the fields in several of our test problems (e.g.\ the RT instability, MHD rotor, SB cluster, protostellar core collapse) and dramatically over-smooths shock jumps in others (e.g.\ the Zeldovich and Toth problems), leading to systematically incorrect solutions. But if $\alpha_{\rm max}^{B}\ll 0.1$, the method is incapable of properly capturing strong, magnetically-dominated shocks. The lower limit is less important; $\alpha_{\rm min}^{B}\ll 0.1$ is important to prevent excess diffusion that suppresses field growth in e.g.\ the proto-stellar disk problem, but $\alpha_{\rm min}^{B}>0$ greatly reduces the noise and post-shock oscillations that seed low-order SPH errors. 

This is sufficient for SPMHD. However, in running a large suite of shocktube tests (at the suggestion of the referee), it became clear that our default implementation of artificial viscosity (described in detail in \paperone\ and taken from the ``inviscid SPH'' prescription in \citealt{cullen:2010.inviscid.sph}) is not ideal for some MHD problems. Most dramatically, in the \citet{brio:1988.shocktube} shocktube, with small neighbor number (``SPH-lo''), adopting the prescription from \citet{cullen:2010.inviscid.sph} (or the slightly modified forms in \citealt{hu:2014.psph.galaxy.tests} or \paperone) leads to very large oscillations in the post-shock velocity over the domain where the internal energy is large. As shown in \paperone, on pure-hydro problems the method behaves well. The issue in MHD appears to be insufficient viscosity in regions with sub-sonic noise when the accelerations are primarily transverse. The simplest solution is to enforce a constant artificial viscosity, but this seriously degrades the performance of SPH on other problems (hence the reason for these switches). After some experimentation we adopt the following compromise between excessive diffusion and noise: the functional form of the artificial viscosity follows the hydro case in \paperone\ (Eq.~F16), except we replace the sound speed with the fast magnetosonic speed and increase the minimum viscosity from $\alpha_{\rm min}=0.02$ to $\alpha_{\rm min}=0.05$. As well, the dimensionless parameter $\alpha_{0,\,i}$ is set by
\begin{align}
\nonumber \alpha_{\rm tmp} &= 
\begin{cases}
	{\displaystyle 0\ \ \ \ \  \hfill { ((d[\nabla\cdot{\bf v}]/dt)_{i} \ge 0\ , \ \ \ {\rm or}\ \ \ \  (\nabla\cdot{\bf v})_{i} \ge 0)}} \\
	{\displaystyle \, \, }\\
	{\displaystyle \frac{\alpha_{\rm max}\,|(d[\nabla\cdot{\bf v}]/dt)_{i}|}{|(d[\nabla\cdot{\bf v}]/dt)_{i}| + (\tilde{c}/\tilde{h})^{2}}
	\ \ \ \ \ \hfill { ({\rm otherwise})}} 
\end{cases}\\
\nonumber \\
\alpha_{0,\,i}(t+\Delta t) &= 
\begin{cases}
	{\alpha_{\rm tmp}\ \ \ \ \ \hfill { (\alpha_{\rm tmp} \ge \alpha_{0,\,i}(t))}} \\
	{\displaystyle \, \, }\\
	{\alpha_{\rm tmp} + (\alpha_{0,\,i}(t)-\alpha_{\rm tmp})\,e^{-\beta_{\rm d}\,\Delta t\,|v_{{\rm sig},\,i}|/(2\,\tilde{h}) }}\\
	{\ \ \ \ \ \hfill { (\alpha_{\rm tmp} < \alpha_{0,\,i}(t))}} \\
\end{cases}
\end{align}
where in \paperone\ we took $\tilde{c}=0.7\,c_{s,\,i}$ and $\tilde{h}=f_{\rm kern}\,h_{i}$. Here, we use
\begin{align}
\tilde{h}^{2} &\rightarrow \left( f_{\rm kern}\,h_{i} \right)^2 + \frac{ \| {\bf \tilde{v}}_{i} \|^{2} }{ \| \nabla\otimes {\bf v}_{i} \|^{2} } \\ 
\tilde{c} &\rightarrow 0.2\,\left( c_{s,\,i}^{-1} +  \| {\bf \tilde{v}}_{i} \|^{-1} \right)^{-1}
\end{align}
where $\|{\bf \tilde{v}}\|$ denotes the Frobenius norm of ${\bf \tilde{v}}$ and $\| {\bf \tilde{v}} \| = \| {\bf v} \|$. The prescription from \paperone\ has the effect that the viscosity vanishes very quickly whenever the mean velocity gradient is resolved and/or the compressive accelerations are sub-sonic. With the modifications above, a velocity field which has large fractional noise or rate-of-change of the velocity divergence (relative to the velocity itself) will also ``trigger'' viscosity (even if the flow is sub-sonic or some of the noise is transverse). This is sufficient to significantly reduce the noise in the \citet{brio:1988.shocktube} problem, while having only small effects on almost every other problem here. It does, however, somewhat degrade performance (via larger numerical viscosities) on some problems like the rotating Keplerian disk in \paperone. Moreover, as written here, this term violates Galilean invariance; we find this can be restored with similar (slightly more noisy) results by instead using $\|{\bf \tilde{v}}\| = \alpha_{v}\,{\rm MAX}_{j}(\| {\bf v}_{i}-{\bf v}_{j} \|)$ where ${\rm MAX}_{j}$ refers to the maximum among neighbors and $\alpha_{v}\approx 10$. 

This completes the SPMHD implementation.

\vspace{-0.5cm}
\section{On the Use of Variable Wavespeeds in Divergence-Cleaning Operators}
\label{sec:non.constant.dedner}

In \citet{dedner:2002.divb.cleaning.scheme}, the authors showed that their divergence-cleaning method is numerically stable and guaranteed to reduce $|\nabla\cdot{\bf B}|$. However, strictly speaking, their proofs are valid only if the wave and damping speeds $c_{h}$ and $\tau$ ($\tau\equiv c_{p}^{2}/c_{h}^{2}$ in their notation) are {\em constant in both time and space}. 

In that case, if we neglect any other fluid forces and evolve the system only under the influence of the mixing terms in ${\bf B}$ and $\psi$, the evolution equations for $\psi$ and $\nabla\cdot{\bf B}$ take the form: 
\begin{align}
\label{eqn:dedner0} 0&=\frac{\partial^{2}\psi}{\partial t^{2}} + \frac{1}{\tau}\,\frac{\partial \psi}{\partial t} - c_{h}^{2}\,\nabla^{2}\psi  \\ 
\label{eqn:dedner1} &= \frac{\partial^{2}(\nabla\cdot{\bf B})}{\partial t^{2}} + \frac{1}{\tau}\,\frac{\partial (\nabla\cdot{\bf B})}{\partial t} - c_{h}^{2}\,\nabla^{2}(\nabla\cdot{\bf B}) 
\end{align}
i.e.\ both $\psi$ and $(\nabla\cdot{\bf B})$ obey a damped wave equation. 

However, as discussed in the text, it is desirable to vary $c_{h}$ and $\tau$. Imposing a constant $c_{h}$ equal to some global maximum wavespeed essentially forces all elements (cells/particles) to use a constant, global timestep (a tremendous numerical cost), because they must all satisfy the Courant (CFL) condition for the wave equations above (otherwise $|\nabla\cdot {\bf B}|$ can grow unstably).\footnote{The CFL condition for Eq.~\ref{eqn:dedner0}-\ref{eqn:dedner1} requires {\em all} elements have timesteps $\Delta t_{i} \ll \tau \sim {\rm MIN}(h_{j}) / {\rm MAX}(v_{{\rm sig},\,j})$ (the criterion in $c_{h}$ is slightly less demanding). This is equal to or smaller than the normal minimum timestep for any single element in the simulation, ${\rm MIN}(\Delta t_{i}) \sim {\rm MIN}( h_{i} / v_{{\rm sig},\,i} )$.} Moreover in many problems the ``fastest'' wave speed may reside in regions which have wildly different physical properties and are causally disconnected from others. This also leads to wavespeeds which dramatically exceed local physical signal velocities. 

If we allow $c_{h}$ and $\tau$ to depend on space and time, following the same derivation as in \citet{dedner:2002.divb.cleaning.scheme} gives:
\begin{align}
\label{eqn:psi.eqn.wderiv} \frac{\partial^{2}\psi}{\partial t^{2}} &+ \frac{1}{\tau({\bf x},\,t)}\,\frac{\partial \psi}{\partial t} - c_{h}^{2}({\bf x},\,t)\,\nabla^{2}\psi
=\\
\nonumber & \left(\frac{\partial \psi}{\partial t}+\frac{\psi}{\tau}\right)\,\frac{1}{c_{h}^{2}}\,\frac{\partial c_{h}^{2}}{\partial t} - \psi\,\frac{\partial}{\partial t}\left(\frac{1}{\tau}\right) \\
\label{eqn:b.eqn.wderiv} \frac{\partial^{2}(\nabla\cdot{\bf B})}{\partial t^{2}} &+ \frac{1}{\tau({\bf x},\,t)}\,\frac{\partial (\nabla\cdot{\bf B})}{\partial t} - c_{h}^{2}({\bf x},\,t)\,\nabla^{2}(\nabla\cdot{\bf B}) = \\
\nonumber & \psi\,\nabla^{2}\left(\frac{1}{\tau}\right) + 2\,\nabla\psi\cdot\nabla\left(\frac{1}{\tau}\right) + \\
\nonumber & (\nabla\cdot{\bf B})\,\nabla^{2}(c_{h}^{2}) + 2\left[\nabla(\nabla\cdot{\bf B})\right]\cdot\nabla (c_{h}^{2})
\end{align}
Now the equations have time-and-space dependent coefficients, and source terms dependent on derivatives of $c_{h}$ and $\tau$. And unfortunately, these extra terms cannot be eliminated by simply modifying the original source terms in the \citet{dedner:2002.divb.cleaning.scheme} scheme.\footnote{Consider e.g.\ modifying the \citet{dedner:2002.divb.cleaning.scheme} assumption that the correction term scales as $\partial {\bf B}/\partial t = -\nabla \psi$ with the insertion of an arbitrary function $f$ such that $\partial {\bf B}/\partial t = -f({\bf x},\,t,\psi,\,(\nabla\cdot{\bf B}))\,\nabla \psi$, or adopting alternative operators $\mathcal{D}$, $g$ defined such that $\mathcal{D}(\psi) + g(\nabla\cdot{\bf B}) = 0$ (in their formulation, $g=1$, $\mathcal{D}=c_{h}^{-2}\,\partial/\partial t + \tau^{-1}$). If we do this and attempt to recover Eq.~\ref{eqn:dedner0}, one can show that the correction terms can simply be folded into a new field $\psi^{\prime}$ that obeys Eq.~\ref{eqn:dedner1} with constant coefficients. If one desires $(\nabla\cdot {\bf B})$ to obey a damped wave equation with constant coefficients, and to build the ``correction terms'' out of linear operators acting on $(\nabla \cdot {\bf B})$ and any arbitrary field $\psi$, then the \citet{dedner:2002.divb.cleaning.scheme} scheme is the most general possible solution.} 

So consider the extra terms. First, take $c_{h}=c_{h}(t)$, $\tau=\tau(t)$, i.e.\ the coefficients are spatially constant, but change in time. This corresponds to the original implementation proposed by \citet{dedner:2002.divb.cleaning.scheme}, and most subsequent work. This is generally not a problem. The time-derivative terms appear only in the $\psi$ equation (Eq.~\ref{eqn:psi.eqn.wderiv}), which has no direct physical consequence. Eq.~\ref{eqn:b.eqn.wderiv} for $\nabla\cdot{\bf B}$ remains a damped-wave equation, but with time-dependent coefficients. So long as their time variation is sufficiently slow, one can apply the usual Wentzel-Kramers-Brillouin (WKB) approximation and show that $\nabla\cdot {\bf B}$ still behaves as a damped wave. Because $c_{h}$ is chosen to be a maximum wavespeed in the domain, and $1/\tau$ some maximum damping rate $\sim c_{h}/{\rm MIN}(h_{i})$, the variation of $c_{h}$ and $\tau$ in time (which evolve with the physical properties of the system such as $\rho$, ${\bf B}$, etc.) will always be slow compared to the local evolution timescale for the damping wave (provided the system obeys the CFL condition in the first place), and this is easily satisfied.\footnote{We caution that if the time-variations of $c_{h}$ and $\tau$ are large on a wave-crossing time (e.g.\ $c_{h}^{-1}\,| \partial c_{h} / \partial t | \gtrsim c_{h} / |L|$, where $|L|\sim |\nabla\cdot{\bf B}|/|\nabla(\nabla\cdot {\bf B})| \sim h_{i}$), the WKB approximation is invalid and $|\nabla\cdot {\bf B}|$ can converge to a constant or grow. If for example $c_{h}$ evolves on some dynamical time $t_{\rm dyn}$, the requirement for stable behavior is $c_{h} \gtrsim h_{i}/t_{\rm dyn}$. This is another reason to choose $c_{h}$ to be the maximum (local) wavespeed (which always satisfies this). Choosing a slower wavespeed, even if uniform in space, can de-stabilize the cleaning.}

Similarly, if $c_{h}$ and $\tau$ depend on position, we will recover the desired behavior so long as $c_{h}({\bf x})$ and $\tau({\bf x})$ are {\em sufficiently smooth}. From Eq.~\ref{eqn:b.eqn.wderiv}, the corrections terms do not change the behavior of the system if $|\nabla\tau|/\tau$ and $|\nabla c_{h}|/c_{h}$ are $\lesssim |\nabla(\nabla\cdot{\bf B})|/|\nabla\cdot{\bf B}|$.\footnote{More precisely, a 1D analysis following \citet{dedner:2002.divb.cleaning.scheme} gives the following sufficient (although not strictly necessary) criteria for stability of the divergence-damping: 
\begin{align}
\frac{\partial_{x} (c_{h}^{2}\tau)}{c_{h}^{2}\tau} &< \left| k \right| 
\ , \ \frac{\partial_{x} (c_{h}^{2})}{c_{h}^{2}} < \left(\frac{1}{2}+\frac{1}{|k|^{2}\,c_{h}^{2}\,\tau^{2}}\right)\,\left| k \right|
 \ \ \ \ \ \hfill {\left( k < 0 \right)} \\ 
 \frac{\partial_{x} (c_{h}^{2}\tau)}{c_{h}^{2}\tau} &> -\left| k \right| 
\ , \ \frac{\partial_{x} (c_{h}^{2})}{c_{h}^{2}} > -\left(\frac{1}{2}+\frac{1}{|k|^{2}\,c_{h}^{2}\,\tau^{2}}\right)\,\left| k \right|
 \ \ \ \ \ \hfill {\left( k > 0 \right)} \\ 
 k &\equiv \frac{\partial_{x}(\nabla \cdot {\bf B})}{(\nabla \cdot {\bf B})}\ , \ \ \ \ \ \ \partial_{x}\equiv \frac{\partial}{\partial x}
\end{align}
i.e., the sign of $\partial_{x}[c_{h}^{2}\tau\,(\nabla\cdot{\bf B})]$ (and $\partial_{x}[c_{h}^{2}\,(\nabla\cdot{\bf B})]$) must match the sign of $\partial_{x}(\nabla\cdot{\bf B})$. This is particularly easy to see if we consider just the parabolic term in a 1D case (where any deviation from $B_{x}=B_{0}=$\,constant represents $\nabla\cdot{\bf B}\ne 0$). In this case we obtain the heat equation: $\partial_{t} B_{x} = \partial_{x}( c_{h}^{2}\,\tau\,\partial_{x}\,B_{x})$, and deviations from $B_{0}$ are diffused away, provided the condition above is met. Physically, the stability condition corresponds to the requirement that the spatial dependence of $c_{h}$ and $\tau$ does not introduce new local extrema in $c_{h}^{2}\,\tau\,\nabla\cdot{\bf B}$ that are not present in $\nabla\cdot{\bf B}$. These conditions are obviously satisfied (in 3D) if we have $|\nabla c_{h}^{2}|/c_{h}^{2}$ and $|\nabla c_{h}^{2}\,\tau|/c_{h}^{2}\,\tau$ much less than $|\nabla(\nabla\cdot{\bf B})|/|\nabla\cdot{\bf B}|$, our (more restrictive) criterion above.}

We expect and show in the text that (since it is a local numerical error term) $|\nabla\cdot{\bf B}|/|\nabla(\nabla\cdot{\bf B})| \sim h_{i}$; so our choice of $c_{h}$ and $\tau$ (which depend on local physical properties) should guarantee this condition is satisfied wherever the flow is resolved.\footnote{Note that $c_{h}$ appears in two places: in the Riemann problem (the first-step update of the normal component of ${\bf B}$) and in the source term for $\psi$. In the former, we are solving the one-dimensional Riemann problem in operator-split fashion, so the only states that matter are the two faces -- we can either enforce a constant $c_{h}$ within each Riemann problem (our default approach described in the text) or explicitly solve for the two-$c_{h}$ system (described in \S~\ref{sec:appendix:twowave.bbar}; since the waves propagate away from the discontinuity at the face into media with constant $c_{h}$ on either ``side,'' there is no instability).} However, caution is needed when the flow is poorly-resolved, especially in multi-phase, turbulent systems (where physical properties vary rapidly). By linking $c_{h,\,i}$ and $\tau_{i}$ to the maximum signal velocity among interacting neighbors, as opposed to speeds at $i$ alone, we maintain smoothness even with kernel-scale noise. 

As a result, we confirm in our tests that $|\nabla\cdot{\bf B}|$ does not grow unstably to levels that swamp the physical solutions. However, one could probably improve the cleaning (especially in noisy, multiphase flows) by doing more to ensure $c_{h}({\bf x})$ and $\tau({\bf x})$ remain sufficiently smooth. For example, we have experimented with calculating $c_{h,\,i}$ and $\tau_{i}$ following our standard method in a first particle sweep, then computing a kernel-weighted average $\langle c_{h} \rangle_{i}$, $\langle \tau \rangle_{i}$ on a second sweep; this ensures smoothness on a super-kernel scale. Alternatively, we have considered an effective slope-limiter which limits the magnitude of $c_{h,\,i}^{2}$ and $\tau_{i}$ such that among interacting neighbors, the particles which are local (kernel-scale) extrema in $\nabla\cdot{\bf B}$ remain so in $c_{h}^{2}\,\nabla\cdot{\bf B}$ and $c_{h}^{2}\,\tau\,\nabla\cdot{\bf B}$. Although there are hints of minor improvement on a couple of problems, we do not find these make a large difference and have not experimented with them extensively. However, the issue merits investigation in future work.

\vspace{-0.5cm}
\section{Cosmological Integration of Divergence-Cleaning}
\label{sec:cosmological}

The modifications in our code necessary for cosmological integrations are described in detail in \paperone. These are, for the most part, unchanged. As described therein, it is easy to show that if we define appropriate co-moving units, the cosmological expansion is automatically handled, and the Riemann problem is locally unchanged, provided we convert into physical variables before solving it. This is identical in MHD. For the magnetic field ${\bf B}$, we consider the co-moving field ${\bf B}_{c} = a^{2}\,{\bf B}$, where $a=1/(1+z)$ is the usual scale factor. This is invariant under pure adiabatic expansion in a Hubble flow in ideal MHD. Since we evolve the conserved variable $(V{\bf B})$, and the volume/length units are also co-moving, we simply have $(V{\bf B})_{c} = a^{-1}\,(V\,{\bf B})$. For the divergence-cleaning field $\psi$, the ``correct'' co-moving units are slightly more ambiguous; how $\psi$ behaves in a Hubble-flow expansion actually depends on the ratio of timescales (whether, for example $c_{s}>v_{A}$, because $\psi$ depends on the fastest wavespeed). We have experimented with simply assuming $\psi_{c} = a^{3}\,\psi$ (the appropriate choice in the typical cosmological case, when either $c_{s}$ or the inter-particle fluid velocity $|{\bf v}|$ is much larger than $v_{A}$), or $\psi_{c}=a^{5/2}\,\psi$ (appropriate if $v_{A}$ dominates), or explicitly solving for the evolution terms from expansion; in practice we find this makes no detectable difference to any problem, since the physical $\psi$-field growth and decay time to respond to evolving $\nabla\cdot {\bf B}$ ($\sim h/c_{h}$, where $h$ is a resolution element and $c_{h}$ the fastest wave speed) is always vastly shorter than the cosmological expansion/Hubble time.

\vspace{-0.5cm}
\section{The Damping Speed}
\label{sec:damping.speed}

In \S~\ref{sec:methods}, we note that the divergence-wave is damped with a source term $(m\psi)_{i}/\tau_{i}$, where usually $\tau_{i} \equiv h_{i}/(\sigma_{p}\,c_{\tau,\,i})$ with $c_{\tau,\,i}$ the damping wave speed. There is no {\em a priori} obvious choice for this speed, and the ability of the scheme to damp divergence does not depend sensitively on the choice, provided it satisfies the conditions in \S~\ref{sec:non.constant.dedner}. Of course, much too large a value compared to other characteristic speeds in the problem will mean that $\psi$ cannot grow (and therefore cannot remove divergences), while much too small a value can lead to a ``lag'' in the response of the cleaning to $\divB$. 

We have therefore considered a variety choices: 
\begin{align}
c_{\tau,\,i} &= \frac{1}{2}\,v_{{\rm sig},\,i}^{\rm MAX} \\
c_{\tau,\,i} &= v_{f\psi,\,i} \equiv \sqrt{ c_{s,\,i}^{2} + v_{A,\,i}^{2} + \left(\frac{2\,\psi_{i}}{v_{{\rm sig},\,i}^{\rm MAX}} \right)^{2} } \\
c_{\tau,\,i} &= v_{\rm fastest} \equiv {\rm MAX}_{i}\left[ \frac{1}{2}\,v_{{\rm sig},\,i}^{\rm MAX}\, , \, \left( c_{s,\,i}^{2} + v_{A,\,i}^{2} \right)^{1/2} \right] \\ 
c_{\tau,\,i} &= f_{\rm kern}\,h_{i}\,\tau^{-1}_{\rm fastest} \\
& \tau^{-1}_{\rm fastest} \equiv {\rm MAX}_{i}\left[ \frac{v_{{\rm sig},\,i}^{\rm MAX}}{2\,f_{\rm kern}\,h_{i}}\, , \, \frac{1}{f_{\rm kern}\,h_{i}}\,\left( c_{s,\,i}^{2} + v_{A,\,i}^{2} \right)^{1/2} \right] \\
\label{eqn:ctau} c_{\tau,\,i} &= {\rm MAX}\left[ \frac{1}{2}\,v_{{\rm sig},\,i}^{\rm MAX}\, , \, v_{f\psi,\,i} \, , \, \epsilon_{h}\,v_{\rm fastest}  \right]
\end{align}

The first is the standard signal velocity $v_{{\rm sig},\,i}^{\rm MAX}/2$ (this is the default choice in SPH MHD). We find this works well in every problem here, although in some cases it produces excess dissipation of the fields because it operates too slowly for some subset of particles when $\psi$ is very large (since the speed at which $\psi$ can induce changes in ${\bf B}$ is not accounted for with this velocity).

The second ($v_{f\psi,\,i}$) is the particle-based fastest-possible magnetosonic speed, with the additional term $2\,\psi_{i}/v_{{\rm sig},\,i}^{\rm MAX}$. This term can be thought of as representing the ``potential magnetic energy'' in the $\psi$ field -- while usually negligible compared to the Alfven speed $v_{A}$, it is certainly possible, on some problems, that $\psi$ grows until it reaches values $|\psi_{i}| \gg |{\bf B}| / |v_{\rm sig}|$ -- this is clearly in the limit where damping of the $\psi$ field is desired. Therefore we find this choice works much better than the pure magnetosonic speed. This choice works comparably well to $v_{{\rm sig},\,i}^{\rm MAX}$, but is less than ideal in some cases where the particles have super-sonic local approach velocities (not accounted for here). 

The third choice ($v_{\rm fastest}$) is similar to the default choice in \citet{dedner:2002.divb.cleaning.scheme}, namely the fastest wavespeed and/or signal velocity in the entire domain. This works well on idealized test problems -- including almost all of the tests in this paper; however, it makes little sense for high-dynamic range problems like cosmological or galaxy/star formation simulations. In those cases the medium is highly multi-phase, so there is a huge range of local fastest wavespeeds, often in regions which are not even in causal contact. We therefore find that this produces too-efficient damping (hence less-efficient cleaning) in the slowly-evolving regions of these problems. 

The fourth choice ($\tau_{\rm fastest}$) is similar to the maximum wavespeed $v_{\rm fastest}$, but instead sets $\tau_{i}$ directly to the minimum across all particles (independent of whether they have small/large local volumes $h_{i}$). In a uniform-grid code this is identical to $v_{\rm fastest}$. Here we find it works comparably well, again in idealized test problems, but has the same problems in inherently multi-scale problems.

The final choice, therefore, represents our best attempt at a ``compromise'' between these. We take the maximum of either the signal velocity $v_{{\rm sig},\,i}^{\rm MAX}/2$, the local magnetosonic speed (plus $\psi$) $v_{f\psi,\,i}$, and some multiple with $\epsilon_{h}\ll 1$ of $v_{\rm fastest}$. We adopt this as our default for all problems here, with $\epsilon_{h}=0.01$. However we stress that {\em all} of our qualitative results are robust to any of the choices above for $c_{\tau,\,i}$ -- we find only minor quantitative differences (in many of the test problems here, these are completely indistinguishable).

\vspace{-0.5cm}
\section{Additional Flux-Limiters for Divergence-Cleaning Operations}
\label{sec:limiters}

In \paperone\ we describe our slope-limiting procedure for reconstruction; the same is used here. However, in a couple of cases (e.g.\ the isolated galaxy disk and Santa Barbara cluster), some additional flux limiters greatly help in improving numerical stability. These are specific to the divergence-cleaning ($\psi$) terms, so do not directly affect our reconstruction of the physical quantities. They will alter how efficiently the $\divB$ errors are cleaned; while this is important, it is not the only consideration for stability.

In particular, numerical instability can arise owing to the ``mixing'' terms between $\psi$ and $B$ which appear in the updated values $\bar{B}_{x,\,ij}^{\prime}$ and $\bar{\psi}_{ij}$ in the Riemann problem (see Eq.~\ref{eqn:b.normal}). In the limit where there is large particle disorder (e.g.\ poorly-resolved, turbulent shocks) and the particles are being rapidly re-arranged by non-MHD forces, or if boundary particles are only able to find a couple of neighbors (if vacuum boundaries, which are not recommended for MHD, are used) then one can have $|\psi| \gg v_{\rm sig}^{\rm MAX}\,|{\bf B}_{i}|$ and the implicit instantaneous update of $\bar{B}$ ($\sim (\psi_{L}-\psi_{R})/ c_{h}$) in the Riemann problem can be unstable (since small residuals in $\psi$ which are not completely damped by the time particles locally re-arrange can lead to large changes in $\bar{B}$); this is related to the discussion in \S~\ref{sec:non.constant.dedner}.

One solution would be to simply drop these terms. However, this prevents the divergence-cleaning from acting across single-particle discontinuities, which (as discussed in the text) can lead to incorrect jumps. We find a more accurate, robust, and flexible solution, which works well for all problems in this paper, is to simply apply an additional set of limiters for $\psi$. 

In the Riemann problem, we modify Eq.~\ref{eqn:b.normal} to be:
\begin{align}
\bar{B}_{x,\,ij}^{\prime} &= \frac{1}{2} \left( B_{x,\,L}^{\prime} + B_{x,\,R}^{\prime} \right) + \frac{\alpha_{\psi,\,ij}}{2\,\tilde{c}_{h,\,ij}}\,\left( \psi_{L} - \psi_{R} \right) \\ 
\bar{\psi}_{ij} &= \frac{1}{2} \left( \psi_{L} + \psi_{R} \right) + \bar{\psi}_{ij}^{B} \\ 
\nonumber \bar{\psi}_{ij}^{B} &\equiv  \alpha_{\psi,\,ij}\,\frac{\tilde{c}_{h,\,ij}}{2}\,\left( B_{x,\,L}^{\prime} - B_{x,\,R}^{\prime} \right) \\ 
\alpha_{\psi,\,ij} &\equiv {\rm MIN}\left[ 1,\ 
\alpha_{\psi}^{0}\,\frac{\tilde{c}_{h,\,ij}\,|B_{x,\,L}^{\prime} + B_{x,\,R}^{\prime} |}{|\psi_{L} - \psi_{R}|} 
\right]
\end{align}
This is a flux-limiter on the implicit 1D Riemann problem between $B^{\prime}$ and $\psi$ at a discontinuity (or equivalently, we can think of it as limiting the slope of the $\psi$ discontinuity). The coefficient $\alpha_{\psi}^{0}$ should be $<1$ for stability; the precise value is (like all limiters) set by a balance between stability and diffusion. Our experiments prefer $\alpha_{\psi}^{0}=0.75$. 
For the source terms for ${\bf B}$ which are $\propto (V\nabla\psi)^{\ast}_{i}$ in Eq.~\ref{eqn:source.discrete}, we also modify
\begin{align}
(V\nabla\psi)^{\ast}_{i} &\equiv -\sum_{j} \bar{\psi}_{ij}\,{\bf A}_{ij} 
\rightarrow (V\nabla\psi)^{\ast}_{i,\,0} + \alpha_{\psi,\,i}^{B}\,(V\nabla\psi)^{\ast}_{i,\,B} \\ 
 (V\nabla\psi)^{\ast}_{i,\,0} &\equiv -\sum_{j}\frac{\left( \psi_{L} + \psi_{R} \right)_{ij}}{2}{\bf A}_{ij} \\
 (V\nabla\psi)^{\ast}_{i,\,B} &\equiv -\sum_{j}\bar{\psi}_{ij}^{B}\,{\bf A}_{ij} \\ 
\alpha_{\psi,\,i}^{B} &\equiv {\rm MIN}\left[ 1,\ \frac{10\,\zeta^{2}}{|(V\nabla\psi)^{\ast}_{i,\,B}|^{2}} \right] \\
\zeta &\equiv {\Bigl |} \frac{d(V{\bf B})_{i}}{dt_{0}}{\Bigr|}^{2} + {\Bigl |} 0.1\,\frac{0.5\,v_{{\rm sig},\,i}\,(V{\bf B})_{i}}{h_{i}} {\Bigr|}^{2}
\end{align}
where $d(V{\bf B})_{i}/dt_{0}$ represents the value of $d(V{\bf B})_{i}/dt$ calculated for element $i$ including all other fluxes and source terms {\em except} $(V\nabla\psi)^{\ast}_{i,\,B}$. Similarly, in the source term for $\psi$ (Eq.~\ref{eqn:source.discrete}), $d\psi/dt \propto (V\divB)^{\ast}_{i}\,c_{h,\,i}^{2}$, we limit the effective value of $(V\divB)^{\ast}_{i}$ allowed to a maximum $=100\,V_{i}\,|{\bf B}_{i}| / h_{i}$. The pre-factor is of course arbitrary but should be $\gg 1$. Finally, we check each timestep whether $|\psi_{i}| > \alpha\,v_{\rm max}\,|{\bf B}_{i}|$ with $\alpha = 10 \gg 1$ and $v_{\rm max} = {\rm MAX}(0.5\,v_{{\rm sig},\,i},\,v_{\rm fastest},\,\sqrt{c_{s,\,i}^{2}+v_{A,\,i}^{2}})$; if it exceeds this value, we impose $d\psi/dt={\rm MAX}(0,\,d\psi/dt)$ ($\psi<0$) or $d\psi/dt={\rm MIN}(0,\,d\psi/dt)$ ($\psi>0$). This just corresponds to increasing the (already arbitrary) $\psi$-damping rate super-linearly when $\psi$ becomes very large.

Even in the Santa Barbara and galaxy disk problems, these limiters almost never act. Usually, when they do, the particles are in a situation (e.g.\ at vacuum boundaries) where the fluxes are unresolved and should not, in any case, be trusted. Therefore, the fact that this limiting procedure allows somewhat higher $\divB$ errors (by making the $\psi$-based cleaning less aggressive) is a small price to pay for maintaining numerical stability.

\vspace{-0.5cm}
\section{A Two-Wave Formulation of the Divergence Cleaning Terms in the Riemann Problem}
\label{sec:appendix:twowave.bbar}

As discussed in the text, in Eq.~\ref{eqn:b.normal}, the normal component of ${\bf B}$ and the divergence-cleaning term $\psi$ are implicitly updated according to the solution of an independent one-dimensional Riemann problem, before solving the MHD Riemann problem. The solution given there assumes a single wavespeed $\tilde{c}_{h,\,ij}={\rm MAX}\left[ v_{{\rm f},\,L}\, , \,v_{{\rm f},\,R}  \right]$ (the maximum of the fast magnetosonic speed on left and right sides of the problem) for the divergence-cleaning wave at the discontinuity between the left and right states. Following a solution proposed by E.\ Gaburov (private communication), we could instead assume two independent wavespeeds, ${c}_{L} = v_{{\rm f},\,L}$ and ${c}_{R} = v_{{\rm f},\,R}$ on either side of the discontinuity. This yields the solution
\begin{align}
\label{eqn:b.normal.2wave}\bar{B}_{x,\,ij}^{\prime} &= \frac{1}{c_{L}+c_{R}} \left[ c_{L}\,B_{x,\,L}^{\prime} + c_{R}\,B_{x,\,R}^{\prime} +  \psi_{L} - \psi_{R} \right] \\ 
\bar{\psi}_{ij} &= \frac{1}{c_{L}+c_{R}} \left[ c_{R}\,\psi_{L} + c_{L}\,\psi_{R}  + c_{L}\,c_{R}\,\left( B_{x,\,L}^{\prime} - B_{x,\,R}^{\prime} \right)\right] 
\end{align}
This trivially reduces to the solution in Eq.~\ref{eqn:b.normal} for $c_{L}=c_{R}$. We have considered this formulation, instead of the default in the text, for all test problems in this paper. In all cases, the differences are small. The two-wave formulation introduces some additional dissipation and/or grid noise, depending on the problem, however it is also more stable in situations with large particle disorder (essentially because it provides an up-wind weighting of the $\psi$ and normal-${\bf B}$ terms).

\end{appendix}

\end{document}